\PassOptionsToPackage{table}{xcolor}
\documentclass[a4paper,12pt]{article}

\usepackage{jheppub} 

\usepackage[T1]{fontenc} 

\usepackage[percent]{overpic}
\usepackage{slashed}
\usepackage{wrapfig}
\usepackage{tabu}
\usepackage{mathrsfs,amsmath,amssymb,amsthm,amsfonts,tikz,graphicx,accents,hyperref, color}
\usepackage{dsfont,epiolmec, latexsym, stmaryrd, comment}
\usepackage{slashed,ccaption}
\usepackage{mathrsfs, calligra}
\usepackage{leftidx}
\usepackage{import}
\usepackage{multirow}
\usepackage{amsfonts}
\usepackage{pifont}
\usepackage{tabularx}
\usepackage[utf8]{inputenc}
\usetikzlibrary{intersections,calc}
\usepackage{tikz-3dplot}
\usepackage{ifthen}
\usetikzlibrary{arrows}
\usepackage{amsmath}
\usepackage{xcolor}
\usepackage{geometry}

\usepackage{array}
\DeclareFontFamily{OT1}{rsfs}{}

\DeclareFontShape{OT1}{rsfs}{m}{n}{ <-7> rsfs5 <7-10> rsfs7 <10->rsfs10}{} 

\DeclareMathAlphabet{\mycal}{OT1}{rsfs}{m}{n}

\newcommand{\bms}{{$\mathfrak{bms}_3$}}
\newcommand{\hbms}{$\widehat{\mathfrak{bms}}_3$}
\newcommand{\vir}{{$\mathfrak{vir}$}}

\newcommand{\virtwo}{{$\mathfrak{vir}^2$}}
\newcommand{\sltwo}{{\mathfrak{sl}(2,\mathbb{R})}}
\newcommand{\iso}{{\mathfrak{iso}(2,1)}}

\newcommand{\eps}{\varepsilon}

\newcommand{\be}{\begin{equation}}
\newcommand{\ee}{\end{equation}}
\newcommand{\xmark}{\ding{55}}
\tdplotsetmaincoords{48}{112}
\makeatletter \@addtoreset{equation}{section}

\newcommand\snote[1]{\textcolor{blue}{\bf [Sh-J:\,#1]}}
\preprint{IPM/P-2018/071}

\title{\centerline{\boldmath \emph{On Rigidity of 3d Asymptotic Symmetry Algebras}}}

\author[a,b]{A. Farahmand Parsa}
\author[c]{, H. R. Safari}
\author[c]{and M. M. Sheikh-Jabbari}

\affiliation{$^a$ School of Mathematics, Institute for Research in Fundamental Sciences (IPM),\\ P.O. Box: 19395-5746, Tehran, Iran.}
\affiliation{$^b$School of Mathematics, TATA Institute of Fundamental Research (TIFR),\\ 400 005, Mumbai, India.}
\affiliation{$^c$ School of Physics, Institute for Research in Fundamental
Sciences (IPM),\\ P.O.Box 19395-5531, Tehran, Iran}

\emailAdd{a.parsa@ipm.ir}
\emailAdd{hrsafari@ipm.ir}
\emailAdd{jabbari@theory.ipm.ac.ir}

\abstract{
We study rigidity and stability of infinite dimensional algebras which are not subject to the Hochschild-Serre factorization theorem. In particular, we consider algebras appearing as asymptotic symmetries of three dimensional spacetimes, the \bms, $\mathfrak{u}(1)$ Kac-Moody and  Virasoro algebras. We construct and classify the family of algebras which appear as deformations of \bms, $\mathfrak{u}(1)$ Kac-Moody and their central extensions by direct computations and also by  cohomological analysis. The Virasoro algebra appears as a specific member in this family of rigid algebras; for this case stabilization procedure is inverse of the In\"on\"u-Wigner contraction relating Virasoro to \bms\ algebra. We comment on the physical meaning of deformation and stabilization of these algebras and relevance of the family of rigid algebras we obtain.
}
\begin{document}
\maketitle
\section{Introduction and motivations}
Symmetries are cornerstones of any modern physical theory and are associated to groups in mathematical settings. Continuous symmetries are then related to Lie algebras. In physics formulations symmetries are transformations on the space of degrees of freedom and/or parameters of the theory which keep the action or Hamiltonian invariant and, Noether theorem states how to associate a conserved charge to each symmetry. 

Physical theories and in particular quantum field theories are usually analyzed within perturbation theory: we deform the action and analyze the deformed/perturbed theory. These deformations may or may not respect symmetries of the original theory. A relevant question is then whether there is a 
similar notion as deformations at the level of (symmetry) algebras and if yes, whether there is any relation between the ``algebraic deformation theory'' and those of physical theories. As we will review below there is a fairly well established notion of deformation of algebras in the mathematical literature. Moreover, in the context of quantum field theories we have the notion of RG flows and their fixed points in the space of deformation parameters. One can ask whether there is a similar concept as fixed points in the space of algebraic deformations. Again, as we will review below, the answer is affirmative and we have the notion of rigid algebras which are stable against deformations. 

In the case of finite dimensional Lie algebras the above questions may be analyzed in the context of cohomology of algebras where the celebrated Hochschild-Serre factorization theorem sets the stage on how one can deform any given Lie algebra and which algebras are rigid. 

While the stability and rigidity  of the algebras in physics literature are not new \cite{levy1967deformation, levy1968first, Figueroa-OFarrill:1989wmj,mendes1994deformations,Chryssomalakos:2004gk} and have recently been studied in a series of papers {in} the context of ``kinematical algebras'' \cite{Figueroa-OFarrill:2017sfs, Figueroa-OFarrill:2017ycu, Figueroa-OFarrill:2017tcy, Andrzejewski:2018gmz, Figueroa-OFarrill:2018ygf, Figueroa-OFarrill:2018ilb}, physicists are more familiar with the reverse process; the In\"on\"u-Wigner
(IW) contraction \cite{Inonu:1953sp}. The IW contraction was primarily worked through relating the Poincar\'e algebra to the Galilean symmetry algebra, through a non-relativistic limit; i.e.\ through sending speed of light $c$ to infinity and scaling the generators in an appropriate way \cite{levy1965nouvelle, mendes1994deformations}. Similar contraction is also known to relate the AdS or de Sitter symmetry algebras, respectively $\mathfrak{so}(d-2,2)$ or $\mathfrak{so}(d-1,1)$, to Poincar\'e algebra $\mathfrak{iso}(d-2,1)$ \cite{levy1967deformation}, see \cite{gilmore2012lie} for a comprehensive review. Geometrically, this latter may be understood as a particular large (A)dS radius $\ell$ limit where it is expected to recover flat $d-1$ dimensional Minkowski spacetime. The IW contraction relates algebras with the same number of generators, but different structure constants and always results in a non-rigid (non-stable) algebra. 

One may then ask if we can reverse this contraction process in a systematic way. For example, one may deform the Galilean algebra by $1/c$ corrections, or the Poincar\'e algebra by $1/\ell$ corrections and examine the rigid algebra this deformation will end up to. In this case, the expected result is of course what we started before contraction. However, one would like to 
know, given an algebra whether the rigid algebra coming out of its deformations is unique; i.e. whether there are various algebras which upon IW contraction yield the same algebra. The theory of cohomology of algebras answers this question, unambiguously, for finite dimensional Lie algebras: The result of stabilizing an algebra is unique (up to signature of the metric of the algebra, and up to central extensions). As a rule of thumb, the (semi)simple algebra with the same number of generators is generically the answer.

There are, however, infinite dimensional algebras which are of interest from physical viewpoint. The most famous ones are Virasoro and (Kac-Moody)
current algebras, or algebras of diffeomorphisms on certain given manifolds. One may then pose the same question of the deformations here. To our knowledge the mathematical literature on deformations of these algebras, specially when they admit central extensions, is not as vast and established as the finite dimensional cases, see however, \cite{Fialowski:2001me, fialowski2012formal, gao2008derivations, gao2011low, Ecker:2017sen,Ecker:2018iqn}. In this work we plan to study such cases and hopefully enrich the mathematical literature on this issue and provide physical implications and insights.

Recent developments in the context of ``asymptotic symmetry algebras'' have opened a new venue to obtain and look at infinite dimensional algebras in physics, e.g. see \cite{Brown:1986nw, Barnich:2001jy,  Strominger:2017zoo, Campoleoni:2017qot, Barnich:2017ubf, Afshar:2018apx, Concha:2018zeb, Hosseinzadeh:2018dkh}. In a sense the local (gauge) symmetries appearing in gauge field theories, like Maxwell or Yang-Mills theories, or diffeomorphism invariance  of general relativity are also related to infinite dimensional algebras. In a physical theory, however, the role of gauge symmetries are generically to remove the redundancy in the description of the theory. Nonetheless, as recent analyses have revealed, there could be a measure zero, but still infinite dimensional, subset of gauge transformations which are ``non-trivial,'' in the sense that one can associate well-defined charges to them. These charges are given by integrals over spacelike, compact codimension two surfaces of a codimension two-form constructed from the action of the theory, see e.g. \cite{Seraj:2016cym, Compere:2018aar} for reviews. This codimension two surface is usually (but not necessarily) taken to be at the asymptotic region of spacetime  and the charges are associated with gauge transformations which do not die off there. These surface charges are hence customarily called asymptotic symmetries and the gauge transformations called (asymptotically) large gauge transformations. 

The asymptotic symmetry algebra of nontrivial diffeomorphisms has been of special interest. In fact the first example of such symmetries was constructed in early 1960's by Bondi-van der Burg-Metzner and Sachs \cite{Bondi:1962px, Sachs:1962zza, Sachs:1962wk} and usually goes under the name BMS algebra. The BMS algebra discussed in the early examples was associated with  symmetries of four dimensional ($4d$) asymptotic flat spacetime and is usually  denoted as $\mathfrak{bms}_4$ {or its central extension $\widehat{\mathfrak{bms}}_4$ \cite{Barnich:2011ct}}. There is a similar notion in three dimensional ($3d$) flat space \cite{Ashtekar:1996cd}, on which we mainly focus in this paper, and is denoted by \hbms. \hbms\ is the central extension of the \bms \cite{Barnich:2006av, Barnich:2011ct, Oblak:2016eij}. Another famous and classic example of asymptotic symmetries is the one discussed by Brown and Henneaux in 1986 \cite{Brown:1986nw}. The Brown-Henneaux analysis was carried out for asymptotically AdS$_3$ spacetime and revealed two copies of commuting Virasoro algebras (usually called by Left and Right sectors) at a given central charge; the Brown-Henneaux central charge is 3/2 of  AdS$_3$ radius measured in units of $3d$ Newton constant. 

The asymptotic symmetry analysis has been carried out for $3d$ asymptotically (locally) flat, de Sitter or AdS spacetimes with various boundary falloff conditions recovering various classes of algebras, e.g. see \cite{Compere:2009zj, Compere:2014bia, Compere:2014cna, Troessaert:2013fma,  Afshar:2016wfy, Grumiller:2016pqb}. These algebras typically contain a Virasoro algebra and other generators associated with primary fields in this Virasoro algebra, e.g. like a $\mathfrak{u}(1)$ Kac-Moody algebra,  \hbms\ or $2d$ Galilean-Conformal-Algebra (GCA). The most general in the family of AdS$_3$ algebras are $\sltwo\oplus \sltwo$ current algebras and the associated Virasoro algebras built through (twisted) Sugawara construction \cite{Grumiller:2016pqb}. One may then realize many subalgebras of this algebra as asymptotic symmetry algebras with appropriate falloff boundary conditions \cite{Grumiller:2016pqb}. Similar analysis for asymptotically flat case has been carried out yielding $\mathfrak{iso}(2,1)$ current algebras \cite{Grumiller:2017sjh}. There are also other similar algebras coming from surface charge analysis near the horizon of black holes (rather than the asymptotic charges) \cite{Afshar:2016wfy, Afshar:2016uax,Afshar:2017okz} and in the $3d$ higher spin theories \cite{Grumiller:2016kcp, Campoleoni:2017mbt}. 

Given the variety of these algebras and their ``geometric'' realization as asymptotic symmetries, one may ask whether these algebras are rigid and if not, what is the result of the stabilization procedure. While rigidity of Witt and Virasoro algebras was first conjectured  in \cite{fialowski1990deformations}, a sketch of the proof was provided in \cite{fialowski2003global} using cohomological point of view and finally a full proof of rigidity (stability) of Witt and Virasoro algebras was provided in \cite{fialowski2012formal, schlichenmaier2014elementary}. These proofs also 
point to how the usual algebra cohomology arguments have shortcomings when applied to infinite dimensional algebras with central charges. Moreover, in the case of the Brown-Henneaux Virasoro algebra it is known that one can obtain the \hbms\ 
by the In\"on\"u-Wigner (IW) contraction \cite{Barnich:2006av,Barnich:2012rz}. This contraction is an extension of the same IW contraction relating $\mathfrak{so}(2,2)$ and $\iso$, respectively the isometry group of the underlying (asymptotic) AdS$_3$ and flat spacetimes. In this work we examine stability and rigidity of a class of algebras associated with $3d$ asymptotic symmetries and try to give a general  and mathematically solid answer to this question. 

\paragraph{Organization of the paper.} In section \ref{sec:2}, we review algebras related to asymptotic symmetries of $3d$ spacetimes. In section \ref{sec:3}, we review some relevant mathematical literature including the theory of Lie algebra deformations, the notion of cochains and cohomology of algebras, the Serre-Hochschild factorization and spectral sequence theorems, and basic results regarding rigidity of finite dimensional Lie algebras. In sections \ref{sec:4} and \ref{sec:5}, we study deformation and stabilization of centerless \bms, its centrally extended version \hbms\ algebras and other algebras appearing as $3d$ asymptotic symmetries. As we will see addition of central charges significantly alters the stabilization arguments. In the last section we have summarized our results (given in Tables \ref{table1}, \ref{table2}), discussions,  some conjectures about rigidity of the algebras we obtain, outlook and physical implications. {In the appendix we have presented proof of rigidity of algebras obtained from deformation of the \bms, \hbms\ and $\mathfrak{u}(1)$ Kac-Moody algebras.}

\paragraph{Notation.} Following \cite{Oblak:2016eij} we use ``\emph{fraktur} fonts'' e.g. $\mathfrak{bms},~ \mathfrak{vir}\inplus \mathfrak{vir}_{ab}$ for algebras and their centrally extended versions will be denoted by a hat, e.g. $\hat{\mathfrak{g}}$ denotes centrally extended version of the algebra $\mathfrak{g}$. In particular, 
\hbms\ $=\mathfrak{vir}\inplus \mathfrak{vir}_{ab}$, $\mathfrak{u}(1)$ Kac-Moody algebra and its centrally extended versions will be respectively denoted by $\mathfrak{KM}_{\mathfrak{u}(1)}$ and $\widehat{\mathfrak{KM}_{\mathfrak{u}(1)}}$.


\section {Asymptotic symmetries of \textit{3d} spacetimes}\label{sec:2}
{In this section we  review the structure of  asymptotic symmetry algebras appearing in the context of $3d$ gravity.}
Depending on the asymptotic behavior of the metric and the chosen boundary falloff conditions one can get different asymptotic symmetry algebras. The two set of ``standard'' falloff behaviors are the Brown-Henneaux boundary conditions for AdS$_3$ which yields two copies of Virasoro \cite{Brown:1986nw} and the BMS falloff behavior for $3d$ flat space which leads to $\widehat{\mathfrak{bms}}_3$ \cite{Barnich:2006av,Barnich:2011ct, Ashtekar:1996cd}. Here we briefly review these algebras.

\subsection {\textit{3d} flat space asymptotic symmetry algebra}

The centerless asymptotic symmetry  algebra of $3d$ flat spacetime is $\mathfrak{bms}_{3}$ \cite{Ashtekar:1996cd, Barnich:2006av}:
 \begin{equation} 
\begin{split}
 & i[\mathcal{J}_m,\mathcal{J}_n]=(m-n)\mathcal{J}_{m+n}, \\
 &i[\mathcal{J}_m,\mathcal{P}_n]=(m-n)\mathcal{P}_{m+n},\\
 &i[\mathcal{P}_m,\mathcal{P}_n]=0,
\end{split}\label{bms3}
\end{equation}
where $m,n\in \mathbb{Z}$ and it is defined over the field of real numbers $\mathbb{R}$. {The $\mathfrak{bms}_{3}$ is an infinite dimensional algebra with countable basis} and has two sets of generators, ${\cal J}_n$ and ${\cal P}_n$. The former generates the  Witt subalgebra of \bms\ and is the algebra of smooth vector fields on a circle. ${\cal P}_n$  construct an adjoint representation of the Witt algebra and form the ideal part of \bms.  Eq.\eqref{bms3} makes it clear that $\mathfrak{bms}_{3}$ has a semi-direct sum structure:
\begin{equation} \label{bms=witt+ideal}
\mathfrak{bms}_{3}= \mathfrak{witt}\inplus_{ad}\mathfrak{witt}_{ab},
\end{equation}
where the subscript $ab$ is to emphasize ${\cal P}$ being abelian and $ad$ denotes the adjoint action.  
The global part of $\mathfrak{bms}_{3}$ is $3d$ Poincar\'{e} algebra, $\mathfrak{iso}(2,1)$, and is obtained by restricting to $m, n=\pm 1, 0$ in relation (\ref{bms3}).
In physics terminology ${\cal J}$ which are generalizations of the $3d$ Lorentz group are called superrotations and ${\cal P}$ which are generalizations of $3d$ momenta, supertranslation. 

The asymptotic symmetry analysis of $3d$ flat space actually leads to centrally extended version of the algebra, denoted by $\widehat{\mathfrak{bms}}_{3}$:
 \begin{equation}\label{BMS-centrally-extended} 
\begin{split}
 & i[\mathcal{J}_m,\mathcal{J}_n]=(m-n)\mathcal{J}_{m+n}+\frac{c_{JJ}}{12}m^{3}\delta_{m+n,0}, \\
 &i[\mathcal{J}_m,\mathcal{P}_n]=(m-n)\mathcal{P}_{m+n}+\frac{c_{JP}}{12}m^{3}\delta_{m+n,0},\\
 &i[\mathcal{P}_m,\mathcal{P}_n]=0,
\end{split}
\end{equation}
in which $c_{JJ}$ and $c_{JP}$ are called central charges. The central part, besides the $m^3$ piece, may also have a piece proportional to $m$. This latter can be absorbed in a proper redefinition of the generators, therefore, in our analysis we only include the $m^{3}$ term.

For flat space there are of course ``more relaxed'' boundary conditions which yield algebras with more number of fields in the various representations of the Virasoro algebra, the most general being $\mathfrak{iso}(2,1)$ Kac-Moody current algebra \cite{Grumiller:2017sjh}. 

\subsection { \texorpdfstring{AdS$_3$}{AdS} asymptotic symmetry algebra }
Centerless asymptotic symmetry algebra of $3d$ AdS spacetime with Brown-Henneaux boundary conditions is:
\begin{equation} 
\begin{split}
 & i[L_m,L_n]=(m-n)L_{m+n}, \\
  &i[L_m,\Bar{L}_n]=0, \\
 &i[\Bar{L}_m,\Bar{L}_n]=(m-n)\Bar{L}_{m+n},
\end{split}\label{Centerless-AdS3}
\end{equation}
in which $m,\ n\ \in \mathbb{Z}$, and is  defined over the field  $\mathbb{R}$. It is obvious that the algebra \eqref{Centerless-AdS3} has two Witt 
subalgebras. That is, \eqref{Centerless-AdS3} is direct sum of two different Witt algebras,
\begin{equation} 
\mathfrak{witt}^2= \mathfrak{witt}_L \oplus \mathfrak{witt}_R. 
\end{equation}

Centrally extended version of \eqref{Centerless-AdS3} is 
\begin{equation} 
\begin{split}
 & i[L_m,L_n]=(m-n)L_{m+n} +\frac{c}{12} n^3\delta_{m+n,0}, \\
  &i[L_m,\Bar{L}_n]=0, \\
 &i[\Bar{L}_m,\Bar{L}_n]=(m-n)\Bar{L}_{m+n}+\frac{\bar c}{12} n^3\delta_{m+n,0},
\end{split}\label{AdS3}
\end{equation}
which is asymptotic symmetry algebra of asymptotically AdS$_3$ solutions to Einstein-$\Lambda$ theory with AdS radius $\ell$ and Newton constant $G_N$, $c=\bar c=\frac{3\ell}{2G_N}$ \cite{Brown:1986nw}. The algebra \eqref{AdS3}  
will be denoted by \virtwo\ and is direct sum of two different Viraroso algebras,
\begin{equation} 
\mathfrak{vir}^2= \mathfrak{vir}_L \oplus \mathfrak{vir}_R.
\end{equation}
With our previous notation, Virasoro algebra is the centrally extended version of Witt algebra, i.e. \vir$=\widehat{\mathfrak{witt}}$. In other words, Virasoro algebra is Witt algebra extended by the Gel'fand-Fuks global 2-cocycle \cite{gel1969cohomologies}. The global part of \virtwo\ is either $\mathfrak{sl}(2,\mathbb{R})\oplus \mathfrak{sl}(2,\mathbb{R})\simeq \mathfrak{so}(2,2)$  which is the algebra of (global) isometries of AdS$_3$ \cite{Brown:1986nw}, or $\mathfrak{so}(3,1)\simeq \mathfrak{sl}(2;\mathbb{C})$ associated with global isometries of $3d$ de Sitter space (without the factors of $i$ in the commutators, though) \cite{Compere:2014cna}. 

On the AdS$_3$ one can get other asymptotic symmetry algebras, e.g. if we relax the Brown-Henneaux boundary conditions to the ones introduced in \cite{Troessaert:2013fma} we get  two copies of centrally extended $\mathfrak{u}(1)$ Kac-Moody algebras. One can also impose more restricted set of boundary conditions and get a single copy of $\mathfrak{u}(1)$ Kac-Moody algebra \cite{Compere:2013bya}. The latter boundary conditions can be also used for the case of chiral gravity theory \cite{Li:2008dq}. In the most general case one can get two copies of $\mathfrak{sl}(2,\mathbb{R})$ Kac-Moody current algebras \cite{Grumiller:2016pqb} which contains all the previous cases as subalgebras (the Virasoro algebras are related to this current algebra through (twisted) Sugawara construction).

\subsection{Contraction of Virasoro to \texorpdfstring{\hbms}{HBMS3}}\label{sec:2.3-contraction}

The \hbms\ and \virtwo\ algebras are related by a IW contraction \cite{Barnich:2012rz}. To see this, let us introduce


\begin{equation}\label{Vir-BMS}
L_n=\frac12(\mathcal{J}_n+\ell{\mathcal{P}_n}),\ \ \ {\bar{L}}_{-n}=\frac12(\mathcal{J}_n-\ell{\mathcal{P}_n}),
\end{equation}
where $\ell$ is a parameter introduced to facilitate taking the contraction as an $\ell\to\infty$ limit, while keeping $\mathcal{J}_n, \mathcal{P}_n$ fixed. Plugging \eqref{Vir-BMS} into the Virasoro algebra we obtain 
\begin{align}
 & i[\mathcal{J}_{m},\mathcal{J}_{n}]=(m-n)\mathcal{J}_{m+n}+\frac{c+\bar c}{12}n^3\delta_{n+m,0}, \cr
 &i[\mathcal{J}_{m},\mathcal{P}_{n}]=(m-n)\mathcal{P}_{m+n}+\frac{c-\bar c}{12\ell}n^3\delta_{n+m,0},\\
 &i[\mathcal{P}_{m},\mathcal{P}_{n}]=\frac{1}{\ell^2}(m-n)\mathcal{J}_{m+n}+\frac{c+\bar c}{12\ell^2}n^3\delta_{n+m,0}.\nonumber
\end{align} 
If we keep $c_{JJ}=c+\bar c$ and $c_{JP}=(c-\bar c)/\ell$ finite while taking $\ell\to\infty$, the above reduces to the $\widehat{\mathfrak{bms}}_3$ \eqref{BMS-centrally-extended}. In the above IW contraction $1/\ell$ may be viewed as the deformation parameter.

\subsection{\textit{3d} near horizon symmetry algebras
}\label{sec:2.4}

On AdS$_3$ space we have the class of BTZ black holes \cite{Banados:1992wn} and/or their Virasoro descendants. It has been shown in \cite{Afshar:2016kjj,Afshar:2016wfy, Afshar:2017okz} that one can impose appropriate boundary (falloff) conditions on the horizon of these black holes and find a ``near horizon'' (in contrast to asymptotic) symmetry algebra. This algebra has been shown to be two chiral copies of a $\mathfrak{u}(1)$ current algebra, i.e. a centrally extended $\mathfrak{u}(1)$ Kac-Moody algebra, $\widehat{\mathfrak{KM}_{\mathfrak{u}(1)}}$ and is defined through 
\begin{align}\label{centrally-extended-u(1)-KM}
 & i[\mathcal{J}_{m},\mathcal{J}_{n}]=(m-n)\mathcal{J}_{m+n}+\frac{c_{JJ}}{12} n^3\delta_{n+m,0}, \cr
 &i[\mathcal{J}_{m},\mathcal{P}_{n}]=-n\mathcal{P}_{m+n}+\frac{c_{PJ}}{12}n^2\delta_{n+m,0},\\
 &i[\mathcal{P}_{m},\mathcal{P}_{n}]={\frac{c_{PP}}{12}} n\delta_{n+m,0}.\nonumber
\end{align} 
One should note that in the computations of the near horizon symmetry algebras the central charges $c_{JJ}, c_{JP}$ and $c_{PP}$ are not (parametrically) independent.

\section{Lie algebra deformation theory}\label{sec:3}

{\it{Deformation}} of a certain Lie algebra $\mathfrak{g}$ is a modification of its structure constants. Some of such deformations could just be a change of basis which are called trivial deformations. {Nontrivial deformations modify/deform a Lie algebra $\mathfrak{g}$ to another Lie algebra with the same vector space structure. In the case of finite {dimensional} Lie algebras the latter implies that deformation does not change the dimension of the algebra.} 
The concept of deformation of rings and algebras was first introduced in a series of papers by Gerstenhaber \cite{gerstenhaber1964deformation, gerstenhaber1966deformation, gerstenhaber1968deformation, gerstenhaber1974deformation} and by Nijenhuis and Richardson for Lie algebras in \cite{nijenhuis1967deformations}. 
Deformations  introduced by Nijenhuis and Richardson are known as `formal' deformations where a Lie algebra is deformed by a formal power series of some variables (deformation parameters). {If {one considers the deformation only up to the linear term in the power series it is called `infinitesimal' deformation}}. 
While  other kinds of generalized deformations of Lie algebras, like  `versal', `global' and `analytic' deformations {have} also been introduced and studied \cite{fialovski1986deformations, fialowski1988example, fialowski2003global, fialowski2005global, guerrini1998formal}, in this work we only focus on infinitesimal and formal deformations.

A Lie algebra $\mathfrak{g}$ is called rigid or stable if it does not admit any deformation or equivalently if its deformed algebra $\mathfrak{g}_{\varepsilon}$, $\varepsilon$ being the deformation parameter, is  isomorphic to the initial algebra. As deformations, there are some different notions of rigidity. Here, we will focus on the `formal' and `infinitesimal' rigidity. One may consider `analytic,' `global or `versal' rigidity of Lie algebras associated with similar deformations, as e.g. done in \cite{guerrini1999formal, fialowski2005global}. 

{Our main focus in this paper will be on infinite dimensional Lie algebras, nonetheless for setting the stage  and building an intuitive picture we review the more studied and established case of finite dimensional Lie algebras. In {the} following sections (up to section \ref{sec:3.4}) we assume that the Lie algebra is finite dimensional and in section \ref{sec:3.4} discuss which notions are not extendable to infinite dimensional cases. }

\paragraph{Lie algebra structure and $\mathcal{L}_{n}$ space.}
 We call $(\mathfrak{g},[,])$ a Lie algebra in which $\mathfrak{g}$ is a vector space over a field $\mathbb{F}$ with characteristic zero (for example $\mathbb{R}$ or $\mathbb{C}$) equipped with Lie bracket, a bilinear and antisymmetric product function $[,]$,
\begin{equation}
    [,]: \mathfrak{g}\times \mathfrak{g}\longrightarrow \mathfrak{g}.
\end{equation}
The Lie bracket $[,]$ must also satisfy the Jacobi identity,
\begin{equation}
    [g_{i},[g_{j},g_{k}]]+[g_{j},[g_{k},g_{i}]]+[g_{k},[g_{i},g_{j}]]=0,\qquad \forall g_{i} \in \mathfrak{g}. \label{GJ}
\end{equation}
We denote the Lie algebra $(\mathfrak{g},[,])$ by $\mathfrak{g}$ and $\{g_{i}\}$ are the basis elements of $n$ dimensional Lie algebra $\mathfrak{g}$ where $i=1,...,n=dim \mathfrak{g}$ which satisfy the Lie bracket $[,]$ as:
 \begin{equation}
     [g_{i},g_{j}]=f^{k}_{i,j}g_{k},
 \end{equation}
where $f^{k}_{i,j}=-f^{k}_{j,i}$'s are the $\mathbb{F}$-valued structure constants.  The Jacobi identity in terms of structure constants is 
 \begin{equation}
   f^{l}_{i,j}f^{m}_{k,l}+f^{l}_{j,k}f^{m}_{i,l}+f^{l}_{k,i}f^{m}_{j,l}=0,\label{stj} 
 \end{equation}
where there is summation over the repeated indices. 
For an $n$ dimensional Lie algebra $\mathfrak{g}$ the maximum number of independent structure constants are $\frac{1}{2}n^{2}(n-1)$. The space of all independent structure constants can be considered as a $\frac{1}{2}n^{2}(n-1)$ dimensional space, $\mathbb{F}^{\frac{1}{2}n^{2}(n-1)}$, each axis is labeled by one independent structure constant $f^{k}_{i,j}$ with values in field $\mathbb{F}$. The space of all $n$ dimensional Lie algebras over the same field, which is specified with the constrains \eqref{stj} and denoted by $\mathcal{L}_{n}$, is an algebraic subset  
of this space \cite{onishchik41lie}. Intuitively each point on the algebraic subset
$\mathcal{L}_{n}$ corresponds to a certain class of Lie algebras isomorphic to $\mathfrak{g}$ with specific structure constants.    

\paragraph{Formal deformation of Lie algebras.}
A formal one parameter deformation of a Lie algebra ($\mathfrak{g},{[,]}_{0}$) is defined as:
 \begin{equation}\label{eq:defdefor}
     {[g_{i},g_{j}]}_\varepsilon:=\Psi(g_{i},g_{j};\varepsilon)=\Psi(g_{i},g_{j};\varepsilon=0)+{{\psi }_1(g_{i},g_{j})}\varepsilon^{1}+{{\psi }_2(g_{i},g_{j})}\varepsilon^{2}+...,
 \end{equation}
where $ \Psi(g_{i},g_{j};\varepsilon=0)={[g_{i},g_{j}]}_0$, $g_{i},g_{j}$ are basis elements of $\mathfrak{g}_{0}$, $\varepsilon \in \mathbb{F}$ (for example $\mathbb{R}$ or $\mathbb{C}$) is the {\it{deformation parameter}} and functions ${\psi }_i:\ \mathfrak{g}\times \mathfrak{g}\longrightarrow \mathfrak{g}$ are bilinear antisymmetric functions, the 2-\textit{cochains}. 
Intuitively, a formal one parameter deformation can be seen as a continuous curve on the algebraic subset $\mathcal{L}_{n}$ parametrized by the deformation parameter $\varepsilon$. A smooth and analytic curve corresponds to a smooth and analytic deformation \cite{onishchik41lie}.  For every $\varepsilon$, the new Lie algebra ($\mathfrak{g},{[,]}_{\varepsilon}$) should satisfy the Jacobi identity,
\be
[g_{i},[g_{j},g_{k}]_{\varepsilon}]_{\varepsilon}+\text{cyclic permutation of}\ (g_{i},g_{j},g_{k}) =0,\label{Jepsilon}
\ee
 order by order in $\varepsilon$, which leads to infinite sequence of equations among $\psi_{i}$.

 For small $\varepsilon$ the leading deformation is given by ${\psi }_{1}(g_{i},g_{j})$-term and the associated Jacobi identity leads to
\be\label{coe}
  {[g_{i},{\psi }_1(g_{j},g_{k})]}_0+{\psi }_1(g_{i},{[g_{j},g_{k}]}_0)+\text{cyclic permutation of}\ (g_{i},g_{j},g_{k}) =0.
\ee
This relation is known as the $2-$cocycle condition. Its solution, the  $2-$cocycle ${\psi }_1$,  specifies an {\it infinitesimal} deformation of Lie algebra $\mathfrak{g}$. The Jacobi identity for higher orders of $\varepsilon$ should also be checked as integrability conditions of ${\psi }_1$ and may lead to obstructions, which will be discussed later in this section. From now we denote the deformed algebra ($\mathfrak{g},{[,]}_{\varepsilon}$) just by $\mathfrak{g}_{\varepsilon}$. Conversely, given an algebra $\mathfrak{g}$ one can take $\varepsilon \to 0$ limit and obtain $\mathfrak{g}_{0}$.
This procedure is known as {\it{contraction}} of Lie algebras. The contraction and deformation are hence inverse of each other.

Deformation by a $2-$cocycle is trivial if the deformed algebra is isomorphic to the initial algebra, i.e. the deformed and original algebras are related by a redefinition of generators or equivalently a change of basis. If two algebras $\mathfrak{g}$ and $\mathfrak{g}_{\varepsilon}$ are isomorphic to each other, they are related by \begin{equation}
    {[g_{i},g_{j}]}_{\varepsilon}=M_{\varepsilon}{[M^{-1}_\varepsilon (g_{i}),M^{-1}_\varepsilon(g_{j})]}_0,\label{change-basis}
\end{equation}
in which invertible operator $M$ is a linear transformation in vector space $\mathfrak{g}$. 
The operator $M_\varepsilon$ can be expanded as
\begin{equation*}
     M_\varepsilon=I+\varepsilon {\varphi }_1.
\end{equation*}
By inserting this relation into (\ref{change-basis}) one gets
\begin{equation*}
    {[g_{i},g_{j}]}_{\varepsilon}={(I+\varepsilon {\varphi }_1)[(I-\varepsilon{\varphi }_1)g_{i}, (I-\varepsilon {\varphi }_1)g_{j}]}_0,
\end{equation*}
yielding
\begin{equation*}
    {[g_{i},g_{j}]}_\varepsilon={[g_{i},g_{j}]}_0+\bigg( {{\varphi }_1([g_{i},g_{j}])}_0-{[{\varphi }_1(g_{i}),g_{j}]}_0-{[g_{i},{\varphi }_1(g_{j})]}_0\bigg)\varepsilon+O(\varepsilon^2).
\end{equation*}
Next consider
\begin{equation}
    {\psi }_1(g_{i},g_{j})={{\varphi }_1([g_{i},g_{j}])}_0-{[{\varphi }_1(g_{i}),g_{j}]}_0-{[g_{i},{\varphi }_1(g_{j})]}_0.\label{cob}
\end{equation}
One can readily check that this satisfies the 2-cocycle condition \eqref{coe}. In fact \eqref{cob} shows ${\psi}_1$ is a 2-coboundary if ${\varphi }_1$ is a 1-cochain. When ${\psi }_1$ is a 2-coboundary the deformation \eqref{eq:defdefor} is called trivial, meaning that the deformation is just a redefinition of basis elements.
\subsection{Relation of deformation theory and cohomology of a Lie algebra}\label{sec:3.1}

We start with the definition of the Chevalley-Eilenberg complex and differential.
A vector space $\mathbb{V}$ is called a $\mathfrak{g}$-module if {there exists}
a bilinear map $\omega:\ \mathfrak{g}\times \mathbb{V}\longrightarrow \mathbb{V}$  
for all $x\in \mathbb{V}$ and $g_{1},g_{2} \in \mathfrak{g}$ with the property $\omega([g_{1},g_{2}],x)=\omega(g_{1},\omega(g_{2},x))-\omega(g_{2},\omega(g_{1},x))$,  \cite{fuks2012cohomology}. In this setting, the Jacobi identity of the Lie bracket implies that a Lie algebra $\mathfrak{g}$ with the adjoint action is a $\mathfrak{g}$-module. 
{A $p$-cochain $\psi$ is a $\mathbb{V}$-valued (as $\mathfrak{g}$-module),  
 bilinear and completely antisymmetric function which is defined as:
\begin{align*}
    \psi: \underbrace{\mathfrak{g}\times\cdots\times\mathfrak{g}}_{p \, \ times}&\longrightarrow \mathbb{V}\\
    (g_1,\cdots ,g_p)&\longmapsto \psi(g_1,\cdots ,g_p);\,\,\,\,\,\ 0 \leq p \leq dim(\mathfrak{g}).
\end{align*}
Suppose ${\mathcal{C}}^p(\mathfrak{g};\mathbb{V})$ is the space of $\mathbb{V}$-valued $p$-cochains on $\mathfrak{g}$. One can then define the cochain complex ${\mathcal{C}}^*(\mathfrak{g};\mathbb{V})=\oplus^{dim(\mathfrak{g})}_{p=o}{\mathcal{C}}^p(\mathfrak{g};\mathbb{V})$ which is known as the \textbf{\textit{Chevalley-Eilenberg complex}}.

The \textbf{\textit{Chevalley-Eilenberg differential}} or equivalently \textbf{\textit{coboundary operator}} ``$d$'' is a linear map defined as \cite{ChevalleyEilenberg, MR0054581}
\begin{align*}
d_{p}:\ {\mathcal{C}}^{p} (\mathfrak{g} ;\mathfrak{g})&\longrightarrow{\mathcal{C}}^{p+1}\left(\mathfrak{g};\mathfrak{g}\right),\\
     \psi &\longmapsto d_{\mathrm{p}}\psi,
\end{align*}
and $p+1$-cochain $d_{{p}}\psi$ is given by:
   \begin{align}\label{p-cochain}
     (d_{p}\psi )\left(g_0,\dots ,g_{p}\right)\equiv&\sum_{0\leq {i}{<}{j}\leq {p}}{({{1}})^{{i}{+}{j}{-}{1}}} {\psi }\left(\left[{{g}}_{{i}},{{g}}_{{j}}\right],{{g}}_{{0}},{\dots },\widehat{{{g}}_{{i}}},{\dots },\widehat{{{g}}_{{j}}},{\dots },{{g}}_{{p}{+1}}\right)\cr {+}&
    \sum_{{1}\leq {i}\leq {p}{+1}}{({{-}{1}})^{{i}}}\left[{{g}}_{{i}},{\psi }\left({{g}}_{0},{\dots },\widehat{{{g}}_{{i}}}, {\dots },{{g}}_{p}\right)\right],
\end{align}
the hat denotes omission. One can check that $d_{p}\circ d_{{p}{-}{1}}{=0.\ }$
A $p$-cochain ${\psi}$  is called \textbf{\textit{$p$-cocycle}} if  ${d}_p{\psi}=0$, and a \textbf{\textit{$p$-coboundary}} if $\ {\psi}={d }_{p-1}{\varphi}$.
By the property $d_{{p}}\circ d_{{p}{-}{1}}{=0\ }$ one concludes that every $p$-coboundary is also a $p$-cocycle. With this definition one can check that 2-cocycle condition \eqref{coe} is just $d_{2} \psi_{1}=0$ where $\psi$ is a $\mathfrak{g}$-valued $2-$cochain and $d_{2}$  given in \eqref{p-cochain}, and the relation \eqref{cob} is $2-$coboundary condition $\psi_{1}=d_{1}\varphi_{1}$ where $\varphi_{1}$ is a $\mathfrak{g}$-valued $1-$cochain.

{One defines $Z^{p}(\mathfrak{g};\mathbb{V})$ as space of $p$-cocycles which is kernel of differential $d$ as
\begin{equation}
    Z^{p}(\mathfrak{g};\mathbb{V})=\{\psi\in\mathcal{C}^p(\mathfrak{g};\mathbb{V})|d_{p}\psi=0 \}.
\end{equation}
$Z^{2}(\mathfrak{g};\mathfrak{g})$ is hence the space of all $\mathfrak{g}-$valued $2-$cocycles which satisfy the relation \eqref{coe}. 
One also defines $B^{p}(\mathfrak{g};\mathbb{V})$ as space of $p$-coboundaries as
\begin{equation}
    B^{p}(\mathfrak{g};\mathbb{V})=\{\psi\in\mathcal{C}^p(\mathfrak{g};\mathbb{V})|\psi=d_{p-1}\varphi \,\,\, \text{for some}\ \ \varphi\ \text{in}\ \  \mathcal{C}^{p-1}(\mathfrak{g};\mathbb{V}) \}.
\end{equation}
$B^{2}(\mathfrak{g};\mathfrak{g})$ is therefore the space of all $\mathfrak{g}-$valued $2-$cocycles which are also $2-$coboundary which means its elements satisfy both relations \eqref{coe} and \eqref{cob}.  $p^{th}\ cohomology$ space of $\mathfrak{g}$ with coefficients in $\mathbb{V}$ is then defined as quotient of the space of  $p$-cocycles $ Z^{p}(\mathfrak{g};\mathbb{V})$ to the space of  $p$-coboundaries $ B^{p}(\mathfrak{g};\mathbb{V})$ as:
 \begin{equation}
     {\mathcal{H}}^p(\mathfrak{g};\mathbb{V}):=Z^{p}(\mathfrak{g};\mathbb{V})/B^{p}(\mathfrak{g};\mathbb{V})=\text{Ker}({\ d }_p)/\text{Im}({\ d }_{p-1}).
 \end{equation}
 \paragraph{Interpretations of cohomology spaces}
 \begin{itemize}
     \item  ${\mathcal{H}}^{0}(\mathfrak{g};\mathbb{V})$ is defined as: 
  \begin{equation}
     \mathcal{H}^{0}(\mathfrak{g};\mathbb{V})=\text{Inv}_{\mathfrak{g}}(\mathbb{V})=\{x\in \mathbb{V}|g.x=0,\,\, \forall g\in \mathfrak{g}\},\label{H01}
 \end{equation}
and is the space of invariants \cite{Roger:2006rz}.
\item ${\mathcal{H}}^{1}(\mathfrak{g};\mathbb{V})$ can be interpreted as exterior derivations of Lie algebra $\mathfrak{g}$ with values in $\mathbb{V}$. In the case ${\mathcal{H}}^{1}(\mathfrak{g};\mathfrak{g})$ the map $\varphi: \mathfrak{g} \longrightarrow \mathfrak{g}$ is called a derivation if it satisfies the Leibniz rule $\varphi([g_{1},g_{2}])=[\varphi(g_{1}),g_{2}]+[g_{1},\varphi(g_{2})]$.
\item ${\mathcal{H}}^{2}(\mathfrak{g};\mathbb{V})$ classifies deformations or isomorphic classes of extensions of the Lie algebra $\mathfrak{g}$ with respect to $\mathbb{V}$. Two particular cases of interest are 
 \begin{itemize}
     \item[i.]  $\mathbb{V}=\mathbb{R}$ (or $\mathbb{C}$) where ${\mathcal{H}}^{2}(\mathfrak{g};\mathbb{R})$ classifies (global) central extensions of algebra $\mathfrak{g}$. 
      \item[ii.] $\mathbb{V}=\mathfrak{g}$ where ${\mathcal{H}}^{2}(\mathfrak{g};\mathfrak{g})$, the second adjoint cohomology, classifies all infinitesimal deformations of algebra $\mathfrak{g}$.  \end{itemize}
 \end{itemize}
 
Therefore,  to classify infinitesimal deformations of a given Lie algebra $\mathfrak{g}$ one has to compute second adjoint cohomology $\mathcal{H}^2(\mathfrak{g};\mathfrak{g})$. Not all infinitesimal deformations integrate to a formal (finite) deformation; there could be obstructions.  We will return to the concept of integrability and obstructions later in this section.

\paragraph{Definition \cite{fialowski2012formal}.} A Lie algebra ($\mathfrak{g},{[,]}_{0}$) is formally rigid, if and only if its every formal deformation  is a trivial deformation. 
Intuitively rigidity of Lie algebra $\mathfrak{g}$ means each Lie algebra ${\mathfrak{g}}_{\varepsilon}$ which is close to $\mathfrak{g}$, is isomorphic to it.  {In the physics literature rigid algebras are also called stable algebras.}
 \paragraph{Theorem 3.1.}\label{2.1} { If $\mathcal{H}^2(\mathfrak{g};\mathfrak{g})=0$ then Lie algebra ($\mathfrak{g},{[,]}_{0}$) is infinitesimally and formally rigid \cite{nijenhuis1966cohomology, goze2006lie, richardson1967rigidity}.}
In fact, for finite dimensional Lie algebras the condition $\mathcal{H}^2(\mathfrak{g};\mathfrak{g})=0$ means that these Lie algebras are rigid in any sense {e.g. infinitesimally, formally, analytically, \dots } \cite{gerstenhaber1964deformation, gerstenhaber1966deformation, gerstenhaber1968deformation,gerstenhaber1974deformation, nijenhuis1967deformations, fialowski-Mc:2005}. For infinite dimensional Lie algebras, however,  $\mathcal{H}^2(\mathfrak{g};\mathfrak{g})=0$ means that theses are just infinitesimally and formally rigid \cite{fialowski1988example}.
That is, the second adjoint cohomology $\mathcal{H}^{2} (\mathfrak{g};\mathfrak{g})$ classifies infinitesimal deformations of Lie algebra $\mathfrak{g}$; if it is zero from last theorem one concludes that $\mathfrak{g}$ is rigid and does not admit a nontrivial deformation. Otherwise, one has found nontrivial infinitesimal deformations and then verify possible obstructions to make sure that these deformations are also formal deformations.

As an intuitive picture, recall the  $\mathcal{L}_{n}$, the set of all $n$ dimensional Lie algebras which is an algebraic subset in the space of structure constants; each point of this set denotes (an isomorphism class of) a certain Lie algebra. Consider the point $\mathfrak{g}$ with the coordinate $(f^{1}_{1,2},f^{2}_{1,2},...,f^{n}_{n-1,n})$. So the space of all $\mathfrak{g}$-valued $2-$cocycles, $\mathcal{Z}^2(\mathfrak{g};\mathfrak{g})$, is the tangent space to $\mathcal{L}_{n}$ in point $\mathfrak{g}$. The space of all $\mathfrak{g}$-valued $2-$coboundaries, $\mathcal{Z}^2(\mathfrak{g};\mathfrak{g})$, is a subspace of the tangent space on point $\mathfrak{g}$. If $\mathcal{H}^2(\mathfrak{g};\mathfrak{g})=0$, the elements of $\mathcal{Z}^2(\mathfrak{g};\mathfrak{g})$ and $\mathcal{B}^2(\mathfrak{g};\mathfrak{g})$  are the same, so the point or equivalently $\mathfrak{g}$ is rigid. However, if $\mathcal{H}^2(\mathfrak{g};\mathfrak{g})\neq 0$, there is at least one direction in the tangent space which can take the algebra $\mathfrak{g}$ to another Lie algebra $\mathfrak{g}_{\varepsilon}$ which is not isomorphic to $\mathfrak{g}$. In fact, nontrivial infinitesimal deformations which are elements of $\mathcal{H}^2(\mathfrak{g};\mathfrak{g})$ are directions where the algebra $\mathfrak{g}$ along them can be deformed to another algebra $\mathfrak{g}_\eps$. 

\paragraph{Integrability conditions and obstructions.} 
As pointed out earlier, to have a  formal deformation \eqref{eq:defdefor}, we need the  corresponding nontrivial infinitesimal deformation to be integrable, to be valid to all orders in the deformation parameter. To first few orders in $\varepsilon$, \eqref{Jepsilon}  leads to 
\begin{subequations}\label{infs}
\begin{align}   &[g_{i},[g_{j},g_{k}]_{0}]_{0}+\text{cyclic permutation of}\ (g_{i},g_{j},g_{k})=0,\label{infs-a}\\
    &d_{2}\psi_{1}=0,\label{infs-b}\\
    &d_{2}\psi_{2}=-\frac{1}{2}[\![\psi_{1},\psi_{1}]\!],\label{infs-c}\\
   & d_2\psi_3=-[\![\psi_1,\psi_2]\!],\label{infs-d}
\end{align}
\end{subequations}
where we used the definition of the {Chevalley-Eilenberg differential} $d_{2}$ in \eqref{p-cochain} and the double-bracket is the Nijenhuis and Richardson bracket \cite{nijenhuis1967deformations} defined as 
$$\frac{1}{2}[\![\psi_{r},\psi_{s}]\!](g_{i},g_{j},g_{k}):=\psi_{r}(g_{i},\psi_{s}(g_{j},g_{k}))+\text{cyclic permutation of}\ (g_{i},g_{j},g_{k}).
$$
The zeroth order in $\varepsilon$, \eqref{infs-a}, is nothing but the Jacobi relation for the undeformed algebra and is hence satisfied by definition. The second equation \eqref{infs-b} is the $2-$cocycle condition \eqref{coe} for $\psi_{1}$ and its solutions provides nontrivial \emph{infinitesimal} deformations. Eq.\eqref{infs-c} would then guarantee that there are no obstructions in viewing $\psi_1(g_i,g_j)$ as the first order term of a formal deformation $\Psi(g_i,g_j;\eps)$ which admits a power series expansion in $\eps$. Of course one should continue the reasoning in higher orders of $\eps$. It is readily seen that for $\eps^3$ level one should satisfy \eqref{infs-d} and so on. The sequence of relations will be stopped if there is an obstruction. 

From cohomological point of view, one can verify that all obstructions are in the space ${\mathcal{H}}^{3}(\mathfrak{g};\mathfrak{g})$.  If ${\mathcal{H}}^{3}(\mathfrak{g};\mathfrak{g})=0$ then there are no obstructions \cite{nijenhuis1967deformations}. The latter is a sufficient condition for integrability of ${\mathcal{H}}^{2}(\mathfrak{g};\mathfrak{g})$'s elements. However, the converse is not true and absence of obstructions does not mean ${\mathcal{H}}^{3}(\mathfrak{g};\mathfrak{g})$ is vanishing.  As a result if ${\mathcal{H}}^{2}(\mathfrak{g};\mathfrak{g})\neq 0$ while ${\mathcal{H}}^{3}(\mathfrak{g};\mathfrak{g})=0$ we have a formal  (formal power series) deformation and the deformation parameter $\varepsilon$ need not be taken to be small. 

Given a nontrivial 2-cocycle $\psi_{1}$ or equivalently given ${\mathcal{H}}^{2}(\mathfrak{g};\mathfrak{g})\neq 0$, some different situations may happen:
\begin{itemize}
\item [i.] While there is no obstruction, all the other functions $\psi_{r}$ ($r\geq 2$) have trivial solution $0$. This means that there is a formal deformation which has only linear term in power series. {As we will see except some specific cases, which are examples of case v. below, the deformations we will find are examples of this case}.
\item [ii.] There are no obstructions but other functions $\psi_{r}$ ($r\geq 2$) also have nontrivial solutions. This means that there is a formal power series in which $\psi_{r}$'s can be seen as Taylor coefficients of a function $\Psi(g_{i},g_{j};\varepsilon)$.
\item [iii.] There is no obstruction but for the set of functions $\psi_{r}$ ($r\geq 2$), \eqref{infs} lead to different solution sets. That is, there are at least two different curves in the $\mathcal{L}_{n}$ space which have the same  first derivative with respect to $\varepsilon$ in the initial point. 
\item [iv.] There is no obstruction up to a specific order $s$. It means that there are only solutions for functions $\psi_{r}$ where $r=2,...,s$ and for $r>s$ equations do not have any solution. We then call $\psi_{1}$ is integrable up to order $s$ \cite{goze2006lie}.
\item [v.] Obstructions start from second order and there is no solution for $\psi_{2}$ and other $\psi_{r}$ where $r\geq2$. As we saw in theorem \ref{2.1} the condition ${\mathcal{H}}^{2}(\mathfrak{g};\mathfrak{g})=0$ is a sufficient condition for the algebra $\mathfrak{g}$ to be rigid. We can also see the meaning of the latter from the relations \eqref{infs}. The condition ${\mathcal{H}}^{2}(\mathfrak{g};\mathfrak{g})=0$ means \eqref{infs-a} does not have any solution. Therefore, other equations which use $\psi_1$ as an input do not have any solution either. However, there are cases  where $\psi_{1}$ exists but not other $\psi_{r}$ where $r\geq 2$. These deformations are not integrable and do not admit a power series as \eqref{eq:defdefor}. An example of such rigid algebra with nontrivial linear (infinitesimal) deformation can be found in \cite{richardson1967rigidity}. {In our analysis of \bms\ or \hbms\ algebra deformations we will find some other examples of this case.}
\end{itemize}

\paragraph{Relation between deformation and contraction.}
As mentioned  the contraction procedure is inverse of deformation. In fact by taking the $\varepsilon\to 0$ limit one can return  to the  original algebra $\mathfrak{g}$ from the  deformed algebra. Physically, deformation can be interpreted as reaching to a ``corrected'' theory from a nonexact one while contraction is getting a certain limit of this corrected theory in which deformation parameter appears as some physical parameter or in some specific examples as  a fundamental constant of nature, e.g. see \cite{mendes1994deformations}. All finite dimensional semisimple Lie algebras are rigid in the sense that they do not admit any deformation so they can be viewed as symmetry algebras of a more fundamental or an undeformed physical theory. On the other hand a non-semisimple Lie algebra (like Galilean algebra) as the symmetry algebra of a physical theory admits deformations (to Poincar\'e algebra, where speed of light $c$ is the deformation parameter and a constant of nature). Conversely, a relativistic theory can be reduced to a non-relativistic theory through a contraction (obtained through sending $c$ to infinity). There are  different approaches to contract a certain Lie algebra. The concept of contraction as limiting process on a Lie algebra was first introduced by Segal in \cite{segal1951class} and then by In\"on\"u and Wigner in \cite{Inonu:1953sp}, which is more known among physicists. Further discussions may be found in  \cite{patera1992graded, weimar1995contractions, gilmore2012lie}.

\paragraph{Examples of simple Lie algebras and their  contraction.} $SO(n+1)$ is the isometry group of $n$ dimensional sphere $S^{n}$ of an arbitrary radius $R$. This sphere may be embedded into the $n+1$ dimensional flat space $\mathbb{R}^{n+1}$. The commutation relations of its algebra are
\begin{equation}
    [s_{ij},s_{kl}]=s_{il}\delta_{jk}+s_{jk}\delta_{il}-s_{ik}\delta_{jl}-s_{jl}\delta_{ik},\label{so(n)}
\end{equation}
where $0\leq i,j,k,l\leq n+1$. In the limit of $R\to\infty$, as (IW) contraction, the above algebra reduces to the algebra of Euclidean group $E(n)$ or $ISO(n)$, with the associated algebra:
\begin{equation}
\begin{split}
   & [r_{ij},r_{kl}]=r_{il}\delta_{jk}+r_{jk}\delta_{il}-r_{ik}\delta_{jl}-r_{jl}\delta_{ik},\\
  & [r_{ij},p_{k}]=p_{i}\delta_{jk}-p_{j}\delta_{ik},\\
  & [p_{i},p_{j}]=0,\label{iso(n)}
\end{split}
\end{equation}
where one has replaced $r_{ij}=s_{ij}$ and $s_{i,n+1}=(1/R) p_{i}$ for $0\leq i,j,k,l \leq n$. 

The next example is the isometry group of  $n$ dimensional hyperboloid $H^{n}$ radius $R$, $SO(n,1)$. The commutation relations of $\mathfrak{so}(n,1)$ is the same as \eqref{so(n)}. A similar limit again brings us to Euclidean group $E(n)$ or $ISO(n)$ with the same commutation relations as $\mathfrak{iso}(n)$. By the Weyl unitarity trick  two algebras $\mathfrak{so}(n+1)$ and $\mathfrak{so}(n,1)$ are mapped to each other. Therefore,  both of $\mathfrak{so}(n+1)$ and $\mathfrak{so}(n,1)$ algebras  in the  $R\to \infty$ limit can be contracted to $\mathfrak{iso}(n)$. Conversely, this means that by deformation procedure one can reach  $\mathfrak{so}(n+1)$ or $\mathfrak{so}(n,1)$, starting from $\mathfrak{iso}(n)$. This is  very similar to the problem  considered by Levy-Nahas in \cite{levy1967deformation} where it is shown that the only algebras which can be contracted to the Poincar\'{e} algebra $\mathfrak{iso}(3,1)$, are $\mathfrak{so}(4,1)$ and $\mathfrak{so}(3,2)$ which are, respectively the algebras corresponding to dS$_4$ and AdS$_4$ isometries.

\subsection{Rigidity of finite dimensional Lie algebras, Whitehead and Hochschild-Serre theorems}

Here we apply theorem \ref{2.1} to finite dimensional Lie algebras. First, we consider simple and semi-simple Lie algebras. In this case a theorem which has been proven by Whitehead has a very important role

 \paragraph{Theorem 3.2.}\label{2.2} ({\it Whitehead's Theorem}) {\it Let $M$ be a (finite-dimensional) simple module over a semisimple
Lie algebra $\mathfrak{g}$ with nontrivial $\mathfrak{g}$-action. Then $\mathcal{H}^p(\mathfrak{g};M)=0$ for all $p \geq 0$.} 

 As $\mathfrak{g}$ is also a $\mathfrak{g}$-module, the above theorem states that $\mathcal{H}^p(\mathfrak{g};\mathfrak{g})=0$ for all $p \geq 0$. This latter and theorem \ref{2.1} leads to,
 \paragraph{Corollary 3.1.}\label{2.3}  {\it All  semisimple finite dimensional Lie algebras are rigid.} 

 As examples all $\mathfrak{so}(n)\,\,n\geq 3$, $\mathfrak{su}(n)$ and $\mathfrak{sl}(n,\mathbb{R}\,\,\text{or}\,\,\mathbb{C})$ are rigid in the sense that they cannot be deformed to another non-isomorphic algebra. 
 
 \paragraph{Semi-direct sum structure.} There are also some powerful mathematical methods for specific finite dimensional  Lie algebra structures. Many of finite dimensional Lie algebras which appear in physics have semi-direct sum structure in the sense that $\mathfrak{g}$ is a semi-direct sum of two (or possibly more) parts as $\mathfrak{g}=\mathfrak{g}_{1}\inplus\mathfrak{g}_{2}$ in which $\mathfrak{g}_{2}$ is ideal part of $\mathfrak{g}$, e.g. Poincar\'{e} algebra in general $d$ dimensions has this form (actually any non-solvable finite dimensional Lie algebra over a field of characteristic zero can be written as a semi direct sum of a semisimple Lie algebra and its radical ideal. This is known as the Levi decomposition). Let us look at the commutation relations of Poincar\'{e} Lie algebra in $3d$,
  \begin{equation} 
\begin{split}
 & i[\mathcal{J}_m,\mathcal{J}_n]=(m-n)\mathcal{J}_{m+n}, \\
 &i[\mathcal{J}_m,\mathcal{P}_n]=(m-n)\mathcal{P}_{m+n},\\
 &i[\mathcal{P}_m,\mathcal{P}_n]=0,
\end{split}\label{poincare-3D}
\end{equation}
where $m,n=0,\pm1$. This algebra is a semi-direct sum of two parts as: Poincar\'{e}$_{3}=\iso= \sltwo\inplus_{ad}\mathfrak{h}$ in which $\sltwo$ is spanned by $\mathcal{J}$s and $\mathfrak{h}$ is ideal part which is spanned by $\mathcal{P}$s.
For algebras with semi-direct sum structure, there is the classic {\it Hochschild-Serre factorization theorem}  proven in \cite{MR0054581} which makes the calculations more convenient. 
\paragraph{Theorem 3.3.}({\it Hochschild-Serre factorization theorem})\label{2.4} {\it Let $\mathfrak{g}$ be a finite dimensional Lie algebra over the field $\mathbb{F}$ of characteristic 0, and let $M$ be a finite dimensional $\mathfrak{g}$-module. Suppose that $\mathfrak{h}$ is ideal
 of $\mathfrak{g}$ such that $\mathfrak{g}/\mathfrak{h}$ is semi-simple. Then
 $\mathcal{H}^p(\mathfrak{g};M)\cong \sum_{r+s=p}\mathcal{H}^r(\mathfrak{g}/\mathfrak{h};\mathbb{F})\otimes \mathcal{H}^s(\mathfrak{h};M)$.}

As we mentioned $\mathfrak{g}$ is a $\mathfrak{g}$-module and the characteristic of real numbers $\mathbb{R}$ (and also complex number $\mathbb{C}$) is zero. Then, theorem \ref{2.4} implies that
 \begin{equation}
     \mathcal{H}^2(\mathfrak{g};\mathfrak{g})\cong  \mathcal{H}^2(\mathfrak{h};\mathfrak{g}).
 \end{equation}
This means that all nontrivial infinitesimal deformations of $\mathfrak{g}$ are  located in the ideal part of $\mathfrak{g}$. In other words, it is not necessary to analyze the deformation of the full algebra,  the deformation of  the ideal part already yields the most general deformations of $\mathfrak{g}$. In the case of  the Poincar\'{e} Lie algebra in $3d$ the ideal part is the third line in (\ref{poincare-3D}). According to   the {\it Hochschild-Serre factorization theorem} this algebra is deformed to  
\begin{equation} 
\begin{split}
 & i[\mathcal{J}_m,\mathcal{J}_n]=(m-n)\mathcal{J}_{m+n}, \\
 &i[\mathcal{J}_m,\mathcal{P}_n]=(m-n)\mathcal{P}_{m+n},\\
 &i[\mathcal{P}_m,\mathcal{P}_n]=\sigma (m-n)\mathcal{J}_{m+n},\qquad m,n=0,\pm,\ \sigma^2=1,
\end{split}\label{satbilized-3d-Poincare}
\end{equation}
which is the $\mathfrak{so}(2,2)$ algebra, isometry of AdS$_3$ for $\sigma=+1$ or the $\mathfrak{so}(3,1)$ algebra, isometry group of dS$_3$, for $\sigma=-1$. See \cite{Figueroa-OFarrill:2017sfs, Figueroa-OFarrill:2017ycu, Figueroa-OFarrill:2017tcy} for more discussions and examples.

\subsection{Hochschild-Serre spectral sequence }\label{sec:HS-seq}

Although in the case of infinite dimensional Lie algebras the Hochschild-Serre factorization theorem does not work, one can still use Hochschild-Serre spectral sequence method {which works for both finite and infinite cases} and extract information about deformations from that. We focus on certain split abelian extensions of Lie algebras. For a Lie algebra $(\mathfrak{g},[,])$ {with a semi-direct sum structure as $\mathfrak{g}=\mathfrak{g}_{0}\inplus \mathfrak{h}$ where $\mathfrak{h}$ is an abelian ideal and $\mathfrak{g}_{0}\cong \mathfrak{g}/\mathfrak{h}$ is its quotient Lie algebra}, we have the following short exact sequence
\begin{equation}
    0\longrightarrow \mathfrak{h}\longrightarrow \mathfrak{g} \longrightarrow \mathfrak{g}/\mathfrak{h}\cong \mathfrak{g}_{0}\longrightarrow 0,\label{short-exact}
\end{equation}
where arrows show {Lie algebra morphisms. Short exactness means the image of each morphism is equal to the kernel of the next  . For this sequence one obtains the Hochschild-Serre spectral sequence of cochain complexes whose  first terms are
    \begin{equation*} 
\begin{split}\label{chain-complex}
 E_{0}^{p,q}=\mathcal{C}^{p}(\mathfrak{g}_{0},\mathcal{C}^{q}(\mathfrak{h},M)),\ E_{1}^{p,q}=\mathcal{H}^{p}(\mathfrak{g}_{0};\mathcal{C}^{q}(\mathfrak{h},M)),\ 
   E_{2}^{p,q}=\mathcal{H}^{p}(\mathfrak{g}_{0};\mathcal{H}^{q}(\mathfrak{h};M)),\ ...,\ E_{n}^{p,q},... 
\end{split}
\end{equation*}
in which $M$ is a $\mathfrak{g}$-module, $\mathcal{C}^{p}$ as we introduced earlier is the space of $p$-cochains and $E$'s are related to each other by the differential operator $d_{n}^{p,q}:E_{n}^{p,q}\longrightarrow E_{n}^{p+n,q-n+1}$ \cite{MR0054581,fuks2012cohomology}. In some specific cases  the differential function becomes trivial for $n\geq n_0$ (for certain $n_0$) and $E_{n}^{p,q}, \forall n\geq n_0$ are isomorphic to each other and therefore, $E_{n}^{p,q}\cong E_{\infty}^{p,q}$. So for the latter we have \footnote{Note that,  in general, this equality is true modulo extensions but all the terms in our cases are vector spaces and hence those extensions do not appear.}
\begin{equation}
    \mathcal{H}^{n}(\mathfrak{g};M)=\oplus_{p+q=n}E_{\infty}^{p,q}.\label{dec}
\end{equation}
{In this setting by the Hochschild-Serre spectral theorem \cite{MR0054581} we have} 
\begin{equation}
E_{2}^{p,q}=
    \mathcal{H}^{p}(\mathfrak{g}_{0}; \mathcal{H}^{q}(\mathfrak{h},M)).\label{E2-dec}
\end{equation}
 This theorem works for both finite and  infinite  dimensional Lie algebras.
{For those split abelian extensions with the property that the ideal action on $M$ is trivial, Theorem 1.2 in \cite{degrijse2009cohomology} states that 
we always have} $n_{0}=2$ and {therefore} $E_{2}^{p,q}\cong E_{\infty}^{p,q}$. So, combining \eqref{dec} and \eqref{E2-dec} one finds 
\begin{equation}
    \mathcal{H}^{2}(\mathfrak{g};M)=\oplus_{p+q=2}E_{2}^{p,q}.\label{dec2}
\end{equation}

{Note that $\mathfrak{h}$ is an ideal of $\mathfrak{g}$ and hence a $\mathfrak{g}$-module and because it is abelian, as a $\mathfrak{g}$-module its action on itself is trivial. Using the short exact sequence (\ref{short-exact}) we consider 
 $\mathfrak{g}_{0}$ as a $\mathfrak{g}$-module as well. In this way the action of $\mathfrak{h}$ on $\mathfrak{g}_{0}$ is trivial. We conclude that via the above arguments, $\mathfrak{g}_{0}$ and $\mathfrak{h}$ are both $\mathfrak{g}$-modules satisfying conditions of theorem 1.2 in \cite{degrijse2009cohomology}, and one can compute the spaces $\mathcal{H}^{2}(\mathfrak{g};\mathfrak{g}_{0})$ and $\mathcal{H}^{2}(\mathfrak{g};\mathfrak{h})$.} 
 
{The short exact sequence \eqref{short-exact} induces the long exact sequence at the level of cohomologies 
\begin{equation}
    \begin{split} & \cdots \longrightarrow \mathcal{H}^{1}(\mathfrak{g};\mathfrak{g}_{0})\longrightarrow \mathcal{H}^{2}(\mathfrak{g};\mathfrak{h}) \longrightarrow \mathcal{H}^{2}(\mathfrak{g};\mathfrak{g})\longrightarrow \mathcal{H}^{2}(\mathfrak{g};\mathfrak{g}_{0})
     \longrightarrow \mathcal{H}^{3}(\mathfrak{g};\mathfrak{g}_{0}) \longrightarrow \cdots \label{long-exact}
    \end{split}
\end{equation}
One may use the above sequence to get information about  $\mathcal{H}^{2}(\mathfrak{g};\mathfrak{g})$ or even compute it. The long exact sequence \eqref{long-exact} is true for both finite and infinite dimensional Lie algebras with the semi-direct sum structure. In finite dimensional cases as a consequence of Hochschild-Serre factorization theorem we have $ \mathcal{H}^{2}(\mathfrak{g};\mathfrak{g})\cong \mathcal{H}^{2}(\mathfrak{g};\mathfrak{h})$. The latter can be obtained from another long exact sequence 
 \begin{equation}
    \begin{split} & \cdots \longrightarrow \mathcal{H}^{1}(\mathfrak{h};\mathfrak{g})\longrightarrow \mathcal{H}^{2}(\mathfrak{g}_{0};\mathfrak{g}) \longrightarrow \mathcal{H}^{2}(\mathfrak{g};\mathfrak{g})\longrightarrow \mathcal{H}^{2}(\mathfrak{h};\mathfrak{g})
     \longrightarrow \mathcal{H}^{3}(\mathfrak{g}_{0};\mathfrak{g}) \longrightarrow \cdots \label{long-exact2}
    \end{split}
\end{equation}
in which  $\mathcal{H}^{2}(\mathfrak{g}_{0};\mathfrak{g})=\mathcal{H}^{3}(\mathfrak{g}_{0};\mathfrak{g})=0$.
 In the case of infinite dimensional Lie algebras we can still use \eqref{long-exact}. While the sequence has in general infinite terms, in some specific cases one finds that some of terms in \eqref{long-exact} are equal to zero, leading to another  short exact sequence. In such situations we can infer some information about lower cohomologies. As we will discuss in next sections, from \eqref{dec2} and \eqref{long-exact} we can learn about cohomological structure of asymptotic symmetry algebras introduced in section \ref{sec:2}.

\subsection{Deformations and rigidity of infinite dimensional Lie algebras}\label{sec:3.4}

{So far we considered deformation theory of Lie algebras for finite dimensional Lie algebras. One can readily extend the main results and notions discussed earlier, except for the the Hochschild-Serre factorization theorem, to infinite dimensional Lie algebras. Here we will be interested in a specific class of infinite dimensional algebras with countable bases. Such algebras, like \emph{Witt algebra} which is the Lie algebra of  vector fields on circle $S^{1}$ and its universal central extension, Virasoro algebra, have very crucial role in {the} context of quantum field theory and string theory.}

Rigidity and deformation analysis of infinite dimensional Lie algebras are more complicated than finite dimensional Lie algebras. For instance {in spite of existence of semi-direct} sum structures for certain infinite dimensional algebras, like $\mathfrak{bms}_3$, we cannot use the Hochschild-Serre factorization theorem.  Rigidity of Witt and Virasoro algebras {has} been considered in \cite{fialowski1990deformations, guerrini1999formal, fialowski2003global,  fialowski2012formal, schlichenmaier2014elementary} where it is shown that Witt algebra and its universal central extension are formally rigid. Later Fialowski and Schlichenmaier showed that Witt algebra is not globally rigid and global deformation of Witt and Virasoro algebra leads to a specific family of algebras  known as Krichever-Novikov type \cite{fialowski2003global}. Fialowski  {has} also worked on formal deformations of some other infinite dimensional Lie algebras such as vector fields on the line \cite{fialowski1988example}.   
 
\emph{Affine Kac-Moody} Lie algebras are another class of infinite dimensional Lie algebras of interest in theoretical physics. Roger in \cite{hazewinkel2012deformation} has given a theorem stating that affine Kac-Moody Lie algebras are (formally) rigid in the sense that their second adjoint cohomology is zero. 
 
\emph{Schr\"{o}dinger-Virasoro} algebras are also infinite dimensional algebras which appear in nonequilibrium statistical physics \cite{Christe:1993ij, Henkel:2012zz}. Deformations of these families have been studied by Unterberger and Roger. They have found three nontrivial formal deformations for a specific kind of Schr\"{o}dinger-Virasoro algebra known as twisted Schr\"{o}dinger-Virasoro algebra. We refer the reader to \cite{unterberger2011schrodinger} for further analysis of Schr\"{o}dinger-Virasoro algebras and their deformations.    
 
 The concept of contraction has also been considered for some infinite dimensional algebras. For instance, contractions of Kac-Moody and affine Kac-Moody Lie algebras {have} been studied in \cite{majumdar1993inonu, daboul2008gradings}. The second example which was discussed earlier in section \ref{sec:2} is contraction of asymptotic symmetry algebras of (A)dS$_3$ spacetime in $3d$ to asymptotic symmetry algebra of flat spacetime in the same dimension. 
 

\section{Deformations of \texorpdfstring{$\mathfrak{bms}_3$}{BMS3}  algebra}\label{sec:4}

In this section we consider deformations of  $\mathfrak{bms}_{3}$ defined in (\ref{bms3}).
As discussed the Hochschild-Serre factorization theorem is not applicable for infinite dimensional Lie algebras and working with them is more complicated than finite dimensional cases. Here, we first analyze possible deformations of $\mathfrak{bms}_{3}$ algebra by deforming  each commutation relation of $\mathfrak{bms}_{3}$ algebra separately. Of course one should check that in this way we do not miss any possible deformation which may involve two sets of commutators. This will be discussed in section \ref{sec:4.5}. Finally, we study obstructions, which infinitesimal deformations yield formal deformations and which rigid algebras are obtained from deformations of $\mathfrak{bms}_{3}$. In section \ref{sec:cohomology-bms} we establish and reinforce our results of the previous subsections through algebraic cohomology discussions. We have summarized all possible formal deformations of the \bms\ and $\mathfrak{KM}_{\mathfrak{u}(1)}$ algebras in Figure. \ref{Fig-abnu}.

\subsection{Deformation of commutators of two \texorpdfstring{$\mathcal{P}$'s}{PP} }\label{sec:4.1}

For the first step we construct  all deformations of the ideal part of $\mathfrak{bms}_{3}$. As we can see from (\ref{bms3}) the ideal part of $\mathfrak{bms}_{3}$ algebra is spanned by supertranslation generators $\mathcal{P}$s. We deform infinitesimally this ideal part as
\begin{equation} 
i[\mathcal{P}_{m},\mathcal{P}_{n}]=\varepsilon\psi_{1}^{PP}(\mathcal{P}_{m},\mathcal{P}_{n}),\label{deform of ideal}
\end{equation}
where $\varepsilon$ is deformation parameter and  $\psi_{1}^{PP}(\mathcal{P}_{m},\mathcal{P}_{n})$ is a $\mathfrak{bms}_{3}$-valued 2-cocycle and hence admits and expansion as
\begin{equation}\label{psi1-PP}
\psi_{1}^{PP}(\mathcal{P}_{m},\mathcal{P}_{n})=F(m,n)\mathcal{J}_{u(m,n)}+G(m,n)\mathcal{P}_{v(m,n)},
\end{equation}
in which coefficients $F(m,n)$ and $G(m,n)$ are antisymmetric functions while indices $u(m,n)$ and $v(m,n)$ are symmetric functions. For convenience, we can extract the antisymmetric part of $F(m,n)$ (or  $G(m,n)$) as
\begin{equation} 
F(m,n)=(m-n)f(m,n),\qquad f(m,n)=f(n,m).
\end{equation}

The main goal is to find the explicit form of $f, g, u$ and $v$. One may show that the general form of functions $u(m,n)$ and $v(m,n)$ can be chosen as $u(m,n)=v(m,n)=m+n$.
To this end, we recall \eqref{bms=witt+ideal} and that $\mathfrak{witt}$ is related to the adjoint action of vectors on two-tensors on an S$^1$.  Consider tensor densities on circle  $\mathcal{M}(\phi)(d\phi)^{\lambda}$ in which $\lambda\in \mathbb{Z}$ is called the degree of tensor density and $\mathcal{M}(\phi)$ is a periodic function on circle (e.g. see \cite{Oblak:2016eij} and references therein). In this way $\mathcal{J}(\phi)(d\phi)^{-1}$ is the vector field on circle. So one finds the Fourier expansion of $\mathcal{M}(\phi)(d\phi)^{\lambda}$ as
\begin{equation}
    \mathcal{M}(\phi)(d\phi)^{\lambda}=\sum M_{n}e^{in\phi}(d\phi)^{\lambda}.\label{Fourier}
\end{equation}
where $e^{in\phi}(d\phi)^{\lambda}$ are the basis. In the terminology of 2d CFT's, if $\mathcal{J}$ is associated with conformal transformations, then $\lambda$ is the conformal weight of operator ${\cal{M}}$. As an example the basis of Witt algebra are $e^{in\phi}\partial_{\phi}=\mathcal{J}_{n}$. The adjoint action of Witt generators on tensor density $\mathcal{M}(\phi)$ is obtained as
$adj_{\mathcal{J}(\phi_1)}\mathcal{M}(\phi_2)=[\mathcal{J}(\phi_1),\mathcal{M}(\phi_2)]=\big(\mathcal{M}^{'}(\phi_2)\mathcal{J}(\phi_1)+\lambda \mathcal{M}(\phi_2)\mathcal{J}^{'}(\phi_1)\big)\delta(\phi_1-\phi_2)$ \cite{Oblak:2016eij}. One can then write the above in terms of its Fourier basis and see that the final result has the form 
$i[\mathcal{J}_{m},\mathcal{M}_{n}]=(\lambda m-n)\mathcal{M}_{m+n}$. That is, dealing with periodic tensors on a circle  $u(m,n), v(m,n)$ appearing in \eqref{psi1-PP} should be $m+n$.

To determine functions $f, g$ one has to consider the Jacobi identities (the 2-cocycle conditions \eqref{coe} or equivalently \eqref{infs-b}). The only Jacobi identities which can lead to some relations for $f, g$, are $[\mathcal{P}_{m},[\mathcal{P}_{n},\mathcal{P}_{l}]]+\text{cyclic permutations}=0$ and  $[\mathcal{P}_{m},[\mathcal{P}_{n},\mathcal{J}_{l}]]+\text{cyclic permutations}=0$. The former just contains a relation with linear terms in $\varepsilon$ for $f(m,n)$, yielding
\begin{equation}
(n-l)(m-n-l)f(n,l)+ (l-m)(n-l-m)f(l,m)+ (m-n)(l-m-n)f(m,n)=0, \label{f-first}
\end{equation}
while from the second, one gets
\begin{equation}\label{fg-first}
\begin{split}
&(n-l)(m-n-l)f(m,l+n)+ (l-m)(n-l-m)f(n,l+m)+(m-n)(l-m-n)f(m,n)=0,  \\ 
&(n-l)(m-n-l)g(m,l+n)+ (l-m)(n-l-m)g(n,l+m)+(m-n)(l-m-n)g(m,n) =0.
\end{split}
\end{equation}
As we will show below the equations for $f$ in \eqref{fg-first} and \eqref{f-first} {are the same} and $f,g$ should satisfy the same equation. So, we only focus on $f$ and on \eqref{fg-first}.
We now tackle (\ref{fg-first}). Our goal is to find the most general form of $f(m,n), g(m,n)$. 
As the first step we will prove a very important proposition.
\paragraph{Proposition.}
The most general solution of (\ref{fg-first}) is
  \begin{equation}
f(m,n)=\text{constant}.
 \end{equation}
We prove this proposition by proving some lemmas.
  \paragraph{Lemma 1.}
 For $m\neq n$
 \begin{equation}
 f(m,n)=f(m,n+m).
 \end{equation}
  \begin{proof}
 One puts $l=m$ into the relation (\ref{fg-first}) and finds:
 \begin{equation}
 (n-m)f(m,n+m)+(m-n)f(m,n)=0,
 \end{equation}
 then choosing $m\neq n$ one finds :
 \begin{equation}
   f(m,n)=f(m,n+m).  
 \end{equation}
 \end{proof}
 
 \paragraph{Lemma 2.}
 For $n\neq 0$
 \begin{equation}
 f(0,n)=f(0,1).
 \end{equation}
  \begin{proof}
 By insertion of $l=1, m=0$ into the relation (\ref{fg-first}) one finds:
  \begin{equation}
 (-n-1)f(0,n+1)+f(n,1)+(n)f(0,n)=0.
   \end{equation}
 From Lemma 1 one can show that $f(1,n)=f(1,0)$ (by $n$ times using Lemma 1). So we have:
  \begin{equation}
 (-n-1)f(0,n+1)+f(0,1)+(n)f(0,n)=0.
 \end{equation}\label{f-induction}
We  prove the lemma by induction. For $n=1$ the statement holds $f(0,1)=f(0,1)$. We suppose for $n=k$ the statement holds which means $f(0,1)=f(0,k)$. We use the latter assumption and (\ref{f-induction}) to reach
 \begin{equation}
(k+1)f(0,k+1)=(k+1)f(0,1)\rightarrow f(0,k+1)=f(0,1).
\end{equation}
 \end{proof}
  \paragraph{Lemma 3.} For $m\neq n$ we have
  \begin{equation}
 f(m,n)=f(0,n).
 \end{equation}
 \begin{proof}
By inserting $l=-m$ in (\ref{fg-first}) and we get:
\begin{equation}
(n+m)(2m-n)f(m,n-m)+(-2m)(n)f(n,0)+(m-n)(-2m-n)f(m,n)=0.\label{lemma 3}
\end{equation}
From Lemma 1 we infer that $f(m,n)=f(m,n-m)$. By putting the latter into (\ref{lemma 3}) we get
\begin{equation}
 f(m,n)=f(n,0).
\end{equation}
\end{proof}
Combining results of Lemmas 2 and 3 we conclude $f(m,n)=f(0,1)=\text{constant}=f$ and proof of the proposition is completed. It is immediate to see that this solution also solves \eqref{f-first}. So, by solving the $2-$cocycle condition  we have found that $\psi_{1}^{PP}(\mathcal{P}_{m},\mathcal{P}_{n})=f(m-n)\mathcal{J}_{m}+g(m-n)\mathcal{P}_{m}$ in which $f,g$ are some arbitrary (complex or real) constants, i.e.
\begin{equation}
 i[\mathcal{P}_{m},\mathcal{P}_{n}]=\varepsilon_{1}(m-n)\mathcal{J}_{m+n}+\varepsilon_{2}(m-n) \mathcal{P}_{m+n},
\end{equation}
where $\varepsilon_{1}=\varepsilon f$ ($\varepsilon_{2}=\varepsilon g$). So we have the theorem
\paragraph{Theorem 4.1}{\it The most general infinitesimal deformation of $\mathfrak{bms}_{3}$ ideal part is 
 \begin{equation} 
i[\mathcal{P}_{m},\mathcal{P}_{n}]=\varepsilon_{1}(m-n)\mathcal{J}_{m+n}+\varepsilon_{2}(m-n) \mathcal{P}_{m+n}.\label{theorem 4.1}
\end{equation}}
We note that special $\varepsilon_1=0$ or $\varepsilon_2=0$  cases, corresponding to  
$i[\mathcal{P}_{m},\mathcal{P}_{n}]=\varepsilon_2(m-n)\mathcal{P}_{m+n}$ or $i[\mathcal{P}_{m},\mathcal{P}_{n}]=\varepsilon_1(m-n)\mathcal{J}_{m+n}$ are both isomorphic to the general case of \eqref{theorem 4.1}. This means that  by a proper redefinition of the generators they can be mapped onto each other and the algebra becomes the direct sum of two Witt algebras as $\mathfrak{witt}_{L}\oplus \mathfrak{witt}_{R}$. Since the case $\varepsilon_{1}=0$ and the general form are isomorphic, without any loss of generality only consider
 $i[\mathcal{P}_{m},\mathcal{P}_{n}]=\varepsilon_1(m-n)\mathcal{J}_{m+n}$. 
$|\varepsilon_1|$ may be absorbed into the normalization of ${\cal P}_n$ but its sign will remain: $i[\mathcal{P}_{m},\mathcal{P}_{n}]=\sigma (m-n)\mathcal{J}_{m+n}, \sigma^2=1$. Therefore, there are two choices for this deformation. These two choices parallels the two choices for the stabilization of $\mathfrak{iso}(2,1)$ into $\mathfrak{so}(2,2)$ and $\mathfrak{so}(3,1)$ discussed in \eqref{satbilized-3d-Poincare}.

So far we have found a nontrivial infinitesimal deformation of $\mathfrak{bms}_{3}$ ideal part, showing that the $\mathfrak{bms}_{3}$ algebra is not, at least, ``infinitesimally rigid'' and $\mathcal{H}^2(\mathfrak{bms}_{3};\mathfrak{bms}_{3})\neq 0$. One can show that in an appropriate basis deformed algebra is just the direct sum of two Witt algebras $\mathfrak{witt}_{L}\oplus \mathfrak{witt}_{R}$ and so not isomorphic to $\mathfrak{bms}_{3}$. This is of course expected  from contraction procedure which was introduced in section \ref{sec:2.3-contraction}. Here, we have shown there is a unique deformation in $\mathfrak{bms}_{3}$ ideal part corresponding to that contraction. 
To establish that $\mathfrak{bms}_{3}$ is not also formally rigid one should check the integrability conditions or probably obstructions for this infinitesimal deformation, which will be discussed in section \ref{sec:integrability-bms}.

\subsection{ Deformation of commutators of \texorpdfstring{$[\mathcal{J},\mathcal{P}]$}{JP} }\label{sec:4.2}

Now, we consider  deformations of commutator of superrotations and supertranslations which is the second line in (\ref{bms3}) without changing other commutators. To this end and as in the previous subsection, we add a $2-$cocycle function:
\begin{equation} 
 i[\mathcal{J}_{m},\mathcal{P}_{n}]=(m-n)\mathcal{P}_{m+n}+\zeta \psi_{1}^{JP}(\mathcal{J}_{m},\mathcal{P}_{n}).\label{JP-deformation}
\end{equation}
The 2-cocycle $\psi_{JP}(\mathcal{J}_{m},\mathcal{P}_{n})$ is linear combination of generators as
\begin{equation}\label{eqdefIK}
\psi_{1}^{JP}(\mathcal{J}_{m},\mathcal{P}_{n})=I(m,n)\mathcal{J}_{m+n}+K(m,n)\mathcal{P}_{m+n},
\end{equation}
where we have fixed the indices of $\mathcal{J}$ and $\mathcal{P}$ to be $m+n$ (\emph{cf.}  discussion around \eqref{Fourier}) and
the coefficients $I(m,n)$ and $K(m,n)$ are arbitrary functions.

To find the explicit form of functions $I(m,n)$ and $K(m,n)$ we check Jacobi identities. Two different Jacobi identities put constraints on $I(m,n)$ and $K(m,n)$. The first Jacobi identity is  $[\mathcal{P}_{m},[\mathcal{P}_{n},\mathcal{J}_{l}]]+\text{cyclic permutations}=0$. Keeping up to first order in $\zeta$ and using the fact that $[\mathcal{P}_{m},\mathcal{P}_{n}]=0$, we get
\begin{equation} 
 -(m-n-l)I(l,n)+(n-l-m)I(l,m)=0.\label{firsteqI}
\end{equation}
 This relation is exactly the same as $2-$cocycle condition \eqref{coe}. It is obvious that this specific Jacobi does not put any constraint on $K(m,n)$. By some easy steps we show $I(m,n)=0$. Let us put $m=l+n$ into (\ref{firsteqI}) to reach
\begin{equation} 
 (2l)I(l,l+n)=0,
\end{equation}
and for $l\neq 0$, one gets
\begin{equation} 
 I(l,q)=0,
\end{equation}\label{result1}
 where $q=l+n$. By insertion of $l=0$ into (\ref{firsteqI}) we get
 \begin{equation} 
 I(0,n)=-I(0,m),\label{result2}
\end{equation}
for $m\neq n$. Suppose $I(0,n)=F(n)$ so from (\ref{result2}) we have $F(n)=-F(m)$ which means that $I(0,n)=0 ,\,\,\,\text{for}\,\,\,\, n\neq 0$. Finally one concludes that
\begin{equation}
    I(m,n)=0 ,\,\,\,\forall\,\, m,n\in \mathbb{Z}.
\end{equation}

We next examine the Jacobi identity $[\mathcal{J}_{m},[\mathcal{J}_{n},\mathcal{P}_{l}]]+\text{cyclic permutations}=0$ and up to first order in $\zeta$ we obtain,
 {\begin{multline}
(n - l) K(m, l + n) + ( m-n-l) K(n, l) + (l-m) K(
   n, l+m) +\\  +
   (l+m-n) K(m,l)+(n-m) K(m+n,l)=0.\label{eq-K}
\end{multline}}
One puts $l=m=0$ into (\ref{eq-K}) to get
\begin{equation} 
 n (K(0, n) - K(0,0))=0.
\end{equation}
The above then yields 
\begin{equation} 
  K(0, n)=\text{constant}.\label{first contraint on K}
\end{equation}
To solve \eqref{eq-K} we note that it is linear in $K$ and hence linear combination of any two solutions is also a solution. Since the coefficients of the $K$'s is first order in $l,m$ or $n$, the solutions should be homogeneous functions of a given degree $N$, i.e. $K(m,n)=\sum_{r=1}^N A_r m^r n^{N-r}$ where we already used \eqref{first contraint on K}. For $N=0,1$ one may readily check that $K(m,n)=\alpha+\beta m $ in which $\alpha,\beta$ are arbitrary real (complex) constants. So, let us focus on $N\geq 2$ case. For $N=2$ one may show that there is a solution of the form
\begin{equation} 
 K(m,n)=\gamma m(m-n).\label{Ksolution}
\end{equation}
where $\gamma$ is an arbitrary constant. This term, however, is not representing a nontrivial deformation and can be absorbed into normalization of ${\cal P}$. To see this consider redefining ${\cal P}$ as
\begin{equation}\label{rescale-P}
   \mathcal{P}_{n}:= N(n)\tilde{\mathcal{P}}_{n},
\end{equation}
for a function $N$ we are free to choose. Replacing this into (\ref{bms3}) one gets
\begin{equation} 
\begin{split}
 & i[\mathcal{J}_{m},\mathcal{J}_{n}]=(m-n)\mathcal{J}_{m+n}, \\
 &i[\mathcal{J}_{m},\tilde{\mathcal{P}}_{n}]=(m-n)\frac{N(m+n)}{N(n)}\tilde{\mathcal{P}}_{m+n},\\
 &i[\tilde{\mathcal{P}}_{m},\tilde{\mathcal{P}}_{n}]=0.
\end{split}\label{Q-normalization}
\end{equation}
If we choose $N$ as
\begin{equation*}
   N(n)=1+\zeta\gamma n+\mathcal{O}(\zeta^2),
\end{equation*}
then the $\gamma$ term can be absorbed into redefinition of generators. 
\paragraph{Higher $N$ and uniqueness of solutions of \eqref{eq-K}.} As an explicit verification with straightforward algebra for generic $N$ shows that there is no other solution.

The most general deformation of \bms\ through deformation of $[\mathcal{J},\mathcal{P}]$ commutator is hence
\begin{equation} 
\begin{split}
 & i[\mathcal{J}_{m},\mathcal{J}_{n}]=(m-n)\mathcal{J}_{m+n}, \\
 &i[\mathcal{J}_{m},\mathcal{P}_{n}]=-(n+bm+a)\mathcal{P}_{m+n},\\
 &i[\mathcal{P}_{m},\mathcal{P}_{n}]=0,
\end{split}\label{W(a,b)}
\end{equation}
where $a,b \in \mathbb{C}$ are two independent deformation parameters. The Lie algebra (\ref{W(a,b)}) is known as $W(a,b)$ algebra and some of its properties such as its central extensions has been analyzed in  
\cite{Roger:2006rz}.\footnote{Since  the Witt algebra is rigid any extension of the Witt algebra by its abelian representations is a semi-direct sum $\mathfrak{witt}\inplus \mathfrak{witt}_{ab}$ which is the case for $\mathfrak{bms}_{3}$. This also indicates that we cannot expect exotic extensions of the Witt algebra. Furthermore, in deformations of $[\mathcal{J},\mathcal{P}]$ commutator only elements from the abelain part can appear.
Despite the similarity between the \bms\ case and the kinematical algebras, e.g. see \cite{Figueroa-OFarrill:2017sfs, Figueroa-OFarrill:2017ycu, Figueroa-OFarrill:2017tcy}, this result already shows the difference between the infinite and finite dimensional algebras and that the former does not obey Hochschild-Serre factorization theorem.}  

\paragraph{More on $W(a,b)$ algebras.}
As we mentioned before, ${W}(a,b)$ algebra which is an extension of the Witt algebra, has been studied in different papers \cite{gao2011low,ovsienko1996extensions}. It has a semi-direct sum structure as $W(a,b)\cong \mathfrak{witt}\inplus \mathcal{P}(a,b)$ in which $\mathcal{P}(a,b)$, with $a, b$ being arbitrary real (or complex) constants, is called tensor density module. Here we briefly discuss two interesting questions about these algebras: 1. What is the physical interpretations of parameter $a,b$? 2.  Are there  special points in $(a,b)$ parameter space? 

To answer the first question we rewrite the algebra as 
\begin{equation} 
\begin{split}
 & i[\mathcal{J}_{m},\mathcal{J}_{n}]=(m-n)\mathcal{J}_{m+n}, \\
 &i[\mathcal{J}_{m},\mathcal{P}_{n}]=-\left(n+(1-h)m+a\right)\mathcal{P}_{m+n},\\
 &i[\mathcal{P}_{m},\mathcal{P}_{n}]=0,\label{cwh}
\end{split}
\end{equation}
where $h=1-b$. These commutation relations are very familiar in context of $2d$ conformal field theories. Consider a primary field ${\cal P}(\phi)$ of conformal weight $h$, with Fourier modes ${\cal P}_n$ and assume that $P(\phi)$ has the quasi-periodicity property
\be\label{Fourier-expand-a}
{\cal P}(\phi+2\pi)=e^{2\pi i a} {\cal P}(\phi),\qquad {\cal P}(\phi)=\sum_n {\cal P}_n e^{i(n+a)\phi}.
\ee
The usual conformal transformation for this conformal primary field can then be recast as $[{\cal J},{\cal P}]$ commutator in \eqref{cwh}. Here we are assuming that ${\cal P}_n$ satisfy a commuting (abelian) algebra . The above argument makes it clear that the range of $a$ parameter which yields independent algebras is $a\in[-1/2,1/2]$. However, one may show that the negative and positive values of $a$ parameter are related by a $\mathbb{Z}_2$ parity transformation which takes $\phi\to 2\pi-\phi$, i.e. a $\mathbb{Z}_2$ inner automorphism of the $W(a,b)$ algebra: ${\cal J}_n\to -{\cal J}_{-n},\ {\cal P}_n\to -{\cal P}_{-n}$. Therefore, the independent range for $a$ parameter is $[0,1/2]$, as depicted in Figure. \ref{Fig-abnu}.   

The deformation which moves us along the $b$ direction in the parameter space, given the physical interpretation above, can be understood as an RG flow in the presumed $2d$ CFT dual to the $3d$ flat space (which realizes \bms \ as its symmetry algebra). A flow from $b=-1$ $(h=2)$ to $b=0$ $(h=1)$ is what is expected from comparing the asymptotic symmetry and the ``near horizon'' symmetry analysis. We shall comment on this latter further in the discussion section.


As for the second question, the definition of $W(a,b)$ algebra implies that \bms$=W(0,-1)$ and the $\mathfrak{u}(1)$ Kac-Moody algebra is $W(0,0)$. These two points are special in the sense that one can deform the algebra in two different ways while in a generic point of $a,b$ space allowed deformations moves us only within the $W(a,b)$ family.\footnote{By $W(a,b)$ we may mean a specific algebra for a given $a,b$ parameters, or the ``$W(a,b)$ family,'' which means all algebras for generic but different values of $a,b$.\label{footnote3}} As discussed, one can deform the ideal part of \bms\ to $\mathfrak{witt}\oplus \mathfrak{witt}$, while one can also keep the ideal part intact and change $(a,b)$ from $(0,-1)$ value. As for the $\mathfrak{u}(1)$ Kac-Moody case, one can check that one cannot deform the ideal part. However, besides deforming the $[{\cal J},{\cal P}]$ part and moving in $(a,b)$ plane, one is also allowed to deform the algebra in the $[{\cal J},{\cal J}]$ part (by adding a $(m-n) {\cal P}_{m+n}$ 2-cocyle). We have depicted the parameter space of $W(a,b)$ algebras in Figure. \ref{Fig-abnu}, which shows as far as the infinitesimal deformations are concerned, there are no other special points.\footnote{We note that when the central extensions of $W(a,b)$ is concerned and as discussed in \cite{gao2011low}, while for generic $(a,b)$ there is always one central extension in the Witt part of the algebra, there are other special points in $a,b$ space with more possibilities for central extensions. Points $(0,-1)$, $(0,0)$, $(0,1)$ and $(1/2,0)$ are special in the sense that they admit other central extensions \cite{gao2011low}.}

\subsection {Deformation of commutators of two \texorpdfstring{$\mathcal{J}$'s}{JJ} }\label{sec:4.3}

We finally consider deformations of $[\mathcal{J},\mathcal{J}]$ part of $\mathfrak{bms}_{3}$. As it was mentioned the Witt algebra is rigid in the sense that it cannot be deformed by a $\mathcal{J}$ valued 2-cocycle. We can however add the 2-cocycle which is linear combination of $\mathcal{P}$s as
\begin{equation}
i[\mathcal{J}_{m},\mathcal{J}_{n}]=(m-n)\mathcal{J}_{m+n}+\eta\psi_{1}^{JJ}(J_{m},J_{n})=
(m-n)\mathcal{J}_{m+n}+\eta(m-n)h(m,n)\mathcal{P}_{m+n}\label{psi-JJ}
\end{equation} 
in which $\psi_{JJ}(J_{m},J_{n})$ is a 2-cocycle and $h(m,n)$ is a symmetric function. We then insert (\ref{psi-JJ}) into the Jacobi identity $[\mathcal{J}_{m},[\mathcal{J}_{n},\mathcal{J}_{l}]]+\text{cyclic}=0$ which in first order in $\varepsilon$ yields
\begin{multline} 
(n-l)(m-n-l)[h(m,l+n)+h(n,l)]+ (l-m)(n-l-m)[h(n,l+m)+h(l,m)]+\\
 (m-n)(l-m-n)[h(l,m+n)+h(m,n)]=0.\label{h-eq}
\end{multline}
One can then readily check that any $h$ of the form
\be\label{JJ-h-Z}
h(m,n)=Z(m)+Z(n)-Z(m+n),
\ee
for any arbitrary function $Z$, provides a solution to \eqref{h-eq}. One may argue that this is in fact the most general solution. To this end, we note that \eqref{h-eq} is linear in $h$ and has quadratic coefficients in $l,m$ or $n$. So, a generic solution for $h(m,n)$ is expected to be a polynomial of homogeneous degree $N$:
\begin{equation*}
    h(m,n)=\sum_{r=0}^N A_{r} m^r n^{N-r},\qquad A_r=A_{N-r}.\label{power series}
\end{equation*}
Subtracting the general solution \eqref{JJ-h-Z}, without loss of generality, one can use the ansatz $h(m,n)=mn \sum_{r=1}^{N-1} A_r m^r n^{N-r}$. Plugging this into \eqref{h-eq} one can see that the equation is only satisfied for $A_r=0$.  

One can, however, show deformations of the form  \eqref{JJ-h-Z} are trivial deformations, as they can be reabsorbed into the redefinition of generators:
\begin{equation} 
\begin{split}
 &  \mathcal{J}_{m}:=\tilde{\mathcal{J}}_{m}+Z(m)\tilde{\mathcal{P}}_{m}, \\
 &\mathcal{P}_{m}:=\tilde{\mathcal{P}}_{m},
\end{split}\label{Z-redefinition}
\end{equation}
where $\tilde{\mathcal{J}}_{m}$ and $\tilde{\mathcal{P}}_{m}$ satisfy  $\mathfrak{bms}_{3}$ commutation relations (\ref{bms3}).

\paragraph{$\mathfrak{u}(1)$ Kac-Moody algebra, ${\mathfrak{KM}}_{\mathfrak{u}(1)}$; an example of deformation of $[\mathcal{J},\mathcal{J}]$ part.}
The $\mathfrak{u}(1)$ Kac-Moody algebra, ${\mathfrak{KM}}_{\mathfrak{u}(1)}$, is defined through 
\begin{equation} 
\begin{split}
 & i[\mathcal{J}_{m},\mathcal{J}_{n}]=(m-n)\mathcal{J}_{m+n}, \\
 &i[\mathcal{J}_{m},\mathcal{P}_{n}]=-n\mathcal{P}_{m+n},\\
 &i[\mathcal{P}_{m},\mathcal{P}_{n}]=0.
\end{split}\label{Kac-Moody}
\end{equation}
One may verify that the $[\mathcal{J},\mathcal{J}]$ part can be deformed by the term $\nu (m-n)\mathcal{P}_{m+n}$. One can then check that the latter is infinitesimal and also formal deformation. This deformation has been found by \cite{Roger:2006rz}. In the next subsections, we discuss the most general deformations of $\mathfrak{u}(1)$ Kac-Moody algebra.


\subsection{ Integrability conditions and obstructions}\label{sec:integrability-bms}

So far we found all infinitesimal deformations of $\mathfrak{bms}_{3}$ which consist of three independent classes: two of them deform the $[\mathcal{J},\mathcal{P}]$ part of the algebra while keeping the ideal part $[\mathcal{P},\mathcal{P}]=0$. This leads to $W(a,b)$ algebra. The last case deforms the ideal part and leads to $\mathfrak{witt}\oplus \mathfrak{witt}$. Here we study integrability conditions of these infinitesimal deformations.    

There are three different approaches to check integrability conditions: 
\begin{itemize}
\item[1.] Direct method: One can consider the entire infinite sequence of relations \eqref{infs} and directly verifies their solutions or probable obstructions. 
\item[2.] ${\cal H}^3$ method: We mentioned that all obstructions are located in $\mathcal{H}^{3}(\mathfrak{bms}_{3};\mathfrak{bms}_{3})$. It can be computed e.g. using the Hochschild-Serre spectral sequence, \emph{cf}. section \ref{sec:HS-seq}. If $\mathcal{H}^{3}(\mathfrak{bms}_{3};\mathfrak{bms}_{3})$ vanishes there is no obstruction. For further discussions we refer the reader to  section \ref{sec:3}.
\item[3.] ``A quick test'': One may examine if an infinitesimal deformation is a formal one by promoting the linear (infinitesimal) deformation $\psi_{1}(g_{i},g_{j})$ to $\Psi(g_{i},g_{j};\varepsilon)$ and check whether it satisfies the Jacobi identity or not. If one finds that the linear term in Taylor expansion of $\Psi(g_{i},g_{j};\varepsilon)$ satisfies the Jacobi \eqref{GJ} one concludes it is also a formal deformation of algebra. This is basically checking whether we are dealing with case \texttt{i.} discussed in section \ref{sec:3.1}.
\end{itemize}
Since in most of the known cases we are dealing with case \texttt{i.}of section \ref{sec:3.1} and since it is very convenient and handy to check this, we will focus on case 3. above and we will comment on the cohomology considerations  in section \ref{sec:cohomology-bms}.

Let us start with the deformation of ideal part of $\mathfrak{bms}_{3}$ which is deformed by $2-$cocycle
$\psi^{PP}_{1}=(m-n)\mathcal{P}_{m+n}$. As discussed in the end of section \ref{sec:4.1}, one may ignore the $(m-n)\mathcal{J}_{m+n}$ deformation as it does not lead to a nonisomorphic infinitesimal deformation. Two Jacobi identities put constraints on the form of deformation function $\Psi(g_{i},g_{j};\varepsilon)$. The first Jacobi is  
\begin{equation*}
    [\mathcal{P}_{m},[\mathcal{P}_{n},\mathcal{P}_{l}]]+\text{cyclic permutations}=0,
\end{equation*}
which in terms of $\Psi(\mathcal{P}_{m},\mathcal{P}_{n};\varepsilon)=(m-n)\tilde{g}(m,n;\varepsilon)\mathcal{P}_{m+n}$ leads to
\begin{equation}
  \begin{split}
       &(n-l)(m-n-l)\tilde{g}(m,n+l;\varepsilon)\tilde{g}(n,l;\varepsilon)+(l-m)(n-l-m)\tilde{g}(n,l+m;\varepsilon)\tilde{g}(l,m;\varepsilon)\\
      & +(l-n-m)(m-n)\tilde{g}(m,n;\varepsilon)\tilde{g}(l,m+n;\varepsilon)=0. 
  \end{split}
\end{equation}
One can check that the solution of linear term $\psi^{PP}_{1}(\mathcal{P}_{m},\mathcal{P}_{n})=g(m-n)\mathcal{P}_{m+n}$ ($g$ is an arbitrary constant) satisfies the above equation. The second Jacobi $[\mathcal{P}_{m},[\mathcal{P}_{n},\mathcal{J}_{l}]]+\text{cyclic permutations}=0$ leads to the constraint 
\begin{equation}   
       (n-l)(m-n-l)\tilde{g}(m,l+n;\varepsilon)+(l-m)(n-l-m)\tilde{g}(n,m+l;\varepsilon)
       +(l-n-m)(m-n)\tilde{g}(m,n;\varepsilon)=0, 
\end{equation}
which is exactly the same as \eqref{fg-first} so it has the same solution $\tilde{g}(m,n;\varepsilon)=\text{constant}=g$. Therefore,  $\psi^{PP}_{1}(\mathcal{P}_{m},\mathcal{P}_{n})=g(m-n)\mathcal{P}_{m+n}$ is also a formal deformation and moreover, as argued in previous subsection, this deformation is unique. 

The same procedure may be repeated for linear (infinitesimal) deformation $\psi^{JP}_{1}(\mathcal{J}_{m},\mathcal{P}_{n})=(\alpha + \beta m){\cal P}_{m+n}$. The only Jacobi to consider is  $[\mathcal{J}_{m},[\mathcal{J}_{n},\mathcal{P}_{l}]]+\text{cyclic permutations}=0$ which for the formal deformation function $\Psi(\mathcal{J}_{m},\mathcal{P}_{n};\varepsilon)=\tilde{K}(m,n;\varepsilon)\mathcal{P}_{m+n}$ leads to,
\begin{equation}\label{eq-Ktild}
(m-l) X(l+m,n;\varepsilon) - X(l,n;\varepsilon) X(m, n+l;\varepsilon)+ X(l,m+n;\varepsilon) X(m,n;\varepsilon)=0,
\end{equation}
where $X(m,n;\varepsilon)=(m-n)+\tilde{K}(m, n;\varepsilon)$. One can check that 
\begin{equation}
\tilde{K}(m,n;\varepsilon)=\alpha +\beta m,\label{K-tilde-generic}
\end{equation}
provides a solution to \eqref{eq-Ktild}, and that, if $X(m,n)$ is a solution to \eqref{eq-Ktild}, so as 
$$
Y(m,n)=X(m,n)\frac{N(m+n)}{N(n)},
$$
for an arbitrary function $N(n)$. We had argued in the previous subsection that the unique solution to the linearized equation (up to the rescaling by $N(n)$) is of the form \eqref{K-tilde-generic}. Therefore, \eqref{K-tilde-generic} is also a unique all orders solution to  \eqref{eq-Ktild}, if we require solutions to be smoothly connected to solutions to linearized equation.
In summary,  the infinitesimal deformation $\psi^{JP}_{1}(\mathcal{J}_{m},\mathcal{P}_{n})=(\alpha + \beta m){\cal P}_{m+n}$ is also a formal deformation. In this way we have found that all three infinitesimal deformations are integrable and there is no obstruction. 

\subsection{Classification of \texorpdfstring{$\mathfrak{bms}_3$}{BMS3} and 
$\mathfrak{KM}_{\mathfrak{u}(1)}$ deformations}\label{sec:4.5}

In sections \ref{sec:4.1}, \ref{sec:4.2} and \ref{sec:4.3},  we classified all ``infinitesimal'' deformations of the $\mathfrak{bms}_3$ algebra, by deforming each of $[{\cal P}, {\cal P}]$, $[{\cal P}, {\cal J}]$ and $[{\cal J}, {\cal J}]$ separately. This led to only two nontrivial deformations. Then in section \ref{sec:integrability-bms} we showed that
both of these two cases are integrable and one can obtain two class of ``formal'' deformations, the $W(a,b)$ algebra and $\mathfrak{witt}\oplus \mathfrak{witt}$ algebra.

One may wonder if in this process we could have missed cases which involve simultaneous deformations  of the three classes of the commutators of the algebra. With the experience gained in the previous analysis here we tackle this question. The most general possible deformation of the $\mathfrak{bms}_3$ algebra takes the form:
\begin{equation} 
\begin{split}
 & i[\mathcal{J}_{m},\mathcal{J}_{n}]=(m-n)\mathcal{J}_{m+n}+\eta (m-n)h(m,n)\mathcal{P}_{m+n}, \\
 &i[\mathcal{J}_{m},\mathcal{P}_{n}]=(m-n)\mathcal{P}_{m+n}+\zeta K(m,n)\mathcal{P}_{m+n}+\kappa I(m,n) {\cal J}_{m+n},\\
 &i[\mathcal{P}_{m},\mathcal{P}_{n}]=\varepsilon_{1}(m-n)f(m,n)\mathcal{J}_{m+n}+\varepsilon_{2}(m-n)g(m,n)\mathcal{P}_{m+n}.\label{general-deformation}
\end{split}
\end{equation}
The Jacobi identities in the first order in deformation parameters for $f(m,n)$, $K(m,n)$ and $h(m,n)$ yield exactly equations as we found in the previous subsections. For $I(m,n)$ one finds a relation similar to \eqref{eq-K} and for $g(m,n)$ we find
\begin{multline}\label{g-I-mixed}
(n-l)(m-n-l)g(m,l+n)+ (l-m)(n-l-m)g(n,l+m)+(m-n)(l-m-n)g(m,n) \\ +\frac{\kappa}{\varepsilon_2}\left[(n+l-m)I(l,n)- (m+l-n) I(l,m)\right]=0.
\end{multline}
The separate equation we have for $I(m,n)$, yields $I(m,n)=\tilde{\alpha}+\tilde{\beta}m$ for some constants $\tilde{\alpha},\tilde{\beta}$. We note the above equation has the following structure: it is linear in $I, g$ and the $I$ terms come with coefficients first order in $m,n,l$ while the $g$ terms come with second order coefficients. Since $g, I$ are expected to be polynomials of positive powers in their arguments, if $I$ is a monomial of degree $p$, $g$ should be a monomial of degree $p+1$. On the other hand we know $p$ is either 0 or 1 and we may check these two cases directly in  \eqref{g-I-mixed} and verify that it does not have any solution expect for $I(m,n)=0$. Therefore, $g(m,n)$, also, satisfies the same equation as \eqref{fg-first} and hence our method of deforming each commutator separately, to linear order, captures all possible infinitesimal deformations.

One may  discuss integrability of the most general deformations \eqref{general-deformation}. To this end we can simply drop the expansion parameters $\eta,\cdots ,\varepsilon_2$ and analyze the Jacobi identities. Since $I(m,n)=0$ at linear order, it will remain zero to all orders. If we for instance turn on $g(m,n)$ and $K(m,n)$ simultaneously, we observe that although there can be an infinitesimal deformation, higher order Jacobi equations for this deformation are not integrable. The same feature happens when we consider the deformation induced by $f(m,n)$ and $K(m,n)$; the Jacobi allows one to make simultaneous  infinitesimal deformations in  the $[\mathcal{J},\mathcal{P}]$  part by $K(m,n)$ and ideal part by $f(m,n)$ and $g(m,n)$. One should hence analyze integrability of these infinitesimal deformations. The answer, as mentioned above, is that the only integrable deformations  are constructed from the  infinitesimal deformations  induced by $K(m,n)$ or $g(m,n)$ and $f(m,n)$. One then realizes that although the simultaneous deformation induced by  $K(m,n)$ and $g(m,n)$ is an infinitesimal deformation, it does not integrate to a formal deformation. There is a similar structure when we have simultaneous infinitesimal deformation induced by  $K(m,n)$ and $f(m,n)$, which is isomorphic to the infinitesimal deformation induced by $K(m,n)$ and $g(m,n)$, and is not integrable.To summarize, all infinitesimal deformations of $\mathfrak{bms}_{3}$ are those induced by $K(m,n)$ and $g(m,n)$.

In fact the latter can be considered as an example of the case v. in section \ref{sec:3.1} where we have infinitesimal deformation which is not integrable in higher order analysis ($r\geq 0$).}
We can summarize the above analysis in the following  theorem: 
\paragraph{Theorem 4.2} {\it The most general formal deformations of \bms\ are either $\mathfrak{witt} \oplus \mathfrak{witt}$ or $W(a,b)$ algebras}. 

We note that, as discussed earlier, the $\mathfrak{witt} \oplus \mathfrak{witt}$ has two options: the subalgebra generated by ${\cal P}_{0,\pm1},\ {\cal J}_{0,\pm 1}$ can be either $\mathfrak{so}(2,2)$ or $\mathfrak{so}(3,1)$.
\begin{figure}
\begin{center}
\begin{tikzpicture}[scale=4,tdplot_main_coords,>=latex, x={(1,-0.5,0)}]
        %
    \filldraw[fill opacity=0.50, draw=white, fill={rgb:red,60;green,110;blue,0}]
    (0.5,1.3,0)--(0.5,-1.3,0)--(0,-1.3,0) --(0,1.3,0)--cycle;

 \filldraw[fill opacity=0.50, draw=white, fill={rgb:red,245;green,80;blue,0}]
    (0.5,0,-1.3)--(0,0,-1.3)--(0,0,1.3) --(0.5,0,1.3)--cycle;
    
    \draw[thick,->] (0,0,0)--(0,0,1.5) node[anchor=north east]{$\nu$}; 
    \draw[thick,->] (1/2,0,0)--(1,0,0) node[anchor=north east]{$a$}; 

    
    \filldraw [color=red!70!black,fill=red!70!black]  (0,-1,0)  ++(-0.7pt,-0.7pt) rectangle ++(1.5pt,1.5pt);  \node[align=center] at (0,-1,0.15) (ori) {$W(0,-1)=\mathfrak{bms}_{3}$ \\\text{}};

\filldraw [orange] (0,-2,0.8) ++(-1pt,-1pt) -- ++(2.5pt,0pt) -- ++(-1.2pt,2.1pt) -- cycle;
\node[align=center] at (0,-2,0.5) (ori) {$\mathfrak{witt}\oplus \mathfrak{witt}$ \\\text{}};
\draw[->,help lines,shorten >=4pt] (0,-1,0) .. controls (0,-0.9) and (-0.5,0.1) .. (0,-2,0.8);

    
    \filldraw[ultra thick] (0.5,0,0) circle (0.4pt) ++ (0,-0.14,0.1);
    \node[align=center] at (0.65,-0.1,0) (ori) {$a=\frac{1}{2}$ \\\text{}};

    \filldraw [color=red!70!black,fill=red!70!black]  (0,0,0)  ++(-0.7pt,-0.7pt) rectangle ++(1.5pt,1.5pt);  \node[align=center] at (0,0.6,0.12) (ori) {$W(0,0)=\mathfrak{KM}_{\mathfrak{u}(1)}$\\\text{}};
  

    \filldraw[red] (0.25,0.5,0) circle (0.7pt) ++ (0,-0.14,0.1) ;
    \node[align=center] at (0.48,0.8,0) (ori) {$W(a,b)$\\\text{}};


    \filldraw[red] (0.25,0,-0.5) circle (0.7pt) ++ (0,-0.14,0.1) ;
    \node[align=center] at (0.25,0,-0.8) (ori) {$\mathfrak{KM}{(a,\nu)}$\\\text{}};

    \draw[ thick](-0.5,0,0)--(0,0,0);
     \draw[ thick](0,0,0)--(0.5,0,0);
      \draw[ thick](0.5,0,0)--(1,0,0);
    \draw[ thick] (0,-1.3,0)--(0,-0.45,0);
    \draw[ thick,dashed](0,-0.45,0)--(0,0,0);
    \draw[ thick,->] (0,0,0)--(0,1.4,0) node[anchor=north east]{$b$};
    \draw[ thick] (0,0,-1.3)--(0,0,-0.5);
    \draw[ thick,dashed] (0,0,-0.5)--(0,0,0);
    \draw[ thick](0,0,0)--(0,0,1);
\end{tikzpicture}
\end{center}
\caption{Parametric space of formal deformations of $\mathfrak{bms}_{3}$ and $\mathfrak{KM}_{\mathfrak{u}(1)}$. $W(a,b)$ algebra appears as the formal deformation of both $\mathfrak{bms}_{3}$ and $\mathfrak{KM}_{\mathfrak{u}(1)}$. As discussed below \eqref{Fourier-expand-a} parameter $a$ ranges in $[0,1/2]$ while $b,\nu $ can be any real number. Besides the $W(a,b)$,  $\mathfrak{KM}_{\mathfrak{u}(1)}$
can be deformed to  $\mathfrak{KM}(a,\nu)$ which is spanned by $b=0$ plane. The
\bms\ algebra  can be  deformed  to $\mathfrak{witt}\oplus \mathfrak{witt}$ which again is out of the $(a,b,\nu)$ space displayed above. }\label{Fig-abnu}
\end{figure}
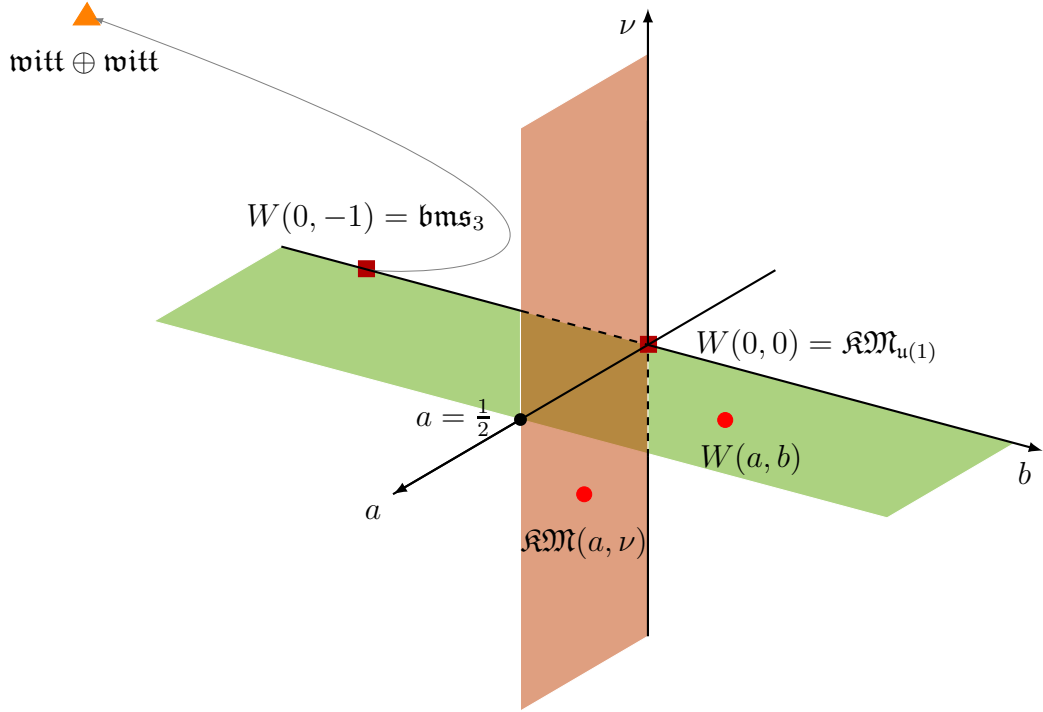

\paragraph{Formal deformations of $\mathfrak{u}(1)$ Kac-Moody algebra ($\mathfrak{KM}_{\mathfrak{u}(1)}$).}
In this part we would like to discuss the most general infinitesimal and formal deformations of $\mathfrak{u}(1)$ Kac-Moody algebra $\mathfrak{KM}_{\mathfrak{u}(1)}$ \eqref{Kac-Moody}. The most general infinitesimal deformation of $\mathfrak{KM}_{\mathfrak{u}(1)}$ is of the same form as \eqref{general-deformation} (but with $m-n$ in the first term on the right-hand-side of the second line  is replaced by $-n$). The Jacobi identities then yield $f(m,n)=0$, $g(m,n)=constant=g$, $I(m,n)=gm$, $K(m,n)=\alpha+\beta m$ and $h(m,n)=constant=h$. One can then show that the infinitesimal deformation induced by $g(m,n)$ and $I(m,n)$ can be absorbed by a proper redefinition of generators and hence 
these are just three different nontrivial infinitesimal deformation of $\mathfrak{KM}_{\mathfrak{u}(1)}$. (None of other combinations of the above leads to a nontrivial deformation.) One can then also show that these three parameter family of infinitesimal deformations have no obstructions and can be integrated into a formal deformation. These results can be summarized in the following theorem:
\paragraph{Theorem 4.3 \cite{Roger:2006rz}} {\it The most general formal deformations of $\mathfrak{KM}_{\mathfrak{u}(1)}$ algebra are either $W(a,b)$ algebras or the $\mathfrak{KM}(a,\nu)$ algebra defined as}
\begin{equation} 
\begin{split}
 & i[\mathcal{J}_{m},\mathcal{J}_{n}]=(m-n)(\mathcal{J}_{m+n}+\nu \mathcal{P}_{m+n}), \\
 &i[\mathcal{J}_{m},\mathcal{P}_{n}]=-(n+a)\mathcal{P}_{m+n},\\
 &i[\mathcal{P}_{m},\mathcal{P}_{n}]=0.
\end{split}\label{Kac-Moody-nu}
\end{equation}


\subsection{Algebraic cohomology considerations}\label{sec:cohomology-bms}
In the previous subsections we classified all nontrivial infinitesimal and formal deformations of the \bms\ algebra by an explicit checking of Jacobi identity and then verifying nontriviality of the deformation by checking if it can be removed by a change of basis. As discussed in section \ref{sec:3}, a similar question can be addressed and analyzed by algebraic cohomology consideration through computing $\mathcal{H}^{2}(\mathfrak{bms}_{3};\mathfrak{bms}_{3})$. The main tools to study the cohomology are the Hochschild-Serre spectral sequence \eqref{E2-dec} and the long and short exact sequences \eqref{long-exact} discussed in section \ref{sec:3}. By the former we obtain  information about $\mathcal{H}^{2}(\mathfrak{bms}_{3};\mathfrak{witt})$ and $\mathcal{H}^{2}(\mathfrak{bms}_{3};\mathcal{P})$ independently where $\mathcal{P}$ and $\mathfrak{witt}$ respectively denote the ideal part and the Witt subalgebra of  $\mathfrak{bms}_{3}$. Note that since $\mathfrak{witt}$ is not a \bms\ module {by the adjoint action}, $\mathcal{H}^{2}(\mathfrak{bms}_{3};\mathfrak{witt})$ is defined by the action used in the short exact sequence \eqref{short-exact-bms} below, as discussed in section \ref{sec:HS-seq}. Note also that given the semi-direct sum structure of the \bms\ algebra from \eqref{bms=witt+ideal} we should not expect 
$\mathcal{H}^{2}(\mathfrak{bms}_{3};\mathfrak{bms}_{3})=\mathcal{H}^{2}(\mathfrak{bms}_{3};\mathfrak{witt}) \oplus \mathcal{H}^{2}(\mathfrak{bms}_{3};\mathcal{P})$. 

One can however get information about $\mathcal{H}^{2}(\mathfrak{bms}_{3};\mathfrak{bms}_{3})$ from the long exact sequence \eqref{long-exact} as one can show that some terms therein  are zero and hence it reduces to a short exact sequence or an isomorphism. In fact as it has been shown in \cite{gao2011low} in the case of $W(a,b)$ algebras, specially for $\mathfrak{bms}_{3}$ which is $W(0,-1)$, $\mathcal{H}^{1}(W(a,b);\mathfrak{witt})$ is equal to zero. One can, however, show that $\mathcal{H}^{2}(\mathfrak{bms}_{3};\mathfrak{witt})$ is not zero and therefore, the long exact sequence does not give useful information about $\mathcal{H}^{2}(\mathfrak{bms}_{3};\mathfrak{bms}_{3})$. Nonetheless, as we will see, direct analysis of $\mathcal{H}^{2}(\mathfrak{bms}_{3};\mathfrak{witt})$ and $ \mathcal{H}^{2}(\mathfrak{bms}_{3};\mathcal{P})$ reveals the structure of $\mathcal{H}^{2}(\mathfrak{bms}_{3};\mathfrak{bms}_{3})$.    
To this end, we follow the Hochschild-Serre spectral sequence method (\emph{cf}. section \ref{sec:HS-seq}) and consider the following short exact sequence of $\mathfrak{bms}_{3}$  
\begin{equation}
    0\longrightarrow \mathcal{P}\longrightarrow \mathfrak{bms}_{3} \longrightarrow \mathfrak{bms}_{3}/\mathcal{P}\cong \mathfrak{witt}\longrightarrow 0.\label{short-exact-bms}
\end{equation}

\paragraph{Computation of  $\mathcal{H}^{2}(\mathfrak{bms}_{3};\mathcal{P})$.}
From \eqref{dec2} and \eqref{E2-dec} we have 
\begin{equation}
\begin{split}
    \mathcal{H}^{2}(\mathfrak{bms}_{3};\mathcal{P})&=\oplus_{p+q=2}E_{2;\mathcal{P}}^{p,q}=E_{2;\mathcal{P}}^{2,0}\oplus E_{2;\mathcal{P}}^{1,1}\oplus E_{2;\mathcal{P}}^{0,2}\\
    &=\mathcal{H}^{2}(\mathfrak{witt};\mathcal{H}^{0}(\mathcal{P};\mathcal{P}))\oplus \mathcal{H}^{1}(\mathfrak{witt};\mathcal{H}^{1}(\mathcal{P};\mathcal{P}))\oplus \mathcal{H}^{0}(\mathfrak{witt};\mathcal{H}^{2}(\mathcal{P};\mathcal{P})),\label{bmsp}
\end{split}
\end{equation}
where the subscript $\mathcal{P}$ in $E_{2;\mathcal{P}}^{p,q}$ denotes we are computing $\mathcal{H}^{2}(\mathfrak{bms}_{3};\mathcal{P})$. We compute the  three terms  above separately. 
$\mathcal{H}^{2}(\mathfrak{witt};\mathcal{H}^{0}(\mathcal{P};\mathcal{P}))$ contains $\mathcal{H}^{0}(\mathcal{P};\mathcal{P})$ which by the definition \eqref{H01} and the fact that the action of $\mathcal{P}$ on $\mathcal{P}$ is trivial, one concludes that $\mathcal{H}^{0}(\mathcal{P};\mathcal{P})=\mathcal{P}$ then $\mathcal{H}^{2}(\mathfrak{witt};\mathcal{H}^{0}(\mathcal{P};\mathcal{P}))=\mathcal{H}^{2}(\mathfrak{witt};\mathcal{P})$.
 On the other hand, as it is obvious from commutators of $\mathfrak{bms}_{3}$, the adjoint action of its $\mathfrak{witt}$ subalgebra on itself is exactly the same as adjoint action of its $\mathfrak{witt}$ subalgebra on   the  ideal part $\mathcal{P}$. This can be seen in first and second line in \eqref{bms3}. For this reason one concludes that $\mathfrak{witt}\cong \mathcal{P}$ as {$\mathfrak{witt}$-modules}. So one gets $\mathcal{H}^{2}(\mathfrak{witt};\mathcal{P})\cong \mathcal{H}^{2}(\mathfrak{witt};\mathfrak{witt})$, and since the Witt algebra is rigid, i.e. $\mathcal{H}^{2}(\mathfrak{witt};\mathfrak{witt})=0$, we conclude that $\mathcal{H}^{2}(\mathfrak{witt};\mathcal{P})=0$.
 Therefore the first commutator in \eqref{bms3} remains intact by deformation procedure.

Next we  analyze $\mathcal{H}^{1}(\mathfrak{witt};\mathcal{H}^{1}(\mathcal{P};\mathcal{P}))$. It  is constructed by $1-$cocycle $\varphi_{1}$ which is defined as a function $ \varphi_{1}: \mathfrak{witt}\longrightarrow \mathcal{H}^{1}(\mathcal{P};\mathcal{P})$. So
the expression of $\varphi_{1}(\mathcal{J}_{m})(\mathcal{P}_{n})$ can be expanded in terms of $\mathcal{P}$'s as $\varphi_{1}(\mathcal{J}_{m})(\mathcal{P}_{n})=\tilde{K}(m,n)\mathcal{P}_{m+n},\label{Cjpp}$
where $\tilde{K}(m,n)$ is an arbitrary function. The deformation of $[\mathcal{J},\mathcal{P}]$ part corresponding to $\varphi_{1}$ is $i[\mathcal{J}_m,\mathcal{P}_n]=(m-n)\mathcal{P}_{m+n}+\tilde{K}(m,n){\mathcal{P}}_{m+n}$.
The Jacobi identity for the above bracket imposes restraints on $\tilde{K}(m,n)$ exactly like the ones on $K(m,n)$ in \eqref{eqdefIK}, so one finds the same result as $\tilde{K}(m,n)=\alpha +\beta m+ \gamma m(m-n)$. As we mentioned earlier the $\gamma$ term can be absorbed by choosing a proper normalization factor. 
 }

The last term we need to considered is  $\mathcal{H}^{0}(\mathfrak{witt};\mathcal{H}^{2}(\mathcal{P};\mathcal{P}))$. We use the definition of $\mathcal{H}^{0}$ 
\begin{equation}
  \mathcal{H}^{0}(\mathfrak{witt};\mathcal{H}^{2}(\mathcal{P};\mathcal{P}))=\{ \psi\in \mathcal{H}^{2}(\mathcal{P};\mathcal{P})| \mathcal{J}\circ \psi=0,\,\,\, \forall \mathcal{J}\in \mathfrak{witt} \},\label{H0P}
\end{equation}
 where $\psi$ is a $\mathcal{P}$-valued  $2-$cocycle. 
 The action ``$\circ$'' of $\mathcal{J}$ on a 2-cocycle $\psi$ is defined as \cite{MR0054581}
\begin{equation}
  (\mathcal{J}_{l}\circ \psi)(\mathcal{P}_{m},\mathcal{P}_{n})=[\mathcal{J}_{l},\psi(\mathcal{P}_{m},\mathcal{P}_{n})]-\psi([\mathcal{J}_{l},\mathcal{P}_{m}],\mathcal{P}_{n})-\psi(\mathcal{P}_{m},[\mathcal{J}_{l},\mathcal{P}_{n}]),\label{hjacobi}
\end{equation}
Expanding $\psi$ in terms of $\mathcal{P}$s as $\psi(\mathcal{P}_{m},\mathcal{P}_{n})=(m-n)g(m,n)\mathcal{P}_{m+n}$, we get the same relation as (\ref{fg-first}) which has the solution $g(m,n)=constant$. 
The above discussion leads to 
\begin{equation}
\begin{split}
    \mathcal{H}^{2}(\mathfrak{bms}_{3};\mathcal{P})&=\mathcal{H}^{1}(\mathfrak{witt};\mathcal{H}^{1}(\mathcal{P};\mathcal{P}))\oplus \mathcal{H}^{0}(\mathfrak{witt};\mathcal{H}^{2}(\mathcal{P};\mathcal{P})),\label{bmsp2}
\end{split}
\end{equation}
which means that turning on deformations with coefficients in $\mathcal{P}$, we can only  deform the $[\mathcal{J},\mathcal{P}]$ part by $K(m,n)$ and  the ideal part $[\mathcal{P},\mathcal{P}]$  by $g(m,n)$ or the combination of these two, when we turn on both of them simultaneously. This is exactly in agreement of our results of direct and explicit calculations in the previous subsections.

\paragraph{Computation of ${\mathcal{H}}^{2}(\mathfrak{bms}_{3};\mathfrak{witt})$.} Again considering the short exact spectral sequence \eqref{short-exact-bms} we have 
\begin{equation}
\begin{split}
     \mathcal{H}^{2}(\mathfrak{bms}_{3};\mathfrak{witt})&=\oplus_{p+q=2}E_{2;\mathfrak{w}}^{p,q}=E_{2;\mathfrak{w}}^{2,0}\oplus E_{2;\mathfrak{w}}^{1,1}\oplus E_{2;\mathfrak{w}}^{0,2}\\
    &=\mathcal{H}^{2}(\mathfrak{witt};\mathcal{H}^{0}(\mathcal{P};\mathfrak{witt}))\oplus \mathcal{H}^{1}(\mathfrak{witt};\mathcal{H}^{1}(\mathcal{P};\mathfrak{witt}))\oplus \mathcal{H}^{0}(\mathfrak{witt};\mathcal{H}^{2}(\mathcal{P};\mathfrak{witt})),\label{bmsj}
\end{split}
\end{equation}
where the subscript $\mathfrak{w}$ denotes we are computing $  \mathcal{H}^{2}(\mathfrak{bms}_{3};\mathfrak{witt})$.

To compute  $\mathcal{H}^{2}(\mathfrak{witt};\mathcal{H}^{0}(\mathcal{P};\mathfrak{witt}))$, we recall the action  of $\mathcal{P}$ on $\mathfrak{witt}$ (which is induced via the short exact sequence  (\ref{short-exact-bms})), it is trivial and  hence $\mathcal{H}^{0}(\mathcal{P};\mathfrak{witt})\cong \mathfrak{witt}$. We then conclude $\mathcal{H}^{2}(\mathfrak{witt};\mathcal{H}^{0}(\mathcal{P};\mathfrak{witt}))\cong  \mathcal{H}^{2}(\mathfrak{witt};\mathfrak{witt})\cong 0,$ where in the last step we used the fact that Witt algebra is rigid \cite{fialowski2012formal, schlichenmaier2014elementary}.

Next, we consider the second term in \eqref{bmsj}, $\mathcal{H}^{1}(\mathcal{J};\mathcal{H}^{1}(\mathcal{P};\mathcal{J}))$ which is constructed by 1-cocycle $\varphi_{2}$ as $\varphi_{2}: \mathfrak{witt}\longrightarrow \mathcal{H}^{1}(\mathcal{P};\mathfrak{witt})$, where 
$\varphi_{2}(\mathcal{J}_{m})(\mathcal{P}_{n})=\tilde{I}(m,l)\mathcal{J}_{m+l}\label{Cjp}$, in which $\tilde{I}(m,l)$ is an arbitrary function. $\varphi_2$ deforms the commutator $[\mathcal{J},\mathcal{P}]$ part as $i[\mathcal{J}_m,\mathcal{P}_n]=(m-n)\mathcal{P}_{m+n}+\tilde{I}(m,n){\mathcal{J}}_{m+n}$. Jacobi identity for the above bracket implies that $\tilde{I}(m,n)$ should solve \eqref{eqdefIK}, so one finds the same result as $\tilde{I}(m,n)=0$. This means that the $[\mathcal{J},\mathcal{P}]$ commutator cannot be deformed by the terms with coefficients {in} $\mathcal{J}$.

We finally compute the last term in \eqref{bmsj} which is $\mathcal{H}^{0}(\mathfrak{witt};\mathcal{H}^{2}(\mathcal{P};\mathfrak{witt}))$.  One can repeat the procedure exactly the same as the previous case to get 
\begin{equation}
  (\mathcal{J}_{l}\circ \psi)(\mathcal{P}_{m},\mathcal{P}_{n})=[\mathcal{J}_{l},\psi(\mathcal{P}_{m},\mathcal{P}_{n})]-\psi([\mathcal{J}_{l},\mathcal{P}_{m}],\mathcal{P}_{n})-\psi(\mathcal{P}_{m},[\mathcal{J}_{l},\mathcal{P}_{n}]),\label{hjacobi-1}
\end{equation}
so we can expand $\psi$ in terms of $\mathcal{J}$s as $\psi(\mathcal{P}_{m},\mathcal{P}_{n})=(m-n)f(m,n)\mathcal{J}_{m+n}$ where $f(m,n)$ is an arbitrary symmetric function. By inserting the latter into \eqref{hjacobi-1} we get the same relation as 
 \eqref{fg-first} which has the solution $f(m,n)=constant$. 

With the above discussion we conclude that 
\begin{equation}
    \mathcal{H}^{2}(\mathfrak{bms}_{3};\mathfrak{witt})\cong \mathcal{H}^{0}(\mathfrak{witt};\mathcal{H}^{2}(\mathcal{P};\mathfrak{witt})),\label{bms-witt}
\end{equation}
i.e. deformations of $\mathfrak{bms}_{3}$ with coefficients in $\mathcal{J}$ are just in ideal part of the algebra and this is in agreement with our results in {the} previous subsection.  
{So far we have  separately computed  $\mathcal{H}^{2}(\mathfrak{bms}_{3};\mathcal{P})$ and  $\mathcal{H}^{2}(\mathfrak{bms}_{3};\mathfrak{witt})$. Combination of the first term in RHS of \eqref{bmsp2} and \eqref{bms-witt} are just isomorphic with the first and second terms in \eqref{bmsp2} and the second term in RHS of \eqref{bmsp2} and \eqref{bms-witt} are  isomorphic. One can then conclude that \eqref{bmsp2} contains the whole of content of \eqref{bms-witt}.}

As summary of the above discussions one may conclude that 
\begin{equation}
    \mathcal{H}^{2}(\mathfrak{bms}_{3};\mathfrak{bms}_{3})\cong \mathcal{H}^{2}(\mathfrak{bms}_{3};\mathcal{P})\cong \mathcal{H}^{1}(\mathfrak{witt};\mathcal{H}^{1}(\mathcal{P};\mathcal{P}))\oplus \mathcal{H}^{0}(\mathfrak{witt};\mathcal{H}^{2}(\mathcal{P};\mathcal{P})).
\end{equation}

One may also ask about $\mathcal{H}^{3}(\mathfrak{bms}_{3};\mathfrak{bms}_{3})$.  One can show that all obstructions are located in the space $\mathcal{H}^{3}$ \cite{nijenhuis1967deformations}. As we discussed, nontrivial infinitesimal deformations lead to formal deformations if there is no obstruction. However, as discussed above when we turn on deformations $K(m,n)$ and $g(m,n)$ together, there are obstructions to integrability of infinitesimal deformation.  So since we have found a specific example of nontrivial infinitesimal deformation in the case of $\mathfrak{bms}_{3}$ which is not integrable, we conclude that $\mathcal{H}^{3}(\mathfrak{bms}_{3};\mathfrak{bms}_{3})\neq 0$.

\paragraph{$\mathfrak{u}(1)$ Kac-Moody algebra, $\mathfrak{KM}_{\mathfrak{u}(1)}$.} We can repeat the above cohomological analysis for $\mathfrak{KM}_{\mathfrak{u}(1)}$ algebra. The results are discussed in \cite{Roger:2006rz}.  
One can consider the short exact sequence for this algebra as 
 \begin{equation}
    0\longrightarrow \mathcal{P}_{\mathfrak{u}(1)} \longrightarrow \mathfrak{KM}_{\mathfrak{u}(1)} \longrightarrow \mathfrak{KM}_{\mathfrak{u}(1)}/\mathcal{P}_{\mathfrak{u}(1)} \cong \mathfrak{witt}\longrightarrow 0,\label{short-exact-Kac}
\end{equation}
where $\mathcal{P}_{\mathfrak{u}(1)}$ denotes the ideal part and $\mathfrak{witt}$ the subalgebra of $\mathfrak{KM}_{\mathfrak{u}(1)}$. Our goal here is to compute $\mathcal{H}^{2}(\mathfrak{KM}_{\mathfrak{u}(1)};\mathfrak{KM}_{\mathfrak{u}(1)})$.
Based on the above sequence 
and the structure of the algebra, as in the \bms\ case, we should not expect $\mathcal{H}^{2}(\mathfrak{KM}_{\mathfrak{u}(1)};\mathfrak{KM}_{\mathfrak{u}(1)})$ to be equal to  $\mathcal{H}^{2}(\mathfrak{KM}_{\mathfrak{u}(1)};\mathfrak{witt})\oplus \mathcal{H}^{2}(\mathfrak{KM}_{\mathfrak{u}(1)};\mathcal{P}_{\mathfrak{u}(1)}).$ Nonetheless, our arguments of the \bms\ case readily extend to the Kac-Moody case yielding
\begin{equation}
    \mathcal{H}^{2}(\mathfrak{KM}_{\mathfrak{u}(1)};\mathfrak{witt})=0.\label{KM-witt}
\end{equation}
We recall that $\mathcal{H}^{1}(W(a,b);\mathfrak{witt})=0$ \cite{gao2011low} which is also true for $\mathfrak{KM}_{\mathfrak{u}(1)}$ as $W(0,0)$.  From the long exact sequence \eqref{long-exact}, and that $\mathcal{H}^{n}(\mathfrak{KM}_{\mathfrak{u}(1)};\mathfrak{witt})=0$ for $n=1,2$, one concludes that
\begin{equation}
\mathcal{H}^{2}(\mathfrak{KM}_{\mathfrak{u}(1)};\mathfrak{KM}_{\mathfrak{u}(1)})= \mathcal{H}^{2}(\mathfrak{KM}_{\mathfrak{u}(1)};\mathcal{P}_{\mathfrak{u}(1)}).    
\end{equation}

Next, we note that $\mathcal{H}^{2}(\mathfrak{KM}_{\mathfrak{u}(1)};\mathcal{P}_{\mathfrak{u}(1)})$  can be decomposed as 
\begin{equation}
\begin{split}
    \hspace*{-5mm}\mathcal{H}^{2}(\mathfrak{KM}_{\mathfrak{u}(1)};\mathcal{P}_{\mathfrak{u}(1)})
     \cong
     \mathcal{H}^{2}(\mathfrak{witt};\mathcal{H}^{0}(\mathfrak{u}(1);\mathcal{P}_{\mathfrak{u}(1)}))\oplus \mathcal{H}^{1}(\mathfrak{witt};\mathcal{H}^{1}(\mathcal{P}_{\mathfrak{u}(1)};\mathcal{P}_{\mathfrak{u}(1)}))\oplus \mathcal{H}^{0}(\mathfrak{witt};\mathcal{H}^{2}(\mathfrak{u}(1);\mathcal{P}_{\mathfrak{u}(1)})).\nonumber
\end{split}
\end{equation}
As in the \bms\ case, the last term can be argued to be zero and hence
\begin{equation}
\begin{split}
     \mathcal{H}^{2}(\mathfrak{KM}_{\mathfrak{u}(1)};\mathfrak{KM}_{\mathfrak{u}(1)})\cong 
     \mathcal{H}^{2}(\mathfrak{witt};\mathcal{H}^{0}(\mathcal{P}_{\mathfrak{u}(1)};\mathcal{P}_{\mathfrak{u}(1)}))\oplus \mathcal{H}^{1}(\mathfrak{witt};\mathcal{H}^{1}(\mathcal{P}_{\mathfrak{u}(1)};\mathcal{P}_{\mathfrak{u}(1)})),
\end{split}
\end{equation}
where the first and second term in right hand side are respectively associated to deformations of $[\mathcal{J},\mathcal{P}]$ part and $[\mathcal{P},\mathcal{P}]$ part of $\mathfrak{KM}_{\mathfrak{u}(1)}$ algebra. The former leads to $W(a,b)$ and the latter to \eqref{Kac-Moody-nu}.

One can then show that all  nontrivial deformations of $\mathfrak{KM}_{\mathfrak{u}(1)}$ algebra discussed above have no obstructions and are integrable.  However, as mentioned in section \ref{sec:3.1}, absence of obstructions does not mean $\mathcal{H}^{3}(\mathfrak{KM}_{\mathfrak{u}(1)};\mathfrak{KM}_{\mathfrak{u}(1)})=0$.




 
 \section{Deformations of  \texorpdfstring{$\widehat{\mathfrak{bms}_3}$}{BMS3} and  \texorpdfstring{$\widehat{\mathfrak{KM}_{\mathfrak{u}(1)}}$}{KM-u1} algebras}\label{sec:5}

 The second real cohomology 
 of $\mathfrak{bms}_{3}$ algebra, ${\mathcal{H}}^2(\mathfrak{g};\mathbb{R})$, is two dimensional \cite{gao2008derivations} in the sense that it admits two independent nontrivial Gelfand-Fucks 2-cocycles which are \cite{Barnich:2011ct}:
 \begin{equation}
     \psi_{1}({\mathcal{J}_{m},\mathcal{J}_{n}})=\frac{c_{JJ}}{12}m^{3}\delta_{m+n,0},\,\,\,\  \psi_{2}({\mathcal{J}_{m},\mathcal{P}_{n}})=\frac{c_{JP}}{12}m^{3}\delta_{m+n,0}.\,\,\,\
 \end{equation}
Having two independent Gelfand-Fucks 2-cocycles means that we are extending the algebra by two independent central identity elements \cite{Oblak:2016eij}. We denote the central extension of $\mathfrak{bms}_{3}$ algebra with $\widehat{\mathfrak{bms}}_{3}$.
The commutation relations of $\widehat{\mathfrak{bms}}_{3}$ are (see \cite{Oblak:2016eij} and references therein):
\begin{equation} 
\begin{split}
 & i[\mathcal{J}_{m},\mathcal{J}_{n}]=(m-n)\mathcal{J}_{m+n}+\frac{c_{JJ}}{12}m^{3}\delta_{m+n,0}, \\
 &i[\mathcal{J}_{m},\mathcal{P}_{n}]=(m-n)\mathcal{P}_{m+n}+\frac{c_{JP}}{12}m^{3}\delta_{m+n,0},\\
 &i[\mathcal{P}_{m},\mathcal{P}_{n}]=0.\label{5.2}
\end{split}
\end{equation}
The central charges $c_{JJ}$ and $c_{JP}$ are  arbitrary real numbers. Algebras with different nonzero values of central charges $c_{JJ}$ and $c_{JP}$, are cohomologous, i.e they are isomorphic to each other. Here we take the viewpoint that the global central extensions are deformations of the algebra by addition of the unit elements to the algebra, one unit element for each 2-cocycle. The latter means that if deformation procedure leads to $\widehat{\mathfrak{bms}}_{3}$ algebra with different central charges than we started with, we do not view it as a nontrivial deformation of the initial algebra.

Deformations of $\widehat{\mathfrak{bms}}_{3}$ algebra
can be analyzed as we did in the previous section, by  deforming  each commutator of $\widehat{\mathfrak{bms}}_{3}$ algebra separately. As we will see, however, presence of central extensions makes some of the deformations which were trivial become nontrivial, and conversely some of the nontrivial deformations may become trivial.

\subsection{Classification of 2-cocycles of \texorpdfstring{$\widehat{\mathfrak{bms}}_3$}{BMS3} algebra}\label{sec:5.1}

\paragraph{Deformation of commutators of two $\mathcal{P}$’s.}  Since we added new generators as central terms to $\mathfrak{bms}_{3}$ algebra, the most general deformations of the algebra also involves the central term. So the commutation relations of deformed $\widehat{\mathfrak{bms}}_{3}$ algebra are:
\begin{equation} 
\begin{split}
 & i[\mathcal{J}_{m},\mathcal{J}_{n}]=(m-n)\mathcal{J}_{m+n}+\frac{c_{JJ}}{12}m^{3}\delta_{m+n,0}, \\
 &i[\mathcal{J}_{m},\mathcal{P}_{n}]=(m-n)\mathcal{P}_{m+n}+\frac{c_{JP}}{12}m^{3}\delta_{m+n,0},\\
 &i[\mathcal{P}_{m},\mathcal{P}_{n}]=\varepsilon\tilde{\psi}_{1}^{PP}(\mathcal{P}_{m},\mathcal{P}_{n}),\label{deform of ideal-ce2}
\end{split}
\end{equation}
where $\varepsilon$ is the deformation parameter and $\tilde{\psi}_{1}^{PP}(\mathcal{P}_{m},\mathcal{P}_{n})$ is a $2-$cocycle which may be expanded as a linear combination of generators,
\begin{equation}
   \tilde{\psi}_{1}^{PP}(\mathcal{P}_{m},\mathcal{P}_{n})= (m-n)g(m,n)\mathcal{P}_{m+n}+(m-n)f(m,n)\mathcal{J}_{m+n}+\frac{{c}_{PP}}{12} X(m)\delta_{m+n,0},\label{hat-psiPP}
\end{equation}
in which $X(m)$ is an arbitrary function and ${c}_{PP}$ is a new central charge. As in the case of centerless $\mathfrak{bms}_{3}$, we should consider two Jacobi identities $[\mathcal{P}_{m},[\mathcal{P}_{n},\mathcal{P}_{l}]]+\text{cyclic}=0$ and $[\mathcal{P}_{m},[\mathcal{P}_{n},\mathcal{J}_{l}]]+\text{cyclic}=0$. The former leads to two independent relations one with coefficients in $\mathcal{J}$, which is the same as \eqref{fg-first}, and another for the central part. 
The second Jacobi identity leads to two independent equations for $f(m,n)$ and $g(m,n)$ which are the same as \eqref{fg-first}, and a new equation for central part which is \begin{align}
\big[{c}_{PP}\big((l-m)X(n)+(n-l)X(m)\big)+l^3(m-n)\big(c_{JJ}f(m,n)+c_{JP}g(m,n)\big)\big]\delta_{m+n+l,0}=0.
 \label{C_JJ+C_JP+C_PP}
    \end{align}

As discussed in section \ref{sec:4}, solutions to \eqref{fg-first} are $f(m,n),\,\, g(m,n)=\text{constant}$, and \eqref{C_JJ+C_JP+C_PP} yields 
\begin{equation}
{c}_{PP}\big((2n+m)X(m)-(2m+n)X(n)\big)=
 (f {c}_{JJ} + g{c}_{JP})(m-n)(m+n)^{3}.\label{tildec(JJ)}
\end{equation}
The only nontrivial solution of (\ref{tildec(JJ)}) is
\begin{equation}
c_{PP}=f {c}_{JJ}+g {c}_{JP},\qquad X(m)=m^{3}. \label{PP-solutions}
\end{equation}
So the most general deformations of $\widehat{\mathfrak{bms}}_{3}$ ideal part are 
\begin{equation} 
\begin{split}
 & i[\mathcal{J}_{m},\mathcal{J}_{n}]=(m-n)\mathcal{J}_{m+n}+\frac{c_{JJ}}{12}m^{3}\delta_{m+n,0}, \\
 &i[\mathcal{J}_{m},\mathcal{P}_{n}]=(m-n)\mathcal{P}_{m+n}+\frac{c_{JP}}{12}m^{3}\delta_{m+n,0},\\
 &i[\mathcal{P}_{m},\mathcal{P}_{n}]=(m-n) (f\mathcal{P}_{m+n}+g\mathcal{J}_{m+n})+ \frac{(f {c}_{JJ} +g {c}_{JP})}{12}m^{3}\delta_{m+n,0},\label{deform-ideal-central}
\end{split}
\end{equation}
where $f,g$ are two arbitrary (deformation) constants. {One can then show that by a proper redefinition of the generators the relation \eqref{deform-ideal-central} is exactly the same as \eqref{AdS3} which has only two independent central charges.}

As a special case, we may study deformations when we do not allow for the central term $c_{PP}$. In this case, we can have $f(m,n)=f,\ g(m,n)=g$ constant deformations, but they are not arbitrary any more, as they are related by $f c_{JJ}+ g c_{JP}=0$. In the special case when $c_{JP}$ or $c_{JJ}$ are zero, this respectively yields $f=0$ or $g=0$. 

\paragraph{Deformation of commutators of $[\mathcal{J},\mathcal{P}]$.}

We then consider deformation of second commutation relation of $\widehat{\mathfrak{bms}}_{3}$ algebra (\ref{5.2}) as:
\begin{equation} 
\begin{split}
 & i[\mathcal{J}_{m},\mathcal{J}_{n}]=(m-n)\mathcal{J}_{m+n}+\frac{c_{JJ}}{12}m^{3}\delta_{m+n,0}, \\
 &i[\mathcal{J}_{m},\mathcal{P}_{n}]=(m-n)\mathcal{P}_{m+n}+\frac{c_{JP}}{12}m^{3}\delta_{m+n,0}+\zeta\tilde{\psi}^{JP}_{1}(\mathcal{J}_{m},\mathcal{P}_{n}),\\
 &i[\mathcal{P}_{m},\mathcal{P}_{n}]=0,\label{c_(JP)commutation}
\end{split}
\end{equation}
where $\tilde{\psi}^{JP}_{1}(\mathcal{J}_{m},\mathcal{P}_{n})$ is a $2-$cocycle and $\zeta$ is the deformation parameter. One can write $\tilde{\psi}^{JP}_{1}(\mathcal{J}_{m},\mathcal{P}_{n})$ in terms of linear combination of generators as 
\begin{equation}
    \tilde{\psi}^{JP}_{1}(\mathcal{J}_{m},\mathcal{P}_{n})=K(m,n)\mathcal{P}_{m+n}+I(m,n)\mathcal{J}_{m+n}+\frac{\hat{c}_{JP}}{12}Y(m)\delta_{m+n,0}.\label{hat-psiJP}
\end{equation}
We have to check two Jacobi identities
$[\mathcal{P}_{m},[\mathcal{P}_{n},\mathcal{J}_{l}]]+\text{cyclic permutations}=0$ and $ [\mathcal{J}_{m},[\mathcal{J}_{n},\mathcal{P}_{l}]]+\text{cyclic permutations} =0$. From the first Jacobi one gets a relation for $I(m,n)$ which is exactly the same as \eqref{firsteqI}, when we just keep linear term in $\zeta$, with the only solution $I(m,n)=0$. The second Jacobi leads to two relations for $K(m,n)$ up to first order in $\zeta$. The first one is exactly the same as \eqref{eq-K} with the only nontrivial solutions $K(m,n)=\alpha+\beta m$. The second equation obtained from the above Jacobi relates $K(m,n)$ to $Y(m)$ as
 \begin{equation}
 \begin{split}
\big[{c_{JP}K(n,-m-n)m^{3}-\hat{c}_{JP}}\big((2m+n)Y(n)+m Y(m+n)\big)\big]-(m\leftrightarrow n)=0.\label{y and k}
 \end{split}
\end{equation}
The above should be solved for $Y$ with $K(m,n)=\alpha+\beta m$. One can immediately see that for $\alpha,\beta, c_{JP}\neq 0$ the above has no solutions. If $c_{JP}=0$, then \eqref{y and k} admits a solution of the form $Y(n)=an^3+bn$ where $a,b$ are two real numbers. This solution, however, as we will see is not integrable. If $\alpha,\beta=0$, i.e. when we turn off the $K$ deformation, then again we can have $Y(n)=an^3+bn$ as a solution. This solution is, however, not a new deformation, it is just the same as the central extension already turned on. To summarize, for $c_{PJ}\neq 0$ case we do not have any nontrivial deformation and for $c_{JP}=0$ as have $K(m)=\alpha+\beta m$ deformations, as discussed in section \ref{sec:4.2}. That is the $W(a,b)$ algebra  does not admit a $c_{JP}$ central extension, in accord with \cite{gao2008derivations,gao2011low}. 

 \paragraph{Deformation of commutators of two $\mathcal{J}$’s.}\label{JJhat}
 
We finally consider deformations of third commutation relations of $\widehat{\mathfrak{bms}}_{3}$ algebra as:
\begin{equation} 
\begin{split}
 & i[\mathcal{J}_{m},\mathcal{J}_{n}]=(m-n)\mathcal{J}_{m+n}+\frac{c_{JJ}}{12}m^{3}\delta_{m+n,0}+\eta \tilde{\psi}^{JJ}_{1}(\mathcal{J}_{m},\mathcal{J}_{n}) \\
 &i[\mathcal{J}_{m},\mathcal{P}_{n}]=(m-n)\mathcal{P}_{m+n}+\frac{c_{JP}}{12}m^{3}\delta_{m+n,0},\\
 &i[\mathcal{P}_{m},\mathcal{P}_{n}]=0,\label{JJ-h-center}
\end{split}
\end{equation}
in which $\eta$ is a deformation parameter and $\tilde{\psi}^{JJ}_{1}(\mathcal{J}_{m},\mathcal{J}_{n})$ is a $2-$cocycle which can be written in terms of linear combination of generators as
\begin{equation}
    \tilde{\psi}^{JJ}_{1}(\mathcal{J}_{m},\mathcal{J}_{n})=(m-n)h(m,n)\mathcal{P}_{m+n}+\frac{1}{12} U(m)\delta_{m+n,0},\label{hat-psiJJ}
\end{equation}
where $U(m)$ is an arbitrary function. We did not include coefficients in $\mathcal{J}$ since the Virasoro algebra is proved to be a rigid algebra \cite{fialowski2012formal}. 

To find the constraints on $h(m,n)$ and $U(m)$ we  consider the Jacobi identity $ [\mathcal{J}_{m},[\mathcal{J}_{n},\mathcal{J}_{l}]]+\text{cyclic}=0$. This Jacobi for \eqref{JJ-h-center} leads to two different equations. One of them is exactly the same as \eqref{h-eq} and its only solution is \eqref{JJ-h-Z}. The second one is related to the central part as
 \begin{equation}
 \begin{split}
      \big[(n-l)\big(c_{JP}h(n,l)m^{3}+U(m)\big)+\text{cyclic}
      \big]\delta_{m+n+l,0}=0,
      \label{h-center}
 \end{split}
\end{equation}
or equivalently
\begin{equation}
\big[c_{JP}\big((2n+m)m^3 h(n,-m-n)+n(m+n)^3 h(m,n)\big)+(2n+m)U(m)-nU(-m-n)\big] - (m\leftrightarrow n)=0.   
\end{equation}
One can readily verify that $h(m,n)=h=\text{constant}$, $U(m)=a+bm+cm^3$ for arbitrary constants $a,b,c$, provides a general solution. One may also show that these are the most general solutions for $c_{JP}\neq 0$. The $a,b$ terms in $U(m)$ may be reabsorbed in a shift of ${\cal J}_m$ and the $c$ term is nothing but a shift of central charge $c_{JJ}$. So, let us focus on the $h$-deformation:
\begin{equation} 
\begin{split}
 & i[\mathcal{J}_{m},\mathcal{J}_{n}]=(m-n)(\mathcal{J}_{m+n}+\nu P_{m+n})+ \frac{c_{JJ}}{12}m^{3}\delta_{m+n,0}, \\
 &i[\mathcal{J}_{m},\mathcal{P}_{n}]=(m-n)\mathcal{P}_{m+n}+\frac{c_{JP}}{12}m^{3}\delta_{m+n,0},\\
 &i[\mathcal{P}_{m},\mathcal{P}_{n}]=0,\label{JJ-JP-nu}
\end{split}
\end{equation}
where $\nu=\eta h$. This algebra has three parameters, $c_{JJ},\ c_{JP}$ and $\nu$. However, if $c_{JP}\neq 0$ one can remove $c_{JJ}$ or $\nu$ by a simple redefinition of generators. To this end consider
\begin{equation} \label{J-shift-P}
\begin{split}
 &  \mathcal{J}_{m}:=\tilde{\mathcal{J}}_{m}+Z\tilde{\mathcal{P}}_{m}, \\
&\mathcal{P}_{m}:=\tilde{\mathcal{P}}_{m}.
\end{split}
\end{equation}
By choosing $Z=\nu$ we can remove $\nu$ term. This does not change the $c_{JP}$ while shifts $c_{JJ}$ to $\tilde c_{JJ}=c_{JJ}-2\nu c_{JP}$.
Alternatively, one may choose $Z=\frac{c_{JJ}}{2c_{JP}}$ and remove $c_{JJ}$ term in favor of the $\nu$ term, by shifting $h$ to $\tilde{\nu}=\nu-\frac{c_{JJ}}{2c_{JP}}$. It is not possible to remove both central charges in the $\nu$ term.  

If $c_{JP}=0$, \eqref{h-center} does not put any new constraint on $h(m,n)$ and we get $U(m)=\hat{c}_{JJ}m^{3}$. In this case by redefinition \eqref{J-shift-P} with $\nu=Z$, $h(m,n)$ deformation can be absorbed and hence there is no nontrivial deformation of $\widehat{\mathfrak{bms}}_{3}$ in the $[{\cal J},{\cal J}]$ sector.


\subsection{Integrability conditions and obstructions}

After classifying nontrivial infinitesimal deformations we now discuss integrability of these infinitesimal deformations and their possible obstructions. As pointed out there are different approaches to the latter. Here we follow the same approach as in section \ref{sec:4.5} and check if the $2-$cocycle $\psi_{1}$ provides a formal deformation. 

We first consider integrability conditions of deformation of the ideal part of $\widehat{\mathfrak{bms}}_{3}$, $\tilde{\psi}_{1}^{PP}(\mathcal{P}_{m},\mathcal{P}_{n})$. The Jacobi identity $ [\mathcal{P}_{m},[\mathcal{P}_{n},\mathcal{J}_{l}]]+\text{cyclic} =0$ leads to a linear relation for each function $f(m,n)$, $g(m,n)$ and $X(m)$ separately with the same solutions we mentioned before and hence the linear order solution is also a solution to higher order equations. The Jacobi $ [\mathcal{P}_{m},[\mathcal{P}_{n},\mathcal{P}_{l}]]+\text{cyclic} =0$  leads to some other relations. The first one is  the same as \eqref{C_JJ+C_JP+C_PP} which is again linear in terms of deformation parameter. Two other relations are 
\begin{align*}
(n-l)(m-n-l)f(n,l)f(m,n+l)+\text{cyclic permutations}&=0, \\ (n-l)(m-n-l)f(n,l)g(m,n+l)+\text{cyclic permutations}&=0,
\end{align*}
which are quadratic in  deformation parameter. One can check that both of these relations are satisfied with the solutions $f(m,n), g(m,n)=\text{constant}$ and hence the 
nontrivial deformation of ideal part is integrable.

The next infinitesimal deformation is obtained through deforming $[\mathcal{J},\mathcal{P}]$ by $\tilde{\psi}_{1}^{JP}(\mathcal{J}_{m},\mathcal{P}_{n})$. To analyze integrability we  study the Jacobi identity $ [\mathcal{J}_{m},[\mathcal{J}_{n},\mathcal{P}_{l}]]+\text{cyclic} =0$. The latter leads to two different relations.  The first relation is 
\eqref{eq-Ktild} with solution $K(m,n)=\alpha+\beta m$, therefore there is no obstruction on integrability of $K(m,n)$. The second relation is related to central part, 
\begin{equation}
 \begin{split}
       &\big[\frac{c_{JP}}{12}\big({K}(n,l;\zeta)m^{3}-{K}(m,l;\zeta)n^{3}\big) +{K}(n,l;\zeta){Y}(m;\zeta)-K(m,l;\zeta){Y}(n;\zeta)-\\
       &\big((l-n){Y}(m;\zeta)+(m-l){Y}(n;\zeta)+(m-n){Y}(m+n;\zeta)\big)\big]\delta_{m+n+l,0}=0.\label{yandk-tilde}
 \end{split}
\end{equation}
As discussed before, when  $c_{JP}\neq 0$ the above has no solutions. For $c_{JP}=0$ it can be seen from \eqref{yandk-tilde} that, although in the infinitesimal level $K(m,n)$ and $Y(m)$ can admit nonzero values, in the higher order (e.g. in second order in $\zeta$), there are obstructions to integrability. Therefore, to have a nontrivial formal deformation one can only turn on $K(m,n)$ or $Y(m)$, and not both. To summarize, all  deformations which we have found for $\widehat{\mathfrak{bms}}_{3}$ are integrable. We note that although we deformed the commutators of $\widehat{\mathfrak{bms}}_{3}$ separately, one can show that simultaneous deformations do not lead to new relations. So, there are only three integrable nontrivial infinitesimal deformations. As the $\mathfrak{bms}_{3}$ case, infinitesimal deformations induced by $f(m,n)$ or by simultaneous deformations  $f(m,n)$ and $K(m,n)$ are equivalent to deformations inducd by $g(m,n)$ and $K(m,n)$. One can also show that, similar to $\mathfrak{bms}_{3}$ case,  simultaneous $g(m,n)$ and $K(m,n)$ infinitesimal deformation  is  not integrable; this latter is another example of the case v. discussed in section \ref{sec:3.1}.

\paragraph{Theorem 5.1} {\it The most general formal deformations of \hbms\ are either $\mathfrak{vir} \oplus \mathfrak{vir}$ or $\widehat{ W(a,b)} $ algebras,  (the latter is a formal deformation when $c_{JP}=0$).}

We note that the $\mathfrak{vir} \oplus \mathfrak{vir}$ has two options: the subalgebra generated by ${\cal P}_{0,\pm1},\ {\cal J}_{0,\pm 1}$ can be either $\mathfrak{so}(2,2)$ or $\mathfrak{so}(3,1)$.

\subsection{Deformations of \texorpdfstring{$\widehat{\mathfrak{KM}_{\mathfrak{u}(1)}}$}{KMu1}}
The commutation relations of  $\widehat{\mathfrak{KM}_{\mathfrak{u}(1)}}$ are 
\begin{equation} 
\begin{split}
 & i[\mathcal{J}_{m},\mathcal{J}_{n}]=(m-n)\mathcal{J}_{m+n}+\frac{c_{JJ}}{12}m^{3}\delta_{m+n,0}, \\
 &i[\mathcal{J}_{m},\mathcal{P}_{n}]=(-n)\mathcal{P}_{m+n}+\frac{c_{JP}}{12}m^{2}\delta_{m+n,0},\\
 &i[\mathcal{P}_{m},\mathcal{P}_{n}]=\frac{c_{PP}}{12}m\delta_{m+n,0}.\label{KM-hat}
\end{split}
\end{equation}
As it is obvious from the above the second real cohomology of {$\mathfrak{KM}_{\mathfrak{u}(1)}$} algebra, {${\mathcal{H}}^2(\mathfrak{KM}_{\mathfrak{u}(1)};\mathbb{R})$}, is three dimensional \cite{gao2008derivations} in the sense that it admits three independent nontrivial 2-cocycles which are:
 \begin{equation}
     \psi_{1}({\mathcal{J}_{m},\mathcal{J}_{n}})=\frac{c_{JJ}}{12}m^{3}\delta_{m+n,0},\,\,\,\  \psi_{2}({\mathcal{J}_{m},\mathcal{P}_{n}})=\frac{c_{JP}}{12}m^{2}\delta_{m+n,0},\,\,\,\ \psi_{3}({\mathcal{P}_{m},\mathcal{J}_{n}})=\frac{c_{PP}}{12}m\delta_{m+n,0}.
 \end{equation}
The most general deformations of the $\widehat{\mathfrak{KM}}_{\mathfrak{u}(1)}$ may be parametrized as
\begin{equation} 
\begin{split}
 & i[\mathcal{J}_{m},\mathcal{J}_{n}]=(m-n)\mathcal{J}_{m+n}+\frac{c_{JJ}}{12}m^{3}\delta_{m+n,0}+\eta (m-n)h(m,n)\mathcal{P}_{m+n}+\frac{\hat{c}_{JJ}}{12}U(m)\delta_{m+n,0}, \\
 &i[\mathcal{J}_{m},\mathcal{P}_{n}]=-n\mathcal{P}_{m+n}+\frac{c_{JP}}{12}m^{2}\delta_{m+n,0}+\zeta K(m,n)\mathcal{P}_{m+n}+\kappa I(m,n) {\mathcal J}_{m+n}+\frac{\hat{c}_{JP}}{12}Y(m)\delta_{m+n,0},\\
 &i[\mathcal{P}_{m},\mathcal{P}_{n}]=\frac{c_{PP}}{12}m\delta_{m+n,0}+(m-n)(\varepsilon_{1}f(m,n)\mathcal{J}_{m+n}+\varepsilon_{2}g(m,n)\mathcal{P}_{m+n})+\frac{\hat{c}_{PP}}{12}X(m)\delta_{m+n,0}.\label{general-deformation-KM-hat}
\end{split}
\end{equation}
Jacobi identities to first order in deformation parameters lead to two different family of relations for the deformation functions. The first set  are exactly the same relations of the $\mathfrak{KM}_{\mathfrak{u}(1)}$ case discussed in section \ref{sec:4.5}. The second set come from central terms. These new relations put constraints on central functions and also new constraints on first family solutions. Here we are not presenting the details and  just summarize the results:

The Jacobi identity $[\mathcal{P}_{m},[\mathcal{P}_{n},\mathcal{P}_{l}]]+\text{cyclic} =0$ does not put any new constraints on functions at the infinitesimal deformation level. The central part of the Jacobi $[\mathcal{P}_{m},[\mathcal{P}_{n},\mathcal{J}_{l}]]+\text{cyclic} =0$ gives two relations from power analysis point of view as
\begin{equation}
\begin{split}
    \{ &\hat{c}_{PP}(nX(m)- mX(n))+c_{PP}(nK(l,m)-m K(l,n))\\
     & +c_{JP}((n+l)^{2}I(l,n)-(m+l)^{2}I(l,m)+(m-n)(l)^{2}g(m,n))\}\delta_{m+n+l,0}=0.
\end{split}
\end{equation}
One can then show that, taking the solutions $I(m,n)=gm$ and $g(m,n)=g$, the term with the coefficient $c_{JP}$ is equal to zero  independently. The first term, by a power series analysis, leads to $X(m)=m$ and $K(m,n)=0$. 
So one then concludes that there is no simultaneous deformation by $I(m,n)$, $g(m,n)$ and $K(m,n)$. Also the above result leads to the conclusion that there is no deformation by $K(m,n)$ independently and by $K(m,n)$ and $h(m,n)$ simultaneously. 

The Jacobi $[\mathcal{J}_{m},[\mathcal{J}_{n},\mathcal{P}_{l}]]+\text{cyclic} =0$ leads to another relation as
\begin{equation}
\begin{split}
     \big[&\hat{c}_{JP}(-lY(m)+lY(n)-(m-n)Y(m+n)) + c_{JJ}(m^{3}I(n,l)-n^{3}I(m,l))\\
     &+c_{PP}l(m-n)h(m,n)\big]\delta_{m+n+l,0}=0,
\end{split}
\end{equation}
in which we used the previous result $K(m,n)=0$. By a power series analysis, one can show that the above is satisfied when $h(m,n)=0$ and $Y(m)=m^{3}$ when $\hat{c}_{JP}=c_{JJ}$. As a result one concludes that the central part does not allow the deformations by $h(m,n)$ individuallyly and by $h(m,n)$, $I(m,n)$ and $g(m,n)$ simultaneously. 
 The final Jacobi $[\mathcal{J}_{m},[\mathcal{J}_{n},\mathcal{J}_{l}]]+\text{cyclic} =0$ just gives the solution $U(m)=m^{3}$ as we expected. 
 
To summarize, we have found that the only nontrivial infinitesimal deformations of $\widehat{\mathfrak{KM}_{\mathfrak{u}(1)}}$ are those induced by $I(m,n)$, $g(m,n)$ and central functions $X(m)=m$, $Y(m)=m^{3}$ and $U(m)=m^{3}$. The central functions $X(m)=m$ and $U(m)=m^{3}$ do not lead to  a nontrivial deformation.  The Jacobi $[\mathcal{P}_{m},[\mathcal{P}_{n},\mathcal{J}_{l}]]+\text{cyclic} =0$ for higher order deformations yields $I$, $g$ and $Y$ deformations are not integrable. {So, we conclude that
\paragraph{Theorem 5.2} {\it Centrally extended $u(1)$ Kac-Moody algebra $\widehat{\mathfrak{KM}_{\mathfrak{u}(1)}}$ is formally rigid.}}

 {One may verify that this is as another example of the case v. in section \ref{sec:3.1}.}\footnote{While we have shown  $\widehat{\mathfrak{KM}_{\mathfrak{u}(1)}}$ with generic central charges does not admit any formal deformation, for special case of $c_{PP}=c_{JP}=0$,  $\widehat{\mathfrak{KM}_{\mathfrak{u}(1)}}$ can be formally deformed to $\widehat{W(a,b)}$. } 
\subsection{Algebraic cohomology considerations}\label{sec:cohomology-bms-hat}

We  can extend our discussion in the subsection \ref{sec:cohomology-bms} to  $\widehat{\mathfrak{bms}}_{3}$. {As in the $\mathfrak{bms}_{3}$ case, $\mathfrak{vir}$ is not a $\widehat{\mathfrak{bms}}_{3}$ module by the adjoint action and $\mathcal{H}^{2}(\widehat{\mathfrak{bms}}_{3};\mathfrak{vir})$ is defined by the action induced from the short exact sequence \eqref{short-exact-vir}, as it was discussed in \ref{sec:HS-seq}. Moreover, as in the case without central extension, we should not expect
$\mathcal{H}^{2}(\widehat{\mathfrak{bms}}_{3};\widehat{\mathfrak{bms}}_{3})$ to be equal to $\mathcal{H}^{2}(\widehat{\mathfrak{bms}}_{3};\mathfrak{vir})\oplus \mathcal{H}^{2}(\widehat{\mathfrak{bms}}_{3};\mathfrak{vir}_{ab})$.} As in the 
\bms\ case, computation of each of these two factors would be helpful in computing the former.
To this end we employ the Hochschild-Serre spectral sequence theorem. 
The $\widehat{\mathfrak{bms}}_{3}$ has semi direct 
sum structure as $\widehat{\mathfrak{bms}}_{3}\cong\mathfrak{vir}\inplus_{ad}\mathfrak{vir}_{ab}$ where $\mathfrak{vir}$ is spanned by  $\mathcal{J}$ generators plus a unit central generator and $\mathfrak{vir}_{ab}$ is the ideal part which is spanned by $\mathcal{P}$ plus another unit central generator.
The short exact sequence for the above is  
\begin{equation}
    0\longrightarrow \mathfrak{vir}_{ab}\longrightarrow \widehat{\mathfrak{bms}}_{3} \longrightarrow \widehat{\mathfrak{bms}}_{3}/\mathfrak{vir}_{ab}\cong \mathfrak{vir}\longrightarrow 0.\label{short-exact-vir}
\end{equation}

\paragraph{Computation of $\mathcal{H}^{2}(\widehat{\mathfrak{bms}}_{3};\mathfrak{vir}_{ab})$.}
Using theorem 1.2 in \cite{degrijse2009cohomology} and Hochschild-Serre spectral sequence theorem we get
\begin{equation}
\begin{split}
     \mathcal{H}^{2}(\widehat{\mathfrak{bms}}_{3};\mathfrak{vir}_{ab})&=\oplus_{p+q=2}E_{2;\mathfrak{vir}_{ab}}^{p,q}=E_{2;\mathfrak{vir}_{ab}}^{2,0}\oplus E_{2;\mathfrak{vir}_{ab}}^{1,1}\oplus E_{2;\mathfrak{vir}_{ab}}^{0,2}\cr
     &= \mathcal{H}^{2}(\mathfrak{vir};\mathcal{H}^{0}(\mathfrak{vir}_{ab},\mathfrak{vir}_{ab}))\oplus \mathcal{H}^{1}(\mathfrak{vir};\mathcal{H}^{1}(\mathfrak{vir}_{ab},\mathfrak{vir}_{ab}))\oplus\mathcal{H}^{0}(\mathfrak{vir};\mathcal{H}^{2}(\mathfrak{vir}_{ab},\mathfrak{vir}_{ab})),\nonumber
\end{split}
\end{equation}
where we defined $E_{2;\mathfrak{vir}_{ab}}^{p,q}=\mathcal{H}^{p}(\mathfrak{vir};\mathcal{H}^{q}(\mathfrak{vir}_{ab},\mathfrak{vir}_{ab}))$.
 
The first term we should consider is $E_{2;\mathfrak{vir}_{ab}}^{2,0}=\mathcal{H}^{2}(\mathfrak{vir};\mathcal{H}^{0}(\mathfrak{vir}_{ab},\mathfrak{vir}_{ab}))$. From the definition of $\mathcal{H}^{0}$ (subsection \ref{sec:cohomology-bms}) one gets $\mathcal{H}^{0}(\mathfrak{vir}_{ab},\mathfrak{vir}_{ab})=\mathfrak{vir}_{ab}$ because the action of $\mathfrak{vir}_{ab}$ on $\mathfrak{vir}_{ab}$ is trivial. Then one has to compute $E_{2}^{2,0}=\mathcal{H}^{2}(\mathfrak{vir};\mathfrak{vir}_{ab})$. The latter is just the same deformation as \eqref{JJ-h-center}. As we mentioned earlier the latter just leads to a shift in $c_{JJ}$ so it does not lead to a nontrivial infinitesimal deformation and therefore $E_{2}^{2,0}=\mathcal{H}^{2}(\mathfrak{vir};\mathfrak{vir}_{ab})=0$. 
 
 $E_{2;\mathfrak{vir}_{ab}}^{1,1}=\mathcal{H}^{1}(\mathfrak{vir};\mathcal{H}^{1}(\mathfrak{vir}_{ab},\mathfrak{vir}_{ab}))$ is constructed by the map    $\hat{\varphi_{1}}: \mathfrak{vir}\longrightarrow \mathcal{H}^{1}(\mathfrak{vir}_{ab},\mathfrak{vir}_{ab})$.
The expression of $\hat{\varphi_{1}}(\hat{\mathcal{J}}_{m})(\hat{\mathcal{P}}_{n})$ can be written in terms of $\hat{\mathcal{P}}$ as $ \hat{\varphi_{1}}(\hat{\mathcal{J}}_{m})(\hat{\mathcal{P}}_{n})=\tilde{K}(m,l)\mathcal{P}_{m+l}+\tilde{Y}(m)\delta_{m+n,0},\label{phi-jp}$ in which $\tilde{K}(m,l)$ and $\tilde{Y}(m)$ are arbitrary functions; here we have used the hatted generators to emphasize that they also include the central element. The deformation corresponding to $\hat{\varphi_{1}}$ is similar to the relation \eqref{c_(JP)commutation} when $I(m,n)=0$. The Jacobi identity constraints on $\hat{K}(m,n)$  are \eqref{eq-K} and  \eqref{y and k}, whose solutions has been discussed below \eqref{y and k}.  Moreover,  we 
cannot deform the $\mathcal{J}\mathcal{P}$ part for $c_{JP}\neq 0$ and hence  $E_{2;\mathfrak{vir}_{ab}}^{1,1}=\mathcal{H}^{1}(\mathfrak{vir};\mathcal{H}^{1}(\mathfrak{vir}_{ab},\mathfrak{vir}_{ab}))=0$.
 
The last term we must compute is $E_{2}^{0,2}=\mathcal{H}^{0}(\mathfrak{vir};\mathcal{H}^{2}(\mathfrak{vir}_{ab},\mathfrak{vir}_{ab}))$. One can extend the definition of $\mathcal{H}^{0}$ in \eqref{H0P} for the above and verify that its elements are solutions of the equation
\begin{equation}
  (\hat{\mathcal{J}}_{l}\circ \hat{\psi})(\hat{\mathcal{P}}_{m},\hat{\mathcal{P}}_{n})=[\hat{\mathcal{J}}_{l},\hat{\psi}(\hat{\mathcal{P}}_{m},\hat{\mathcal{P}}_{n})]-\hat{\psi}([\hat{\mathcal{J}}_{l},\hat{\mathcal{P}}_{m}],\hat{\mathcal{P}}_{n})-\hat{\psi}(\hat{\mathcal{P}}_{m},[\hat{\mathcal{J}}_{l},\hat{\mathcal{P}}_{n}]),\label{psi-hat-1}
\end{equation}
where $\hat{\psi}(\hat{\mathcal{P}}_{m},\hat{\mathcal{P}}_{n})$ is a $2-$cocycle.  
The linear expansion of $\hat{\psi}$ in terms of generators is $\hat{\psi}(\hat{\mathcal{P}}_{m},\hat{\mathcal{P}}_{n})=(m-n)g(m,n)\mathcal{P}_{m+n}+X(m)\delta_{m+n,0}$. So by inserting the expansion of $\hat{\psi}$ into \eqref{psi-hat-1} we reach to two relations. The first one only involves $g(m,n)$ and is exactly the same as \eqref{fg-first} with the solution $g(m,n)=constant$. The second equation is exactly \eqref{C_JJ+C_JP+C_PP} when $f(m,n)=0$ and its solution is \eqref{PP-solutions}. By a proper redefinition of generators, one can show that the deformed algebra  is just the direct sum of two Virasoro algebras $\mathfrak{vir}\oplus \mathfrak{vir}$. This means that one can deform the ideal part of the $\widehat{\mathfrak{bms}}_{3}$ with coefficients in $\mathcal{P}$ into direct sum of two Virasoro algebras:
\begin{equation}
    \mathcal{H}^{2}(\widehat{\mathfrak{bms}}_{3};\mathfrak{vir}_{ab})\cong \mathcal{H}^{0}(\mathfrak{vir};\mathcal{H}^{2}(\mathfrak{vir}_{ab},\mathfrak{vir}_{ab})). \label{H2-vir(ab)}
\end{equation}

\paragraph{Computation of $\mathcal{H}^{2}(\widehat{\mathfrak{bms}}_{3};\mathfrak{vir})$.}

As the second case we consider $\mathcal{H}^{2}(\widehat{\mathfrak{bms}}_{3};\mathfrak{vir})$ which can be decomposed as 
\begin{equation}
\begin{split}
     \hspace*{-5mm}\mathcal{H}^{2}(\widehat{\mathfrak{bms}}_{3};\mathfrak{vir})&=\oplus_{p+q=2}E_{2;\mathfrak{vir}}^{p,q}=E_{2;\mathfrak{vir}}^{2,0}\oplus E_{2;\mathfrak{vir}}^{1,1}\oplus E_{2;\mathfrak{vir}}^{0,2}\\
     &= \mathcal{H}^{2}(\mathfrak{vir};\mathcal{H}^{0}(\mathfrak{vir}_{ab},\mathfrak{vir}))\oplus \mathcal{H}^{1}(\mathfrak{vir};\mathcal{H}^{1}(\mathfrak{vir}_{ab},\mathfrak{vir}))\oplus\mathcal{H}^{0}(\mathfrak{vir};\mathcal{H}^{2}(\mathfrak{vir}_{ab},\mathfrak{vir})),
\end{split}
\end{equation}
where $E_{2;\mathfrak{vir}}^{p,q}\equiv\mathcal{H}^{p}(\mathfrak{vir};\mathcal{H}^{q}(\mathfrak{vir}_{ab},\mathfrak{vir}))$.
 
The first term we have to consider is $E_{2;\mathfrak{vir}}^{2,0}=\mathcal{H}^{2}(\mathfrak{vir};\mathcal{H}^{0}(\mathfrak{vir}_{ab},\mathfrak{vir}))$. From the definition of $\mathcal{H}^{0}$ (subsection \ref{sec:cohomology-bms}) one gets $\mathcal{H}^{0}(\mathfrak{vir}_{ab},\mathfrak{vir})=\mathfrak{vir}$ because the action of $\mathfrak{vir}_{ab}$ as an ideal part of the algebra, on $\mathfrak{vir}$, {induced by the short exact sequence \eqref{short-exact-vir},} is trivial. Then, recalling the fact that $\mathfrak{vir}$ algebra is rigid \cite{fialowski2012formal, schlichenmaier2014elementary} one concludes that $E_{2;\mathfrak{vir}}^{2,0}=\mathcal{H}^{2}(\mathfrak{vir};\mathfrak{vir})=0$.

$E_{2;\mathfrak{vir}}^{1,1}=\mathcal{H}^{1}(\mathfrak{vir};\mathcal{H}^{1}(\mathfrak{vir}_{ab},\mathfrak{vir}))$ is constructed by $\hat{\varphi_{2}}$ as 
   $\hat{\varphi_{2}}: \mathfrak{vir}\longrightarrow \mathcal{H}^{1}(\mathfrak{vir}_{ab},\mathfrak{vir})$.
The expression of $\hat{\varphi_{2}}(\hat{\mathcal{J}}_{m})(\hat{\mathcal{P}}_{n})$ can be written in terms of $\hat{\mathcal{J}}$ as $ \hat{\varphi_{2}}(\hat{\mathcal{J}}_{m})(\hat{\mathcal{P}}_{n})=\tilde{I}(m,l)\mathcal{J}_{m+l}+\tilde{Y}(m)\delta_{m+n,0},$ in which $\hat{I}(m,l)$ and $\hat{Y}(m)$ are arbitrary functions. 
 This deformation  is similar to the relation \eqref{c_(JP)commutation} when $K(m,n)=0$. 
 The Jacobi identity yields to \eqref{eqdefIK} for $\hat{I}(m,n)$, so one finds  $\hat{I}(m,n)=0$ and $\hat{Y}(m)=m^{3}$. The latter does not lead to a nontrivial deformation. This means that the $[\mathcal{J},\mathcal{P}]$ commutator cannot be deformed by the terms with coefficient of $\hat{\mathcal{J}}$, hence $E_{2}^{1,1}=0$. 

The last term we compute is $E_{2;\mathfrak{vir}}^{0,2}=\mathcal{H}^{0}(\mathfrak{vir};\mathcal{H}^{2}(\mathfrak{vir}_{ab},\mathfrak{vir}))$. One can extend the definition of $\mathcal{H}^{0}$ in \eqref{H0P} for the above and observe that its elements are solutions to
\begin{equation}
  (\hat{\mathcal{J}}_{l}\circ \hat{\psi})(\hat{\mathcal{P}}_{m},\hat{\mathcal{P}}_{n})=[\hat{\mathcal{J}}_{l},\hat{\psi}(\hat{\mathcal{P}}_{m},\hat{\mathcal{P}}_{n})]-\hat{\psi}([\hat{\mathcal{J}}_{l},\hat{\mathcal{P}}_{m}],\hat{\mathcal{P}}_{n})-\hat{\psi}(\hat{\mathcal{P}}_{m},[\hat{\mathcal{J}}_{l},\hat{\mathcal{P}}_{n}]),\label{psi-hat}
\end{equation}                         
where $\hat{\psi}(\hat{\mathcal{P}}_{m},\hat{\mathcal{P}}_{n})$ is a $2-$cocycle and the hatted objects  denote generators of the Viraroso algebra, i.e. Witt algebra plus central element. The linear expansion of $\hat{\psi}$ in terms of generators is $\hat{\psi}(\hat{\mathcal{P}}_{m},\hat{\mathcal{P}}_{n})=(m-n)f(m,n)\mathcal{J}_{m+n}+X(m)\delta_{m+n,0}$. Inserting the expansion of $\hat{\psi}$ into \eqref{psi-hat} we reach to two relations, the first one just contains $f(m,n)$ exactly the same as \eqref{fg-first} with the solution $f(m,n)=constant$, and the second equation is \eqref{C_JJ+C_JP+C_PP} when $g(m,n)=g=0$ which has solutions as \eqref{PP-solutions}. So we conclude that $\mathcal{H}^{2}(\widehat{\mathfrak{bms}}_{3};\mathfrak{vir})\cong \mathcal{H}^{0}(\mathfrak{vir};\mathcal{H}^{2}(\mathfrak{vir}_{ab},\mathfrak{vir}))$. The deformed algebra is, however, nothing but the direct sum of two Virasoro algebras $\mathfrak{vir} \oplus \mathfrak{vir}$. Note also that the deformation induced by $g(m,n)$ and $X(m)$, $f(m,n)$ and $X(m)$ or $g(m,n)$, $f(m,n)$ and $X(m)$, are not independent deformations and can be mapped to each other by a proper redefinition of generators. The content of $\mathcal{H}^{2}(\widehat{\mathfrak{bms}}_{3};\mathfrak{vir})$ is hence the same as $\mathcal{H}^{2}(\widehat{\mathfrak{bms}}_{3};\mathfrak{vir}_{ab})$ or their combination. In summary, we have shown that 
\begin{equation}
    \mathcal{H}^{2}(\widehat{\mathfrak{bms}}_{3};\widehat{\mathfrak{bms}}_{3})
    \cong \mathcal{H}^{2}(\widehat{\mathfrak{bms}}_{3};\mathfrak{vir}_{ab}),\label{cent-bms-H2}
\end{equation}
where in the last equality we used the fact that $\mathcal{H}^{2}(\widehat{\mathfrak{bms}}_{3};\mathfrak{vir})$, $\mathcal{H}^{2}(\widehat{\mathfrak{bms}}_{3};\mathfrak{vir}_{ab})$ are isomorphic.\footnote{In the above analysis we have assumed generic case where none of central charges $c_{JP}$ or $c_{JJ}$ are zero. As we observed from direct calculations in the previous section, for $c_{JP}=0$ case we can also deform $\widehat{\mathfrak{bms}}_{3}$ to $\widehat{W(a,b)}$. The cohomological analysis also confirms direct calculations of  section \ref{sec:5.1}. {We should also mention that cohomological considerations for $\widehat{\mathfrak{KM}_{\mathfrak{u}(1)}}$ case needs more analysis than what we did for other cases since the conditions of theorem 1.2 in \cite{degrijse2009cohomology} do not hold.}}

\section{Summary and concluding remarks}\label{sec-discussions}

We analyzed the deformation and stability (rigidity) of infinite dimensional algebras recently appeared in the context of asymptotic symmetry analysis of the $3d$ spacetimes. In particular, we focused on the \bms\ algebra and its central extensions and on $\mathfrak{u}(1)$ Kac-Moody algebra and its central extensions. We showed that the \hbms\ algebra is not rigid and it can be deformed to $\widehat{W(a,b)}$ algebra, which is rigid for generic $a,b$ (\emph{cf.} appendix \ref{sec:append}) or to 
\vir $\oplus$\vir\ algebra, which is again rigid, as discussed in the appendix.  The $\widehat{\mathfrak{KM}_{\mathfrak{u}(1)}}$ algebra, which appears as the near horizon symmetry algebra can be deformed to $\widehat{W(a,b)}$ algebra {when $c_{PP}=c_{JP}=0$ in $\widehat{\mathfrak{KM}_{\mathfrak{u}(1)}}$}. We obtained our results by direct and explicit analysis and classification of nontrivial 2-cocycles of these algebras and established them through algebraic cohomology arguments. We have summarized our results in the tables below.

As we mentioned the infinite dimensional algebras with countable basis are not subject to the Hochschild-Serre factorization theorem. However, our analysis revealed that  for some extensions of the 
$\mathfrak{witt}$ algebra with a special abelian ideal, like in $W(a,b)$ case including its special cases  $\mathfrak{bms}_{3}$ and $\mathfrak{KM}_{u(1)}$, the allowed nontrivial deformations happen with 2-cocycles taking values in the ideal part. It is interesting to examine if this ``extended Hochschild-Serre factorization theorem'' works for all ``abelian extension of countable basis infinite dimensional''  algebras. This expectation is also supported by theorem 1.2 in \cite{degrijse2009cohomology}.

As we discussed the \hbms\ algebra is associated with the asymptotic symmetry algebra of $3d$ flat space. If gravity on $3d$ flat space has a $2d$ holographic field theory dual, it is expected that the \hbms\ algebra appears as the (global) symmetry in this field theory too. The \hbms\ may be deformed in two branches, one takes us to $\mathfrak{vir}\oplus \mathfrak{vir}$ and the other taking us to $\widehat{W(a,b)}$ algebra (\emph{cf.} Figure.\ref{Fig-abnu}). 

\begin{table}[ht]
\begin{center}
\begin{tabular}[h]{ |  p{3cm}  | p{8cm} | p{3.5cm} |}
\hline
\cellcolor[HTML]{E0FFFF} Original algebra & \cellcolor[HTML]{E0FFFF} \centering Formal deformations & \cellcolor[HTML]{E0FFFF} Deformed algebra  \\ \hline
 \multirow{2}{*}{$\mathfrak{bms}_{3}$}\cellcolor[rgb]{0.87, 0.8, 0.98} & \cellcolor[rgb]{0.87, 0.8, 0.98}$\psi^{JP}_{1}=(\alpha +\beta m)\mathcal{P}_{m+n}$ & \cellcolor[rgb]{0.87, 0.8, 0.98} \hspace{1.2cm}$W(a,b)$  \\ \cline{2-3}   \multirow{-2}{*}{\cellcolor[rgb]{0.87, 0.8, 0.98}$\mathfrak{bms}_{3}$}
 & \cellcolor[rgb]{0.87, 0.8, 0.98}$\psi^{PP}_{1}=(m-n)(g\mathcal{P}_{m+n}+f\mathcal{J}_{m+n})$ & \cellcolor[rgb]{0.87, 0.8, 0.98}\hspace{0.9cm}$\mathfrak{witt}\oplus \mathfrak{witt} $ \\\hline
  \multirow{2}{*}{$\widehat{\mathfrak{bms}}_{3}$}\cellcolor[HTML]{FFE4E1} & \cellcolor[HTML]{FFE4E1} $\psi^{JP}_{1}=(\alpha +\beta m)\mathcal{P}_{m+n} \,\,(c_{JP}=0)$ & \cellcolor[HTML]{FFE4E1}  \hspace{1.2cm}$\widehat{W(a,b)}$  \\ \cline{2-3} \multirow{-2}{*}{\cellcolor[HTML]{FFE4E1}$\widehat{\mathfrak{bms}}_{3}$} & \cellcolor[HTML]{FFE4E1}$\psi^{PP}_{1}=(m-n)(g\mathcal{P}_{m+n}+f\mathcal{J}_{m+n})+\frac{c_{PP}}{12}(m^{3})$ & \cellcolor[HTML]{FFE4E1}\hspace{0.9cm}$\mathfrak{vir}\oplus \mathfrak{vir} $ \\
 \hline
\multirow{2}{*}{\cellcolor[rgb]{0.87, 0.8, 0.98}$\mathfrak{KM}_{\mathfrak{u}(1)}$} & \cellcolor[rgb]{0.87, 0.8, 0.98}$\psi^{JJ}_{1}\oplus \psi^{JP}_{1}=\nu (m-n)\mathcal{P}_{m+n} \oplus a \mathcal{P}_{m+n}$ & \cellcolor[rgb]{0.87, 0.8, 0.98}\hspace{1cm}$\mathfrak{KM}(a,\nu)$  \\ \cline{2-3} \multirow{-2}{*}{\cellcolor[rgb]{0.87, 0.8, 0.98}$\mathfrak{KM}_{\mathfrak{u}(1)}$}
 & \cellcolor[rgb]{0.87, 0.8, 0.98}$\psi^{JP}_{1}=(\alpha+\beta m)\mathcal{P}_{m+n}$ & \cellcolor[rgb]{0.87, 0.8, 0.98}\hspace{1.2cm}$W(a,b)$  \\
  \hline
\cellcolor[HTML]{FFE4E1} $\widehat{\mathfrak{KM}_{\mathfrak{u}(1)}}$ & \cellcolor[HTML]{FFE4E1} \centering $-$ & \cellcolor[HTML]{FFE4E1}\hspace{1.2cm}$\widehat{\mathfrak{KM}_{\mathfrak{u}(1)}}$ \\
\hline
\end{tabular}
\end{center}
\caption{Infinitesimal and formal deformations and the 2-cocycles of \bms, \hbms, $\mathfrak{KM}_{\mathfrak{u}(1)}$ and $\widehat{\mathfrak{KM}_{\mathfrak{u}(1)}}$ algebras.}
\label{table1}
\end{table}

Deformation of the symmetry algebra may then be associated with symmetries of the perturbed field theory deformed by a relevant deformation. Recalling analysis of \cite{Bagchi:2012xr} one may then associate the former to an irrelevant deformation taking us to a UV fixed point and the latter as an IR deformation connecting us to the near horizon. To be more specific, consider a $3d$ flat space cosmology background. On the asymptotic null infinity we find \hbms\ ($\widehat{W(0,-1)}$) algebra, one can then model moving in toward the horizon inside the spacetime, where we find $\widehat{W(0,0)}$ algebra. As discussed in section \ref{sec:4.2} the $b$ parameter of the $W(a,b)$ family can be associated with the scaling dimension of the operator ${\cal P}$ with which we have deformed the dual theory. Then moving in the $b$ direction corresponds to the RG flow for this operator. In short, the idea is to make a direct connection between algebra and (holographic) dual (conformal) field theory deformations. This viewpoint may prove fruitful in analyzing field theory duals of $3d$ flat spaces.

\begin{table}[t]
\begin{center}
\begin{tabular}{ |  p{3.5cm}  | p{6cm} | p{3.5cm} | p{1.5cm} |}
\hline
\cellcolor[HTML]{E0FFFF}Original algebra & \cellcolor[HTML]{E0FFFF}$\hspace*{1cm}$Formal deformations & \cellcolor[HTML]{E0FFFF}Deformed algebras & \cellcolor[HTML]{E0FFFF}Rigidity \\ \hline
\cellcolor[rgb]{0.87, 0.8, 0.98}$\mathfrak{witt}\oplus \mathfrak{witt} $ & \cellcolor[rgb]{0.87, 0.8, 0.98}\centering $-$ & \cellcolor[rgb]{0.87, 0.8, 0.98}\centering $\mathfrak{witt}\oplus \mathfrak{witt}$ & \cellcolor[rgb]{0.87, 0.8, 0.98} \hspace{0.6cm}$\checkmark$  \\
 \hline
\cellcolor[HTML]{FFE4E1}$\mathfrak{vir}\oplus \mathfrak{vir}$ & \cellcolor[HTML]{FFE4E1}\centering $-$ & \cellcolor[HTML]{FFE4E1}\centering $\mathfrak{vir}\oplus \mathfrak{vir}$ & \cellcolor[HTML]{FFE4E1}\hspace{0.6cm}$\checkmark$  \\
 \hline
\cellcolor[rgb]{0.87, 0.8, 0.98}$W(a,b)$ & \cellcolor[rgb]{0.87, 0.8, 0.98}\centering $-$ & \cellcolor[rgb]{0.87, 0.8, 0.98}\centering $W(a,b)$ & \cellcolor[rgb]{0.87, 0.8, 0.98}\hspace{0.6cm}$\checkmark$ \\
 \hline
\cellcolor[HTML]{FFE4E1}$\widehat{W(a,b)}$ & \cellcolor[HTML]{FFE4E1}\centering $-$ & \cellcolor[HTML]{FFE4E1}\centering $\widehat{W(a,b)}$ & \cellcolor[HTML]{FFE4E1}\hspace{0.6cm}$\checkmark$ \\
 \hline
\cellcolor[rgb]{0.87, 0.8, 0.98}$\mathfrak{KM}(a,\nu)$& \cellcolor[rgb]{0.87, 0.8, 0.98}$\hspace{1.5cm}\psi^{JP}_{1}= m\mathcal{P}_{m+n}$ & \cellcolor[rgb]{0.87, 0.8, 0.98}\centering $W(a,b)$ & \cellcolor[rgb]{0.87, 0.8, 0.98}\hspace{0.6cm}\xmark \\
\hline
\end{tabular}
\end{center}
\caption{Rigidity of the algebras appearing as formal deformations of algebras in Table \ref{table1}. We note that here by $W(a,b)$ or $\widehat{W(a,b)}$ algebra we mean family of $W$-algebras with generic values of the $a,b$ parameters. Deformations can move us within this family on the $a,b$ plane. {For specific points $(a=0,b=0)$ or $(a=0,b=-1)$ which are respectively corresponding to $\mathfrak{KM}_{u(1)}$ and \bms, the algebra admits non-trivial deformations.}}\label{table2}
\end{table}

Here we focused on the algebras and their deformations. We know that there are groups associated with \vir, \hbms\ and $\widehat{\mathfrak{KM}_{\mathfrak{u}(1)}}$ algebras. One may ask how the deformation of algebras appear in the associated groups, e.g. whether there are groups associated with $\widehat{W(a,b)}$ algebras. Another related question is analyzing the (co)adjoint orbits of these groups and algebras and how the deformations affect the coadjoint orbits. It is then desirable to study the bearing of cohomological statements and theorems for these groups and orbits. In particular, one may explore whether a deformed algebra allows for unitary (co)adjoint orbits. The latter is crucial for building Hilbert space of physical theories  invariant under the deformed symmetry algebras. 

Other asymptotic symmetry/surface charge algebras have also been discussed, in particular, for $4d$ flat space we have a  $\mathfrak{bms}_4$ algebra \cite{Bondi:1962px, Sachs:1962wk, Sachs:1962zza, Barnich:2011mi, Barnich:2012rz, Troessaert:2017jcm, mccarthy1978lifting}. While much less is known about classification of central extensions of $\mathfrak{bms}_4$ in comparison to the $3d$ version, one may study deformations and rigidity of $\mathfrak{bms}_4$ and $\widehat{\mathfrak{bms}}_4$ algebras \cite{Barnich:2011ct}. This question is of interest as one would expect that deformation of $\mathfrak{bms}_4$ algebra should yield an algebra associated with asymptotic symmetries of (A)dS$_4$ space. This latter has been studied by usual methods and ideas of asymptotic symmetry analysis since 1960's with little success. Our algebraic method based on deformation theory and cohomology may open a new venue in this direction. Besides $\mathfrak{bms}_4$ there are other algebras like the so-called NHEG algebra \cite{Compere:2015mza, Compere:2015bca,Javadinezhad:2017jnv}. One can also study deformations of these algebras and hope that they are going to yield new infinite dimensional algebras relevant to higher dimensional flat or AdS sapcetimes. We hope to return to these questions in future works.

\section*{Acknowledgement}

M.M.Sh-J. and H.R.S. acknowledge the partial support of Iranian NSF under grant No. 950124. A.F.P would like to thank hospitality of school of physics of IPM and H.R.S. the hospitality of IPM, Isfahan during the course of this project.  
We would like to thank Saeed Azam, Alice Fialowski,  Jose Figueroa-O'Farril, Daniel Grumiller, Masoud Khalkhali, Nansen Petrosyan, Blagoje Oblak, Claude  Roger and Malihe Yousofzadeh for the correspondence and comments on the draft. 



\appendix                                     
\section{On rigidity of  \texorpdfstring{$\mathfrak{witt}\oplus \mathfrak{witt}$}{wittpluswitt}, \texorpdfstring{$\mathfrak{vir}\oplus \mathfrak{vir}$}{virplusvir}, \texorpdfstring{$W(a,b)$}{Wab} and \texorpdfstring{$\widehat{W(a,b)}$}{hWab} algebras}\label{sec:append}

As discussed and summarized in Table \ref{table1} formal deformations of the \bms, \hbms, ${\mathfrak{KM}_{\mathfrak{u}(1)}}$ and $\widehat{\mathfrak{KM}_{\mathfrak{u}(1)}}$ algebras yields $\mathfrak{witt}\oplus \mathfrak{witt}$, $W(a,b)$, $\mathfrak{vir}\oplus \mathfrak{vir}$ and $\widehat{W(a,b)}$ algebras repectively. As such one expects these latter algebras to be rigid. We explore this explicitly in this appendix.

\paragraph{\texorpdfstring{$\mathfrak{witt}_{L}\oplus \mathfrak{witt}_{R}$}{virplusvir} is rigid.}
Consider the most general infinitesimal deformation of $\mathfrak{witt}_{L}\oplus \mathfrak{witt}_{R}$ algebra:
\begin{equation} 
\begin{split}
 & i[L_{m},L_{n}]=(m-n)L_{m+n}+\eta (m-n)h(m,n)\bar{L}_{m+n}, \\
 &i[L_{m},\bar{L}_{n}]=\varepsilon_{1}K(m,n)L_{m+n}+\varepsilon_{2} I(m,n)\bar{L}_{m+n},\\
 &i[\bar{L}_{m},\bar{L}_{n}]=(m-n)\bar{L}_{m+n}+\zeta(m-n)f(m,n)L_{m+n},
\end{split}
\end{equation}
where $f(m,n),h(m,n)$ and $K(m,n), I(m,n)$ are symmetric and arbitrary functions. Also ${\varepsilon}_{i}, \eta$ and $\zeta$ are deformation parameters. The Jacobi identities lead to some independent relations for each of the above functions as well as two equations relating $I(m,n)$ and $h(m,n)$, and $K(m,n)$ and $f(m,n)$ to each other. If we turn on each one of the deformations individually, one can see that there are only trivial solutions.

There are solutions involving simultaneous deformation by two parameters. If we turn on $I,h$ together we find solutions  $I(m,n)=\alpha(n-m)m^{r}$ and $h(m,n)=\alpha(m+n)^{r}$ and if we turn on  $K,f$ together one finds $K(m,n)=\beta(n-m)n^{s}$ and $f(m,n)=-\beta(m+n)^{s}$,  where $r,s$ are arbitrary integers.  Let us consider deformations by $I(m,n)$ and $h(m,n)$. The Jacobi identity leads to  $\tilde{I}(m+n,l)=-h(m,n)$, where  $I(m,n):=(m-n)\tilde{I}(m,n)$. Therefore, $\tilde{I}(m+n,l)=-h(m,n)=-R(m+n)$. Redefining generators as
\begin{equation} 
\begin{split}
   L_{m}:=\tilde{L}_{m}+R(m)\tilde{\bar{L}}_{m}, \qquad 
 \bar{L}_{m}:=\tilde{\bar{L}}_{m},
\end{split}\label{A-redefinition}
\end{equation}
one can show that $\tilde{L}_{m}$ and $\tilde{\bar{L}}_{m}$ satisfy the commutation relations of  $\mathfrak{witt}_{L}\oplus \mathfrak{witt}_{R}$ algebra. This shows that all of the solutions we have derived for $I(m,n)$ and $h(m,n)$ are just trivial deformations so $\mathcal{H}^2(\mathfrak{witt}_{L}\oplus \mathfrak{witt}_{R}; \mathfrak{witt}_{L})=0$. The same result can be obtained for $K(m,n)$ and $f(m,n)$). So we conclude that $\mathcal{H}^2(\mathfrak{witt}_{L}\oplus \mathfrak{witt}_{R}; \mathfrak{witt}_{L}\oplus \mathfrak{witt}_{R})=0$ and therefore $\mathfrak{witt}_{L}\oplus \mathfrak{witt}_{R}$ is infinitesimally and formaly rigid.

\paragraph{Cohomological considerations.} Theorem \ref{2.1} states that an infinite dimensional Lie algebra $\mathfrak{g}$ is called formally and infinitesimally rigid when we have $\mathcal{H}^2(\mathfrak{g};\mathfrak{g})=0$. So we consider the second cohomology of $\mathfrak{witt}_{L}\oplus \mathfrak{witt}_{R}$  algebra with values in adjoint module, namely, 
\begin{equation}
    \mathcal{H}^2(\mathfrak{witt}_{L}\oplus \mathfrak{witt}_{R};\mathfrak{witt}_{L}\oplus \mathfrak{witt}_{R}).\label{2-ads3-cohomo}
\end{equation}
Then one can decompose  the relation (\ref{2-ads3-cohomo}) as \cite{ChevalleyEilenberg}, 
\begin{multline}\label{2-ads3-cohomo-decom}
     \mathcal{H}^2(\mathfrak{witt}_{L}\oplus \mathfrak{witt}_{R};\mathfrak{witt}_{L}\oplus \mathfrak{witt}_{R})=
    \mathcal{H}^2(\mathfrak{witt}_{L}\oplus \mathfrak{witt}_{R}; \mathfrak{witt}_{L})\oplus\mathcal{H}^2(\mathfrak{witt}_{L}\oplus \mathfrak{witt}_{R}; \mathfrak{witt}_{R}).
\end{multline}
{We should note that unlike $\mathfrak{bms}_{3}$ and $\mathfrak{KM}_{\mathfrak{u}(1)}$ and their central extensions in this case both of $\mathfrak{witt}_{R,L}$ are $\mathfrak{witt}_{L} \oplus \mathfrak{witt}_{R}$ module by the adjoint action.} Since the first and second terms in the above are essentially the same, it suffices just to show that
\begin{equation}
     \mathcal{H}^2(\mathfrak{witt}_{L}\oplus \mathfrak{witt}_{R}; \mathfrak{witt}_{L})=0.\label{coho-leftwitt}
\end{equation}
We consider the following exact sequence of Lie algebras:
\begin{equation}
     0\longrightarrow\mathfrak{witt}_{L}\longrightarrow\mathfrak{witt}_{L}\oplus \mathfrak{witt}_{R}\longrightarrow \mathfrak{witt}_{R}\longrightarrow 0.
\end{equation}
The Hochschild-Serre spectral sequence associated to the above exact sequence with coefficients in $\mathfrak{witt}_{L}$ is convergent with the following property \cite{degrijse2009cohomology}:
\begin{equation}
     E_{\infty}^{p,q}\Longrightarrow\mathcal{H}^{p+q}(\mathfrak{witt}_{L}\oplus \mathfrak{witt}_{R}; \mathfrak{witt}_{L}).
\end{equation}
On the other hand, we know (see \cite{Ecker:2018iqn} for more detail)
\begin{equation}
\begin{split}
   & \mathcal{H}^{0}(\mathfrak{witt}_{L};\mathfrak{witt}_{L})=\text{Cent}(\mathfrak{witt}_{L})=0,\\
   &\mathcal{H}^{1,2,3}(\mathfrak{witt}_{L};\mathfrak{witt}_{L})=0.
\end{split}
\end{equation}
In view of the Hochschild-Serre theorem \cite{MR0054581}, this implies that 
\begin{equation}
    E_{2}^{p,q}=\mathcal{H}^{p}(\mathfrak{witt}_{R};\mathcal{H}^{q}(\mathfrak{witt}_{L};\mathfrak{witt}_{L}))=0\,\,\,\text{for} \,\,q=0,1,2,3    \Longrightarrow \ \ E_{\infty}^{p,q}=0\,\,\text{for} \,\,q=0,1,2,3.
\end{equation}
Recalling that
\begin{equation}
     \mathcal{H}^2(\mathfrak{witt}_{L}\oplus \mathfrak{witt}_{R}; \mathfrak{witt}_{L})=\oplus_{p+q=2} E_{\infty}^{p,q}=E_{\infty}^{0,2}\oplus E_{\infty}^{1,1}\oplus E_{\infty}^{2,0}=0.\label{coho-decompos}
\end{equation}
The same result can be obtained for $ \mathcal{H}^2(\mathfrak{witt}_{L}\oplus \mathfrak{witt}_{R}; \mathfrak{witt}_{R})$. So we conclude that
the centerless asymptotic symmetries algebra of (A)dS$_3$ space time in $3d$, the direct sum of two Witt algebras $\mathfrak{witt}_{L}\oplus\mathfrak{witt}_{R}$, is formaly and infinitesimally {\it rigid}.

The same result can also be reached through the long exact sequence \eqref{long-exact} and \eqref{coho-leftwitt} without using \eqref{2-ads3-cohomo-decom}. From  \eqref{coho-leftwitt} and the same result for the second term in \eqref{2-ads3-cohomo-decom} which are  the second and fourth terms in \eqref{long-exact}, one concludes $ \mathcal{H}^2(\mathfrak{witt}_{L}\oplus \mathfrak{witt}_{R};\mathfrak{witt}_{L}\oplus \mathfrak{witt}_{R})=0$ .  

\paragraph{On rigidity of $\mathfrak{vir}_{L}\oplus \mathfrak{vir}_{R}.$}
To establish rigidity of $\mathfrak{vir}_{L}\oplus \mathfrak{vir}_{R}$, we show that $\mathcal{H}^2(\mathfrak{vir}_{L}\oplus \mathfrak{vir}_{R};\mathfrak{vir}_{L}\oplus \mathfrak{vir}_{R})=0$. To this end, { as in the previous case}, we use decomposition, 
\begin{equation}
\begin{split}
    \mathcal{H}^2(\mathfrak{vir}_{L}\oplus \mathfrak{vir}_{R};\mathfrak{vir}_{L}\oplus \mathfrak{vir}_{R})=
    \mathcal{H}^2(\mathfrak{vir}_{L}\oplus \mathfrak{vir}_{R}; \mathfrak{vir}_{L})\oplus\mathcal{H}^2(\mathfrak{vir}_{L}\oplus \mathfrak{vir}_{R}; \mathfrak{vir}_{R}).
\end{split}
\end{equation}
     It just suffices to show that
\begin{equation}
     \mathcal{H}^2(\mathfrak{vir}_{L}\oplus \mathfrak{vir}_{R}; \mathfrak{vir}_{L})=0,
\end{equation}
for which we have the short exact sequence
\begin{equation}
     0\longrightarrow\mathfrak{vir}_{L}\longrightarrow\mathfrak{vir}_{L}\oplus \mathfrak{vir}_{R}\longrightarrow \mathfrak{vir}_{R}\longrightarrow 0.
\end{equation}

The Hochschild-Serre spectral sequence associated to the above exact sequence with coefficients in $\mathfrak{vir}_{L}$ is convergent with the following property \cite{degrijse2009cohomology}:
\begin{equation}
     E_{\infty}^{p,q}\Longrightarrow\mathcal{H}^{p+q}(\mathfrak{vir}_{L}\oplus \mathfrak{vir}_{R}; \mathfrak{vir}_{L}).
\end{equation}
On the other hand, we know (see \cite{Ecker:2017sen, Ecker:2018iqn} for more detail)
\begin{equation}
\begin{split}
   & \mathcal{H}^{0}(\mathfrak{vir}_{L};\mathfrak{vir}_{L})=\mathbb{R}c,\\
   &\mathcal{H}^{1,2}(\mathfrak{vir}_{L};\mathfrak{vir}_{L})=0,
\end{split}
\end{equation}
where $c$ is a central element.
In view of the Hochschild-Serre theorem \cite{MR0054581}, this implies that 
\begin{equation}
    E_{2}^{p,q}=\mathcal{H}^{p}(\mathfrak{vir}_{R};\mathcal{H}^{q}(\mathfrak{vir}_{L};\mathfrak{vir}_{L}))=0\,\,\,\text{for} \,\,q=0,1,2    \Longrightarrow \ \ E_{\infty}^{p,q}=0\,\,\text{for} \,\,q=0,1,2.
\end{equation}
We should just explain one point. In the case $E_{2}^{2,0}=\mathcal{H}^{2}(\mathfrak{vir}_{R};\mathcal{H}^{0}(\mathfrak{vir}_{L}, \mathfrak{vir}_{L}))$ the above leads to  $E_{2}^{2,0}=\mathcal{H}^{2}(\mathfrak{vir}_{R};\mathbb{R})$ but we know that $\mathcal{H}^{2}(\mathfrak{vir}_{R};\mathbb{R})=0$ then we conclude that $E_{2}^{2,0}=0$. Recalling that
\begin{equation}
     \mathcal{H}^2(\mathfrak{vir}_{L}\oplus \mathfrak{vir}_{R}; \mathfrak{vir}_{L})=\oplus_{p+q=2} E_{\infty}^{p,q}=E_{\infty}^{0,2}\oplus E_{\infty}^{1,1}\oplus E_{\infty}^{2,0},\label{coho-decompos-vir}
\end{equation}
we obtain
\begin{equation}
     \mathcal{H}^2(\mathfrak{vir}_{L}\oplus \mathfrak{vir}_{R}; \mathfrak{vir}_{L})=0,
\end{equation}
and consequently 
\begin{equation}
    \mathcal{H}^2(\mathfrak{vir}_{L}\oplus \mathfrak{vir}_{R}; \mathfrak{vir}_{L})\oplus\mathcal{H}^2(\mathfrak{vir}_{L}\oplus \mathfrak{vir}_{R}; \mathfrak{vir}_{R})=0.
\end{equation}
That is, the asymptotic symmetry algebra of (A)dS$_3$, the direct sum of two Virasoro algebras $\mathfrak{vir}_{L}\oplus\mathfrak{vir}_{R}$, is formaly and infinitesimally {\it rigid}.


\paragraph{On rigidity of  \texorpdfstring{$W(a,b)$}{Wab}.} 
Here we discuss that $W(a,b)$ algebra for generic $a,b$ is expected to be  rigid in the sense that it does not admit a nontrivial deformation. {As we mentioned in footnote \ref{footnote3}  and also in caption of Table \ref{table2}, here we are considering the $W(a,b)$ family as the two-parameter family of the algebras. However, the $W(a,b)$ algebra as single algebra with two distinct parameters $a$ and $b$ is not rigid and can be deformed through $(a,b)$ space.} In what follows we show that all members of the $W(a,b)$ family for generic values of $a,b$ parameters are cohomologous to each other. To this end and as before, we deform each commutator separately. 
   \begin{equation} 
\begin{split}
 & i[\mathcal{J}_{m},\mathcal{J}_{n}]=(m-n)\mathcal{J}_{m+n}+\eta (m-n)h(m,n)\mathcal{P}_{m+n}, \\
 &i[\mathcal{J}_{m},\mathcal{P}_{n}]=-(n+bm+a)\mathcal{P}_{m+n}+\zeta K(m,n)\mathcal{P}_{m+n}+\kappa I(m,n) {\cal J}_{m+n},\\
 &i[\mathcal{P}_{m},\mathcal{P}_{n}]=\varepsilon_{1}(m-n)f(m,n)\mathcal{J}_{m+n}+\varepsilon_{2}(m-n)g(m,n)\mathcal{P}_{m+n},\label{W-general-deformation}
\end{split}
\end{equation}
where $h(m,n),f(m,n)$ and $g(m,n)$ are symmetric and $K(m,n),I(m,n)$ are arbitrary functions and $\eta,\zeta,\kappa $and $\varepsilon_{i}$ are arbitrary deformation parameters. We then consider Jacobi identities for generic $a,b$. 

The Jacobi $[\mathcal{P}_{m},[\mathcal{P}_{n},\mathcal{J}_{l}]]$ leads to two different relations one of them for just $f(m,n)$ and the second one which relates $g(m,n)$ and $I(m,n)$ which are respectively
\begin{equation}
\begin{split}
    &(l+m-n)(m+bl+a)f(n,m+l)+(m-n-l)(n+bl+a)f(m,n+l)+\\
    &(m-n)(l-m-n)f(m,n)=0,
\end{split}\label{ff-appendix}
\end{equation}
and 
\begin{equation}
\begin{split}
    \hspace*{-16mm}&(m+b(n+l)+a)I(l,n)-(n+b(l+m)+a)I(l,m)+(n-m)(m+n+bl+a)g(m,n)\\ &+(l+m-n)(m+bl+a)g(n,m+l)+(m-n-l)(n+bl+a)g(m,n+l)=0,\label{W-I,g}
\end{split}
\end{equation}
Consider \eqref{ff-appendix}. Since the parameter $a$ and $b$ are arbitrary and independent, the coefficients of $a$ and $b$ should be equal to zero separately. The relation for $b$ coefficient is 
\begin{equation}
    (l+m-n)f(n,m+l)+(m-n-l)f(m,n+l)=0.
\end{equation}
Choosing $m=n+l$ we get $f(n,2l+n)=0$. Since $l$ is an arbitrary integer ($l\neq 0$), we conclude $f(m,n)=0$. For the second relation we need to have more information about $I(m,n)$, for which we consider the Jacobi identity $[\mathcal{J}_{m},[\mathcal{J}_{n},\mathcal{P}_{l}]]$. This leads to two independent relations for $K(m,n)$ and $I(m,n)$ as
\begin{equation}
\begin{split}
    &-(l+bn+a)K(m,n+l)-(n+l+bm+a)K(n,l)+\\
    &(l+bm+a)K(n,m+l)+(m+l+bn+a)K(m,l)+(n-m)K(m+n,l)=0,
\end{split}\label{K-Jacobi-W}
\end{equation}
and 
\begin{equation}
\begin{split}
   &-(l+bn+a)I(m,n+l)+(m-n-l)I(n,l)+\\
    &(l+bm+a)I(n,m+l)+(m+l-n)I(m,l)+
    (n-m)I(m+n,l)=0,
\end{split}\label{I-Jacobi-W}
\end{equation}
\eqref{K-Jacobi-W} states that $K(m,n)=\alpha +\beta m$ and \eqref{I-Jacobi-W} leads to $I(m,n)=0$, and hence  \eqref{W-I,g} yields $g(m,n)=0$. 

Finally one can check the Jacobi  $[\mathcal{J}_{m},[\mathcal{J}_{n},\mathcal{J}_{l}]]$ which leads to the equation for $h(m,n)$,
\begin{multline}
    (n-l)(m-n-l)h(m,n+l)+(l-n)(n+l+bm+a)h(n,l)\\
    +(l-m)(n-l-m)h(n,l+m)+(m-l)(l+m+bn+a)h(l,m)\\
    +(m-n)(l-m-n)h(l,m+n)+(n-m)(m+n+bl+a)h(m,n)=0.\label{h-Jacobi-W}
\end{multline}
If $h(m,n)$ is expanded as power series as \eqref{power series}, the only solution for general $a,b$ is $h(m,n)=constant=h$. This deformation is, however, a trivial one, as it can be absorbed into redefinition of generators as \eqref{Z-redefinition} which leads to the relation $Z(m)(m+bn+a)-Z(n)(n+bm+a)+(n-m)Z(m+n)=\nu(m-n)$. One can then check that by choosing $Z(m)=-\nu/b$ the above relation is satisfied. So we conclude that for generic $a,b$ we cannot deform the $W(a,b)$ algebra; i.e. the family of $W(a,b)$ algebras is rigid. 

\paragraph{Cohomological considerations.} Unlike the case $\mathfrak{witt}\oplus \mathfrak{witt}$ algebra, $W(a,b)$ is not direct sum of two algebras and 
the Witt part is not an ideal in $W(a,b)$. However, like the \bms\ or $\mathfrak{KM}_{\mathfrak{u}(1)}$ case we can still use the Hochschild-Serre spectral sequence for the $W(a,b)$ algebra. We have the following short exact sequence 
\begin{equation}
     0\longrightarrow \mathcal{P}\longrightarrow W(a,b)\longrightarrow W(a,b)/\mathcal{P}\cong \mathfrak{witt}\longrightarrow 0,
\end{equation}   
where $\mathcal{P}$  and $\mathfrak{witt}$, respectively, denote the ideal part and subalgebra of $W(a,b)$. {Note that since $\mathfrak{witt}$ is not a $W(a,b)$ module by the adjoint action, $\mathcal{H}^2(W(a,b); \mathfrak{witt})$ is defined by the action induced from the above short exact sequence. So, the second adjoint cohomology of $W(a,b)$, $\mathcal{H}^2(W(a,b);W(a,b))$, similarly to the $\mathfrak{bms}_{3}$ and $\mathfrak{KM}_{u(1)}$ cases,  may not be decomposed as the sum of two cohomologies, i.e. $\mathcal{H}^2(W(a,b); \mathcal{P})\oplus\mathcal{H}^2(W(a,b); \mathfrak{witt})$.} 
We will argue below the first factor is zero and hence the computations reduce to finding the second factor, which may again be argued to be zero.

We first consider $\mathcal{H}^2(W(a,b); \mathcal{P})$. One can decompose the latter as
\begin{equation}
\begin{split}
    \mathcal{H}^2(W(a,b); \mathcal{P})&=\oplus_{p+q=2} E_{2;\mathcal{P}}^{p,q}=E_{2;\mathcal{P}}^{0,2}\oplus E_{2;\mathcal{P}}^{1,1}\oplus E_{2;\mathcal{P}}^{2,0}\\
    &=\mathcal{H}^{2}(\mathfrak{witt};\mathcal{H}^{0}(\mathcal{P};\mathcal{P}))\oplus \mathcal{H}^{1}(\mathfrak{witt};\mathcal{H}^{1}(\mathcal{P};\mathcal{P}))\oplus \mathcal{H}^{0}(\mathfrak{witt};\mathcal{H}^{2}(\mathcal{P};\mathcal{P})) ,\label{coho-decompos-W}
\end{split}
\end{equation}
where the subscript $\mathcal{P}$ denotes we are considering $\mathcal{H}^2(W(a,b); \mathcal{P})$. Because of trivial action of $\mathcal{P}$ on $\mathcal{P}$, the first term of the above is $\mathcal{H}^{2}(\mathfrak{witt};\mathcal{P})$ which is exactly determined by the nontrivial solutions of \eqref{h-Jacobi-W}. The latter has just one solution as $h(m,n)=constant$ which can be absorbed by a proper redefinition as we mentioned in previous part. Therefore, $\mathcal{H}^{2}(\mathfrak{witt};\mathcal{P})=0$. Next we  consider $\mathcal{H}^{1}(\mathfrak{witt};\mathcal{H}^{1}(\mathcal{P};\mathcal{P}))$. Its elements are  solutions of  \eqref{K-Jacobi-W} which just leads to a shift in $a$ and $b$ and hence these solutions are trivial deformations in the family of $W(a,b)$ algebras, so $\mathcal{H}^{1}(\mathfrak{witt};\mathcal{H}^{1}(\mathcal{P};\mathcal{P}))=0$. The last term $\mathcal{H}^{0}(\mathfrak{witt};\mathcal{H}^{2}(\mathcal{P};\mathcal{P}))$ just contains solutions of \eqref{W-I,g}, when $I=0$. It does not have any nontrivial solution, so $\mathcal{H}^{0}(\mathfrak{witt};\mathcal{H}^{2}(\mathcal{P};\mathcal{P}))=0$. From the above discussions one gets $\mathcal{H}^2(W(a,b); \mathcal{P})=0$. One can readily repeat the above procedure and using the relations of previous part to find that all terms in $\mathcal{H}^2(W(a,b); \mathfrak{witt})$ are also equal to zero. To arrive the same result one can use the long exact sequence \eqref{long-exact} and the above results; the latter are the second and the fourth term in \eqref{long-exact}. 
As summary we conclude that $\mathcal{H}^2(W(a,b); W(a,b))=0$, i.e. $W(a,b)$ for generic $a,b$ is infinitesimally and formaly rigid. 

\paragraph{On rigidity of $\widehat{W(a,b)}$. }
As it is discussed in \cite{gao2011low} $W(a,b)$ algebra, for generic $a,b$, just admits one central extension whose commutation relations are 
\begin{equation} 
\begin{split}
 & i[\mathcal{J}_{m},\mathcal{J}_{n}]=(m-n)\mathcal{J}_{m+n}+{\frac{c_{JJ}}{12}m^{3}\delta_{m+n,0}\mathcal{J}_{n}}, \\
 &i[\mathcal{J}_{m},\mathcal{P}_{n}]=-(n+bm+a)\mathcal{P}_{m+n},\\
 &i[\mathcal{P}_{m},\mathcal{P}_{n}]=0.\label{W-central}
\end{split}
\end{equation}
In the other words, the space of $\mathcal{H}^2(W(a,b); \mathbb{R})$ for generic $a,b$ is one dimensional. $\widehat{W(a,b)}$ has 
the short exact sequence,
\begin{equation}
     0\longrightarrow\mathcal{P}\longrightarrow \widehat{W(a,b)}\longrightarrow \widehat{W(a,b)}/\mathcal{P}\cong \mathfrak{vir}\longrightarrow 0,
\end{equation}  
where $\mathcal{P}$ and $\mathfrak{vir}$ are respectively ideal part and sub algebra of $\widehat{W(a,b)}$. As the previous case, $W(a,b)$, to compute
$\mathcal{H}^2(\widehat{W(a,b)};\widehat{W(a,b)})$ we focus on $\mathcal{H}^2(\widehat{W(a,b)}; \mathcal{P})$ and $\mathcal{H}^2(\widehat{W(a,b)}; \mathfrak{vir})$. 

One can readily observe that   $\mathcal{H}^2(\widehat{W(a,b)}; \mathcal{P})$  is exactly the same as $\mathcal{H}^2(W(a,b); \mathcal{P})$, which is equal to zero. For $\mathcal{H}^2(\widehat{W(a,b)}; \mathfrak{vir})$, we have found $f(m,n)=I(m,n)=0$. The combination of the latter and using the fact that $\mathfrak{vir}$ algebra is rigid yields $\mathcal{H}^2(W(a,b); \mathfrak{vir})=0$. 
{On the other hand one can show, by direct calculations, that the simultaneous  infinitesimal deformations of \eqref{W-central} do not lead to any new nontrivial infinitesimal deformation.
Therefore, $\mathcal{H}^2(\widehat{W(a,b)};\widehat{W(a,b)})=0$ which means that  $\widehat{W(a,b)}$ algebra is infinitesimally and formaly rigid. 
\bibliographystyle{fullsort.bst}
\providecommand{\href}[2]{#2}\begingroup\raggedright\endgroup


\begin{thebibliography}{10}
	
	\bibitem{levy1967deformation}
	M.~Levy-Nahas, ``Deformation and contraction of {L}ie algebras,'' {\em J. Math.
		Phys.} {\bf 8} (1967), no.~6, 1211--1222.
	
	\bibitem{levy1968first}
	M.~Levy-Nahas and R.~Seneor, ``First order deformations of {L}ie algebra
	representations, e(3) and {P}oincar{\'e} examples,'' {\em Commun. Math. Phys}
	{\bf 9} (1968), no.~3, 242--266.
	
	\bibitem{Figueroa-OFarrill:1989wmj}
	J.~M. Figueroa-O'Farrill, ``{Deformations of the Galilean Algebra},'' {\em J.
		Math. Phys.} {\bf 30} (1989)
	2735.
	
	\bibitem{mendes1994deformations}
	R.~V. Mendes, ``Deformations, stable theories and fundamental constants,'' {\em
		J. Phys. A.} {\bf 27} (1994), no.~24, 8091.
	
	\bibitem{Chryssomalakos:2004gk}
	C.~Chryssomalakos and E.~Okon, ``{Generalized quantum relativistic kinematics:
		A Stability point of view},'' {\em Int. J. Mod. Phys.} {\bf D13} (2004)
	2003--2034,
	\href{http://www.arXiv.org/abs/hep-th/0410212}{{\tt hep-th/0410212}}.
	
	\bibitem{Figueroa-OFarrill:2017sfs}
	J.~Figueroa-O'Farrill, ``{Classification of kinematical Lie algebras},''
	\href{http://www.arXiv.org/abs/1711.05676}{{\tt 1711.05676}}.
	
	\bibitem{Figueroa-OFarrill:2017ycu}
	J.~M. Figueroa-O'Farrill, ``{Kinematical Lie algebras via deformation
		theory},'' {\em J. Math. Phys.} {\bf 59} (2018), no.~6, 061701,
	\href{http://www.arXiv.org/abs/1711.06111}{{\tt 1711.06111}}.
	
	\bibitem{Figueroa-OFarrill:2017tcy}
	J.~M. Figueroa-O'Farrill, ``{Higher-dimensional kinematical Lie algebras via
		deformation theory},'' {\em J. Math. Phys.} {\bf 59} (2018), no.~6, 061702,
	\href{http://www.arXiv.org/abs/1711.07363}{{\tt 1711.07363}}.
	
	\bibitem{Andrzejewski:2018gmz}
	T.~Andrzejewski and J.~Figueroa-O'Farrill, ``{Kinematical Lie algebras in 2+1
		dimensions},'' {\em J. Math. Phys.} {\bf 59} (2018), no.~6, 061703,
	\href{http://www.arXiv.org/abs/1802.04048}{{\tt 1802.04048}}.
	
	\bibitem{Figueroa-OFarrill:2018ygf}
	J.~M. Figueroa-O'Farrill, ``{Conformal Lie algebras via deformation theory},''
	\href{http://www.arXiv.org/abs/1809.03603}{{\tt 1809.03603}}.
	
	\bibitem{Figueroa-OFarrill:2018ilb}
	J.~Figueroa-O'Farrill and S.~Prohazka, ``{Spatially isotropic homogeneous
		spacetimes},''
	\href{http://www.arXiv.org/abs/1809.01224}{{\tt 1809.01224}}.
	
	\bibitem{Inonu:1953sp}
	E.~Inonu and E.~Wigner, ``{On the contraction of groups and their
		representations},'' {\em Proc. Nat. Acad. Sci.} {\bf 39} (1953) 510--524.
	
	\bibitem{levy1965nouvelle}
	J.-M. L{\'e}vy-Leblond, ``Une nouvelle limite non-relativiste du groupe de
	poincar{\'e},'' {\em Ann. Inst. H. Poincar{\'e}} {\bf 3} (1965) 1--12.
	
	\bibitem{gilmore2012lie}
	R.~Gilmore, {\em Lie groups, {L}ie algebras, and some of their applications}.
	\newblock Courier Corporation, 2012.
	
	\bibitem{Fialowski:2001me}
	A.~Fialowski and M.~Penkava, ``{Deformation theory of infinity algebras},''
	{\em J. Algebra.} {\bf 255} (2002) 59--88,
	\href{http://www.arXiv.org/abs/math/0101097}{{\tt math/0101097}}.
	
	\bibitem{fialowski2012formal}
	A.~Fialowski, ``Formal rigidity of the {W}itt and {V}irasoro algebra,'' {\em J.
		Mat. Phys.} {\bf 53} (2012), no.~7, 073501.
	
	\bibitem{gao2008derivations}
	S.~Gao, C.~Jiang, and Y.~Pei, ``The derivations, central extensions and
	automorphism group of the {L}ie algebra \emph{W},''
	\href{http://www.arXiv.org/abs/arXiv:0801.3911v1}{{\tt arXiv:0801.3911v1}}.
	
	\bibitem{gao2011low}
	S.~Gao, C.~Jiang, and Y.~Pei, ``Low-dimensional cohomology groups of the {L}ie
	algebras $\text{W}(a, b)$,'' {\em Commun. Algebra} {\bf 39} (2011), no.~2,
	397--423.
	
	\bibitem{Ecker:2017sen}
	J.~Ecker and M.~Schlichenmaier, ``{The vanishing of the low-dimensional
		cohomology of the {W}itt and the {V}irasoro algebra},''
	\href{http://www.arXiv.org/abs/1707.06106}{{\tt 1707.06106}}.
	
	\bibitem{Ecker:2018iqn}
	J.~Ecker and M.~Schlichenmaier, ``{The low-dimensional algebraic cohomology of
		the Virasoro algebra},''
	\href{http://www.arXiv.org/abs/1805.08433}{{\tt 1805.08433}}.
	
	\bibitem{Brown:1986nw}
	J.~D. Brown and M.~Henneaux, ``{Central Charges in the Canonical Realization of
		Asymptotic Symmetries: An Example from Three-Dimensional Gravity},'' {\em
		Commun. Math. Phys.} {\bf 104} (1986)
	207--226.
	
	\bibitem{Barnich:2001jy}
	G.~Barnich and F.~Brandt, ``{Covariant theory of asymptotic symmetries,
		conservation laws and central charges},'' {\em Nucl. Phys.} {\bf B633} (2002)
	3--82,
	\href{http://www.arXiv.org/abs/hep-th/0111246}{{\tt hep-th/0111246}}.
	
	\bibitem{Strominger:2017zoo}
	A.~Strominger, ``{Lectures on the Infrared Structure of Gravity and Gauge
		Theory},''
	\href{http://www.arXiv.org/abs/1703.05448}{{\tt 1703.05448}}.
	
	\bibitem{Campoleoni:2017qot}
	A.~Campoleoni, D.~Francia, and C.~Heissenberg, ``{Asymptotic Charges at Null
		Infinity in Any Dimension},'' {\em Universe} {\bf 4} (2018), no.~3, 47,
	\href{http://www.arXiv.org/abs/1712.09591}{{\tt 1712.09591}}.
	
	\bibitem{Barnich:2017ubf}
	G.~Barnich, ``{Centrally extended BMS4 {L}ie algebroid},'' {\em JHEP} {\bf 06}
	(2017) 007,
	\href{http://www.arXiv.org/abs/1703.08704}{{\tt 1703.08704}}.
	
	\bibitem{Afshar:2018apx}
	H.~Afshar, E.~Esmaeili, and M.~M. Sheikh-Jabbari, ``{Asymptotic Symmetries in
		$p$-Form Theories},'' {\em JHEP} {\bf 05} (2018) 042,
	\href{http://www.arXiv.org/abs/1801.07752}{{\tt 1801.07752}}.
	
	\bibitem{Concha:2018zeb}
	P.~Concha, N.~Merino, O.~Miskovic, E.~Rodr\'iguez, P.~Salgado-Rebolledo, and
	O.~Valdivia, ``{Extended asymptotic symmetries of three-dimensional gravity
		in flat space},''
	\href{http://www.arXiv.org/abs/1805.08834}{{\tt 1805.08834}}.
	
	\bibitem{Hosseinzadeh:2018dkh}
	V.~Hosseinzadeh, A.~Seraj, and M.~M. Sheikh-Jabbari, ``{Soft Charges and
		Electric-Magnetic Duality},'' {\em JHEP} {\bf 08} (2018) 102,
	\href{http://www.arXiv.org/abs/1806.01901}{{\tt 1806.01901}}.
	
	\bibitem{Seraj:2016cym}
	A.~Seraj, {\em {Conserved charges, surface degrees of freedom, and black hole
			entropy}}.
	\newblock PhD thesis, IPM, Tehran, 2016.
	\newblock
	\href{http://www.arXiv.org/abs/1603.02442}{{\tt 1603.02442}}.
	\newblock
	
	\bibitem{Compere:2018aar}
	A.~Fiorucci and G.~Comp\`ere, {\em {Advanced Lectures in General Relativity}}.
	\newblock PhD thesis, Brussels U., PTM, 2018.
	\newblock
	\href{http://www.arXiv.org/abs/1801.07064}{{\tt 1801.07064}}.
	\newblock
	
	\bibitem{Bondi:1962px}
	H.~Bondi, M.~G.~J. van~der Burg, and A.~W.~K. Metzner, ``{Gravitational waves
		in general relativity. 7. Waves from axisymmetric isolated systems},'' {\em
		Proc. Roy. Soc. Lond.} {\bf A269} (1962)
	21--52.
	
	\bibitem{Sachs:1962zza}
	R.~Sachs, ``{Asymptotic symmetries in gravitational theory},'' {\em Phys. Rev.}
	{\bf 128} (1962) 2851--2864.
	
	\bibitem{Sachs:1962wk}
	R.~K. Sachs, ``{Gravitational waves in general relativity. 8. Waves in
		asymptotically flat space-times},'' {\em Proc. Roy. Soc. Lond.} {\bf A270}
	(1962) 103--126.
	
	\bibitem{Barnich:2011ct}
	G.~Barnich and C.~Troessaert, ``{Supertranslations call for superrotations},''
	{\em PoS} {\bf CNCFG2010} (2010) 010,
	\href{http://www.arXiv.org/abs/1102.4632}{{\tt 1102.4632}}.
	[Ann. U. Craiova Phys.21,S11(2011)].
	
	\bibitem{Ashtekar:1996cd}
	{Ashtekar, Abhay and Bi{\v{c}}{\'a}k, Ji{\v{r}}{\'\i} and Schmidt, Bernd G},
	``{Asymptotic structure of symmetry reduced general relativity},'' {\em Phys.
		Rev.} {\bf D55} (1997) 669--686,
	\href{http://www.arXiv.org/abs/gr-qc/9608042}{{\tt gr-qc/9608042}}.
	
	\bibitem{Barnich:2006av}
	G.~Barnich and G.~Compere, ``{Classical central extension for asymptotic
		symmetries at null infinity in three spacetime dimensions},'' {\em Class.
		Quant. Grav.} {\bf 24} (2007) F15--F23,
	\href{http://www.arXiv.org/abs/gr-qc/0610130}{{\tt gr-qc/0610130}}.
	
	\bibitem{Oblak:2016eij}
	B.~Oblak, {\em {BMS Particles in Three Dimensions}}.
	\newblock PhD thesis, Brussels U., 2016.
	\newblock
	\href{http://www.arXiv.org/abs/1610.08526}{{\tt 1610.08526}}.
	\newblock
	
	\bibitem{Compere:2009zj}
	G.~Comp\`ere and S.~Detournay, ``{Boundary conditions for spacelike and
		timelike warped $\text{AdS}_3$ spaces in topologically massive gravity},''
	{\em JHEP} {\bf 08} (2009) 092,
	\href{http://www.arXiv.org/abs/0906.1243}{{\tt 0906.1243}}.
	
	\bibitem{Compere:2014bia}
	G.~Comp\`ere, M.~Guica, and M.~J. Rodriguez, ``{Two Virasoro symmetries in
		stringy warped $\text{AdS}_{3}$},'' {\em JHEP} {\bf 12} (2014) 012,
	\href{http://www.arXiv.org/abs/1407.7871}{{\tt 1407.7871}}.
	
	\bibitem{Compere:2014cna}
	G.~Comp\`ere, L.~Donnay, P.-H. Lambert, and W.~Schulgin, ``{Liouville theory
		beyond the cosmological horizon},'' {\em JHEP} {\bf 03} (2015) 158,
	\href{http://www.arXiv.org/abs/1411.7873}{{\tt 1411.7873}}.
	
	\bibitem{Troessaert:2013fma}
	C.~Troessaert, ``{Enhanced asymptotic symmetry algebra of AdS$_{3}$},'' {\em
		JHEP} {\bf 08} (2013) 044,
	\href{http://www.arXiv.org/abs/1303.3296}{{\tt 1303.3296}}.
	
	\bibitem{Afshar:2016wfy}
	H.~Afshar, S.~Detournay, D.~Grumiller, W.~Merbis, A.~Perez, D.~Tempo, and
	R.~Troncoso, ``{Soft Heisenberg hair on black holes in three dimensions},''
	{\em Phys. Rev.} {\bf D93} (2016), no.~10, 101503,
	\href{http://www.arXiv.org/abs/1603.04824}{{\tt 1603.04824}}.
	
	\bibitem{Grumiller:2016pqb}
	D.~Grumiller and M.~Riegler, ``{Most general AdS$_{3}$ boundary conditions},''
	{\em JHEP} {\bf 10} (2016) 023,
	\href{http://www.arXiv.org/abs/1608.01308}{{\tt 1608.01308}}.
	
	\bibitem{Grumiller:2017sjh}
	D.~Grumiller, W.~Merbis, and M.~Riegler, ``{Most general flat space boundary
		conditions in three-dimensional Einstein gravity},'' {\em Class. Quant.
		Grav.} {\bf 34} (2017), no.~18, 184001,
	\href{http://www.arXiv.org/abs/1704.07419}{{\tt 1704.07419}}.
	
	\bibitem{Afshar:2016uax}
	H.~Afshar, D.~Grumiller, and M.~M. Sheikh-Jabbari, ``{Near horizon soft hair as
		microstates of three dimensional black holes},'' {\em Phys. Rev.} {\bf D96}
	(2017), no.~8, 084032,
	\href{http://www.arXiv.org/abs/1607.00009}{{\tt 1607.00009}}.
	
	\bibitem{Afshar:2017okz}
	H.~Afshar, D.~Grumiller, M.~M. Sheikh-Jabbari, and H.~Yavartanoo, ``{Horizon
		fluff, semi-classical black hole microstates — Log-corrections to BTZ
		entropy and black hole/particle correspondence},'' {\em JHEP} {\bf 08} (2017)
	087,
	\href{http://www.arXiv.org/abs/1705.06257}{{\tt 1705.06257}}.
	
	\bibitem{Grumiller:2016kcp}
	D.~Grumiller, A.~Perez, S.~Prohazka, D.~Tempo, and R.~Troncoso, ``{Higher Spin
		Black Holes with Soft Hair},'' {\em JHEP} {\bf 10} (2016) 119,
	\href{http://www.arXiv.org/abs/1607.05360}{{\tt 1607.05360}}.
	
	\bibitem{Campoleoni:2017mbt}
	A.~Campoleoni, D.~Francia, and C.~Heissenberg, ``{On higher-spin
		supertranslations and superrotations},'' {\em JHEP} {\bf 05} (2017) 120,
	\href{http://www.arXiv.org/abs/1703.01351}{{\tt 1703.01351}}.
	
	\bibitem{fialowski1990deformations}
	A.~Fialowski, ``Deformations of some infinite-dimensional {L}ie algebras,''
	{\em J. Math. Phys.} {\bf 31} (1990), no.~6, 1340--1343.
	
	\bibitem{fialowski2003global}
	A.~Fialowski and M.~Schlichenmaier, ``Global deformations of the {W}itt algebra
	of {K}richever--{N}ovikov type,'' {\em Commun. Contemp. Math} {\bf 5} (2003),
	no.~06, 921--945.
	
	\bibitem{schlichenmaier2014elementary}
	M.~Schlichenmaier, ``An elementary proof of the vanishing of the second
	cohomology of the {W}itt and {V}irasoro algebra with values in the adjoint
	module,'' in {\em Forum Math.}, vol.~26, no 3, pp.~913--929, De Gruyter.
	\newblock 2014.
	
	\bibitem{Barnich:2012rz}
	G.~Barnich, A.~Gomberoff, and H.~A. Gonzalez, ``{Three-dimensional
		Bondi-Metzner-Sachs invariant two-dimensional field theories as the flat
		limit of Liouville theory},'' {\em Phys. Rev.} {\bf D87} (2013), no.~12,
	124032,
	\href{http://www.arXiv.org/abs/1210.0731}{{\tt 1210.0731}}.
	
	\bibitem{gel1969cohomologies}
	I.~M. Gel'fand and D.~Fuks, ``Cohomologies of {L}ie algebra of tangential
	vector fields of a smooth manifold,'' {\em Funct. Anal. Appl.} {\bf 3}
	(1969), no.~3, 194--210.
	
	\bibitem{Compere:2013bya}
	G.~Comp\`ere, W.~Song, and A.~Strominger, ``{New Boundary Conditions for
		$\text{AdS}_3$},'' {\em JHEP} {\bf 05} (2013) 152,
	\href{http://www.arXiv.org/abs/1303.2662}{{\tt 1303.2662}}.
	
	\bibitem{Li:2008dq}
	W.~Li, W.~Song, and A.~Strominger, ``{Chiral Gravity in Three Dimensions},''
	{\em JHEP} {\bf 04} (2008) 082,
	\href{http://www.arXiv.org/abs/0801.4566}{{\tt 0801.4566}}.
	
	\bibitem{Banados:1992wn}
	M.~Banados, C.~Teitelboim, and J.~Zanelli, ``{The Black hole in
		three-dimensional space-time},'' {\em Phys. Rev. Lett.} {\bf 69} (1992)
	1849--1851,
	\href{http://www.arXiv.org/abs/hep-th/9204099}{{\tt hep-th/9204099}}.
	
	\bibitem{Afshar:2016kjj}
	H.~Afshar, D.~Grumiller, W.~Merbis, A.~Perez, D.~Tempo, and R.~Troncoso,
	``{Soft hairy horizons in three spacetime dimensions},'' {\em Phys. Rev.}
	{\bf D95} (2017), no.~10, 106005,
	\href{http://www.arXiv.org/abs/1611.09783}{{\tt 1611.09783}}.
	
	\bibitem{gerstenhaber1964deformation}
	M.~Gerstenhaber, ``On the deformation of rings and algebras: I,'' {\em Ann. Of
		Math} (1964) 59--103.
	
	\bibitem{gerstenhaber1966deformation}
	M.~Gerstenhaber, ``On the deformation of rings and algebras: {II},'' {\em Ann.
		Of Math} (1966) 1--19.
	
	\bibitem{gerstenhaber1968deformation}
	M.~Gerstenhaber, ``On the deformation of rings and algebras: {III},'' {\em Ann.
		Of Math} (1968) 1--34.
	
	\bibitem{gerstenhaber1974deformation}
	M.~Gerstenhaber, ``On the deformation of rings and algebras: {IV},'' {\em Ann.
		Of Math.} (1974) 257--276.
	
	\bibitem{nijenhuis1967deformations}
	A.~Nijenhuis and R.~Richardson, ``Deformations of {L}ie algebra structures,''
	{\em J. Math. Mech.} {\bf 17} (1967), no.~1, 89--105.
	
	\bibitem{fialovski1986deformations}
	A.~Fialowski, ``Deformations of {L}ie algebras,'' {\em Sbornik: Mathematics}
	{\bf 55} (1986), no.~2, 467--473.
	
	\bibitem{fialowski1988example}
	A.~Fialowski, ``An example of formal deformations of {L}ie algebras,'' in {\em
		Deformation theory of algebras and structures and applications},
	pp.~375--401.
	\newblock Springer, 1988.
	
	\bibitem{fialowski2005global}
	A.~Fialowski and M.~Schlichenmaier, ``Global geometric deformations of current
	algebras as {K}richever-{N}ovikov type algebras,'' {\em Commun. Math. Phys}
	{\bf 260} (2005), no.~3, 579--612.
	
	\bibitem{guerrini1998formal}
	L.~Guerrini, ``Formal and analytic deformations of the {W}itt algebra,'' {\em
		Lett. Math. Phys.} {\bf 46} (1998), no.~2, 121--129.
	
	\bibitem{guerrini1999formal}
	L.~Guerrini, ``Formal and analytic rigidity of the {W}itt algebra,'' {\em Rev.
		Math. Phys.} {\bf 11} (1999), no.~03, 303--320.
	
	\bibitem{onishchik41lie}
	A.~Onishchik and E.~B. Vinberg, ``Lie groups and {L}ie algebras {III},
	structure of {L}ie groups and {L}ie algebras,'' {\em Series: Encyclopaedia of
		Mathematical Sciences} {\bf 41} (1994).
	
	\bibitem{fuks2012cohomology}
	D.~B. Fuks, {\em Cohomology of infinite-dimensional {L}ie algebras}.
	\newblock Springer Science \& Business Media, 2012.
	
	\bibitem{ChevalleyEilenberg}
	C.~Chevalley and S.~Eilenberg, ``Cohomology theory of {L}ie groups and {L}ie
	algebras,'' {\em Trans. Am. Math. Soc.} {\bf 63} (1948) 85--124.
	
	\bibitem{MR0054581}
	G.~Hochschild and J.-P. Serre, ``Cohomology of {L}ie algebras,'' {\em Ann. of
		Math. (2)} {\bf 57} (1953) 591--603.
	
	\bibitem{Roger:2006rz}
	C.~Roger and J.~Unterberger, ``{The Schrodinger-Virasoro Lie group and algebra:
		From geometry to representation theory},'' {\em Ann. Henri Poincar{\'e}} {\bf
		7} (2006) 1477--1529,
	\href{http://www.arXiv.org/abs/math-ph/0601050}{{\tt math-ph/0601050}}.
	
	\bibitem{nijenhuis1966cohomology}
	A.~Nijenhuis and R.~W. Richardson~Jr, ``Cohomology and deformations in graded
	{L}ie algebras,'' {\em Bull. Amer. Math. Soc.} {\bf 72} (1966), no.~1, 1--29.
	
	\bibitem{goze2006lie}
	M.~Goze, ``Lie algebras: Classification, deformations and rigidity,'' {\em
		arXiv preprint math/0611793} (2006).
	
	\bibitem{richardson1967rigidity}
	R.~Richardson, ``On the rigidity of semi-direct products of {L}ie algebras,''
	{\em Pac. J. Math.} {\bf 22} (1967), no.~2, 339--344.
	
	\bibitem{fialowski-Mc:2005}
	A.~Fialowski and M.~de~Montigny, ``Deformations and contractions of {L}ie
	algebras,'' {\em J. Phys. A.} {\bf 38} (2005), no.~28, 6335.
	
	\bibitem{segal1951class}
	I.~E. Segal {\em et al.}, ``A class of operator algebras which are determined
	by groups,'' {\em Duke Math. J.} {\bf 18} (1951), no.~1, 221--265.
	
	\bibitem{patera1992graded}
	J.~Patera, ``Graded contractions of {L}ie algebras, representations and tensor
	products,'' in {\em AIP Conference Proceedings}, vol.~266 no. 1, pp.~46--54,
	AIP.
	\newblock 1992.
	
	\bibitem{weimar1995contractions}
	E.~Weimar-Woods, ``Contractions of {L}ie algebras: generalized
	{I}n{\"o}n{\"u}--{W}igner contractions versus graded contractions,'' {\em J.
		Math. Phys.} {\bf 36} (1995), no.~8, 4519--4548.
	
	\bibitem{degrijse2009cohomology}
	D.~Degrijse and N.~Petrosyan, ``On cohomology of split {L}ie algebra
	extensions,'' {\em J. Lie Theory} {\bf 22} (2012) 1--15,
	\href{http://www.arXiv.org/abs/0911.0545}{{\tt 0911.0545}}.
	
	\bibitem{hazewinkel2012deformation}
	M.~Hazewinkel and M.~Gerstenhaber, {\em Deformation theory of algebras and
		structures and applications}, vol.~247.
	\newblock Springer Science \& Business Media, 2012.
	
	\bibitem{Christe:1993ij}
	P.~Christe and M.~Henkel, ``{Introduction to Conformal Invariance and its
		Applications to Critical Phenomena},'' {\em Lect. Notes Phys. Monogr.} {\bf
		16} (1993) 1--260,
	\href{http://www.arXiv.org/abs/cond-mat/9304035}{{\tt cond-mat/9304035}}.
	
	\bibitem{Henkel:2012zz}
	M.~Henkel and D.~Karevski, ``{Conformal invariance: An introduction to loops,
		interfaces and stochastic {L}oewner evolution},'' {\em Lect. Notes Phys.}
	{\bf 853} (2012)
	pp.1--189.
	
	\bibitem{unterberger2011schrodinger}
	J.~Unterberger and C.~Roger, {\em The {S}chr{\"o}dinger-{V}irasoro Algebra:
		Mathematical Structure and Dynamical {S}chr{\"o}dinger Symmetries}.
	\newblock Springer Science \& Business Media, 2011.
	
	\bibitem{majumdar1993inonu}
	P.~Majumdar, ``In{\"o}n{\"u}--{W}igner contraction of {K}ac--{M}oody
	algebras,'' {\em J. Math. Phys.} {\bf 34} (1993), no.~5, 2059--2065.
	
	\bibitem{daboul2008gradings}
	C.~Daboul, J.~Daboul, and M.~de~Montigny, ``Gradings and contractions of affine
	{K}ac--{M}oody algebras,'' {\em J. Math. Phys.} {\bf 49} (2008), no.~6,
	063509.
	
	\bibitem{ovsienko1996extensions}
	V.~Y. Ovsienko and C.~Roger, ``Extensions of the {V}irasoro group and the
	{V}irasoro algebra by modules of tensor densities on \emph{S},'' {\em Funct.
		Anal. Appl.} {\bf 30} (1996), no.~4, 290--291.
	
	\bibitem{Bagchi:2012xr}
	A.~Bagchi, S.~Detournay, R.~Fareghbal, and J.~Sim\`on, ``{Holography of 3D Flat
		Cosmological Horizons},'' {\em Phys. Rev. Lett.} {\bf 110} (2013), no.~14,
	141302,
	\href{http://www.arXiv.org/abs/1208.4372}{{\tt 1208.4372}}.
	
	\bibitem{Barnich:2011mi}
	G.~Barnich and C.~Troessaert, ``{BMS charge algebra},'' {\em JHEP} {\bf 12}
	(2011) 105,
	\href{http://www.arXiv.org/abs/1106.0213}{{\tt 1106.0213}}.
	
	\bibitem{Troessaert:2017jcm}
	C.~Troessaert, ``{The BMS4 algebra at spatial infinity},'' {\em Class. Quant.
		Grav.} {\bf 35} (2018), no.~7, 074003,
	\href{http://www.arXiv.org/abs/1704.06223}{{\tt 1704.06223}}.
	
	\bibitem{mccarthy1978lifting}
	P.~J. McCarthy, ``Lifting of projective representations of the
	bondi—metzner—sachs group,'' {\em Proceedings of the Royal Society of
		London. A. Mathematical and Physical Sciences} {\bf 358} (1978), no.~1693,
	141--171.
	
	\bibitem{Compere:2015mza}
	G.~Comp\`ere, K.~Hajian, A.~Seraj, and M.~M. Sheikh-Jabbari, ``{Extremal
		Rotating Black Holes in the Near-Horizon Limit: Phase Space and Symmetry
		Algebra},'' {\em Phys. Lett.} {\bf B749} (2015) 443--447,
	\href{http://www.arXiv.org/abs/1503.07861}{{\tt 1503.07861}}.
	
	\bibitem{Compere:2015bca}
	G.~Comp\`ere, K.~Hajian, A.~Seraj, and M.~M. Sheikh-Jabbari, ``{Wiggling Throat
		of Extremal Black Holes},'' {\em JHEP} {\bf 10} (2015) 093,
	\href{http://www.arXiv.org/abs/1506.07181}{{\tt 1506.07181}}.
	
	\bibitem{Javadinezhad:2017jnv}
	R.~Javadinezhad, B.~Oblak, and M.~M. Sheikh-Jabbari, ``{Near-horizon extremal
		geometries: coadjoint orbits and quantization},'' {\em JHEP} {\bf 04} (2018)
	025,
	\href{http://www.arXiv.org/abs/1712.07627}{{\tt 1712.07627}}.
	
\end{thebibliography}

\end{document}